\documentclass[pra,notitlepage,superscriptaddress,showpacs]{revtex4-1}
\usepackage[T1]{fontenc}
\usepackage{amsmath}
\usepackage{amssymb}
\usepackage{graphicx}
\usepackage{esint}
\begin{document}
\title{Control of nonlinear optical phenomena and spatially structured optical
effects in a four-level quantum system near a plasmonic nanostructure}
\author{Hamid Reza Hamedi}
\email{hamid.hamedi@tfai.vu.lt}

\affiliation{Institute of Theoretical Physics and Astronomy, Vilnius University,
Saul\.etekio 3, Vilnius LT-10257, Lithuania}
\author{Vassilios Yannopapas}
\email{vyannop@mail.ntua.gr}

\affiliation{Department of Physics, National Technical University of Athens, Athens
157 80, Greece}

\author{Emmanuel Paspalakis}
\email{paspalak@upatras.gr }

\affiliation{Materials Science Department, School of Natural Sciences, University
of Patras, Patras 265 04, Greece}
\begin{abstract}
We investigate the nonlinear optical response of a four-level
double-V-type quantum system interacting with a pair of weak probe
fields while located near a two-dimensional array of metal-coated
dielectric nanospheres. Such a quantum system contains a V-type
subsystem interacting with surface plasmons,
and another V-type subsystem interacting with the free-space vacuum. A
distinctive feature of the proposed setup is its sensitivity to the
relative phase of the applied fields when placed near the plasmonic
nanostructure. We demonstrate that due to the presence of the plasmonic
nanostructure, the third-order (Kerr-type) susceptibility for one of the laser
fields can be significantly modified while another probe field is
acting. Moreover, the Kerr nonlinearity of the system can
be controlled and even enhanced by varying the distance of the quantum
system from the plasmonic nanostructure.We also show that the
Kerr nonlinearity of such a system can be controlled by adjusting
the relative phase of the applied fields. The results obtained may find
potential applications in on-chip nanoscale photonic devices. We also study the light-matter interaction in the case where one probe field carries an optical vortex, and
another probe field has no vortex. We demonstrate that due to the
phase sensitivity of the closed-loop double V-type quantum system, the linear and nonlinear
susceptibility of the nonvortex probe beam depends on the azimuthal
angle and orbital angular momentum (OAM) of the vortex probe beam.
This feature is missing in open four-level double V-type quantum system interacting
with free-space vacuum, as no quantum interference occurs in this
case. We use the azimuthal dependence of optical susceptibility of
the quantum system to determine the regions of spatially-structured
transmittance.

\end{abstract}
\maketitle

\section{Introduction}

Recently, it has been revealed that nonlinear optical effects
can be significantly modified and eventually enhanced at the nanoscale when
quantum systems are placed near plasmonic nanostructures. The strong
modification of nonlinear effects is attributed to the large enhancement of the applied electric field,
the substantial modification of the spontaneous decay rate, and the
strong exciton-plasmon coupling for quantum systems near plasmonic
nanostructures. Many interesting phenomena have been pointed out in this research area including gain without inversion \cite{Sadeghi2010,Paspalakis2013JPC,Sadeghi2013PRA,Zhao2014prb,PaspalakisPRB2017plasmon,Kosionis2018},
optical transparency and slow light \cite{PaspalakisPRA2012transparenct,PaspalakisPRA2013plasmon,Wang2013SR},
the manipulation of spontaneous emission \cite{PaspalakisPRL2009,Paspalakispra2011Populationdynamics,Paspalakispra2011,Gu2012plasmonic,Gu2012plasmon},
Fano effects in energy absorption \cite{ZhangPRL2006,ArtusoRRB2010,SinghAPL2011,Paspalakis2012JPCc},
optical bistability \cite{MalyshevPRB2011,NugrohoPRB2015,Fernando2018a}, enhanced second-harmonic generation \cite{Singh2013} and
four-wave mixing \cite{Li2012a,Emmanuel2014,Singh2016prb}.

Kerr nonlinearity, which is proportional to the third-order susceptibility, plays a crucial
role in nonlinear and quantum optics. A large third-order nonlinear
susceptibility \cite{KerrSchmidt1996,WangPRLkerr2001,Gong2006,Hamedi2015Kerr}
is of interest as it can be used for the realization of single-photon
nonlinear devices \cite{KerrPRL1997,KashPRL1999}. However, for many
years experimental research on quantum nonlinear optics has been limited due to the weak nonlinear response of the available materials. Recently, modification, and in particular enhancement, of the Kerr nonlinearity near plasmonic nanostructures
have been proposed and analyzed \cite{PaspalakisJMO2014,Chen2015,Terzis2016,Ren2016,Tohari2016,Kosionis2019}.

A particular quantum system with interesting optical response is the four-level
double-V quantum system. When located near a two-dimensional array
of metal-coated dielectric nanosphere, this scheme exhibits quantum
interference in spontaneous emission \cite{Paspalakispra2011}. Namely, it was shown that optical transparency
associated with slow light \cite{PaspalakisPRA2012transparenct} and
the strongly modified Kerr nonlinearity \cite{PaspalakisJMO2014}
appear in this system when interacting with a single weak probe
beam of light near the periodic plasmonic nanostructure. If the system interacts with two laser fields, an extra degree of control can be realized exploiting the extra field as well as the phase difference of the applied fields. The later gives rise to phase dependent optical effects
\cite{PaspalakisPRA2013plasmon,PaspalakisPRB2017plasmon}. However,
the control of Kerr nonlinearity for this quantum system under the interaction with two laser fields has not been yet analyzed.

Growing attention has recently emerged in the generation of twisted light beams due to their potential
application in quantum information processing \cite{VaziriPRL2003},
optical micromanipulation \cite{Woerdemann2013}, biosciences
\cite{Stevenson2010} and microtrapping and alignment \cite{Macdonald2002}.
Such beams of light (the so-called optical vortices) carry orbital
angular momentum (OAM) with helical wavefronts focusing to rings, rather
than points. The interaction of such structured light beams with cold
atoms results in a plethora of interesting effects, such as light-induced-torque
\cite{LembessisPRA2010}, atom vortex beams \cite{Lembessis2014PRA},
entanglement of OAM states of photon pairs \cite{ChenPRA2008}, OAM-based
four-wave mixing \cite{WalkerPRL2012,DingOL2012}, spatially dependent
electromagnetically induced transparency (EIT) \cite{Liang2012,Radwell2015,Sharma2017,HamediOE2018},
vortex slow light and transfer of optical vortices \cite{Ruseckas2007PRA,HamediPRA2018,RuseckasPRA2011,DuttonPRL2004,Ruseckas2013,Hamedi2019a,MorettiPRA2009,Bortman2001,Hamedi2019b}.

An interesting topic is the interplay of quantum systems near
plasmonic nanostructures and optical vortices. The usage of the
optical vortex beam together with a plasmonic nanostructure may result in
a significant modification of optical response for the quantum system
when compared to the case where the quantum system is just in free
space. To the best of our knowledge, a similar analysis on interaction
of quantum plasmonic nanostructures and structured light has not been
reported.

In the present work, we explore the nonlinear optical properties of the four-level
double-V-type quantum system interacting with a pair
of weak probe fields and placed near a two-dimensional array of
metal-coated dielectric nanospheres. The double-V-type system has two V-type subsystems. The upper V-type subsystem is influenced
by its interaction with localized surface plasmons, while the other
V-type subsystem interacts with the free-space vacuum. By means of a density
matrix method, we calculate the linear and nonlinear optical susceptibilities
for one of the laser fields in the presence of the other field and the plasmonic nanosructure.
We demonstrate that the presence of the
plasmonic nanostructure results in significant modification, and even
enhancement, of the third-order nonlinear susceptibility for one of
the probe fields. We find that the nonlinear optical susceptibility
of the quantum system can be controlled through different external
parameters such as the distance of the quantum system from the nanostructure
as well as the relative phase between applied fields.

We also study the interaction of the double V-type system next to the periodic plasmonic nanostructure with a pair of probe laser fields, where one of the laser fields
carries OAM, while the other probe laser field is a nonvortex
beam. In particular, we study the angular dependence of optical susceptibility of the quantum system.
We show that the azimuthally varying linear and nonlinear patterns
can be controlled though different external parameters such as the
distance of the quantum system from the surface of plasmonic nanostructure and
the vorticity of twisted probe beam. Finally, we demonstrate that such a scheme can be used
to distinguish the OAM state of a weak vortex beam by mapping the
absorption of nonvortex probe field in the transverse spatial profile.

\section{Theoretical model and formulation \label{sec:Theor-Model-and-formulation}}

\begin{figure}
\includegraphics[width=0.5\columnwidth]{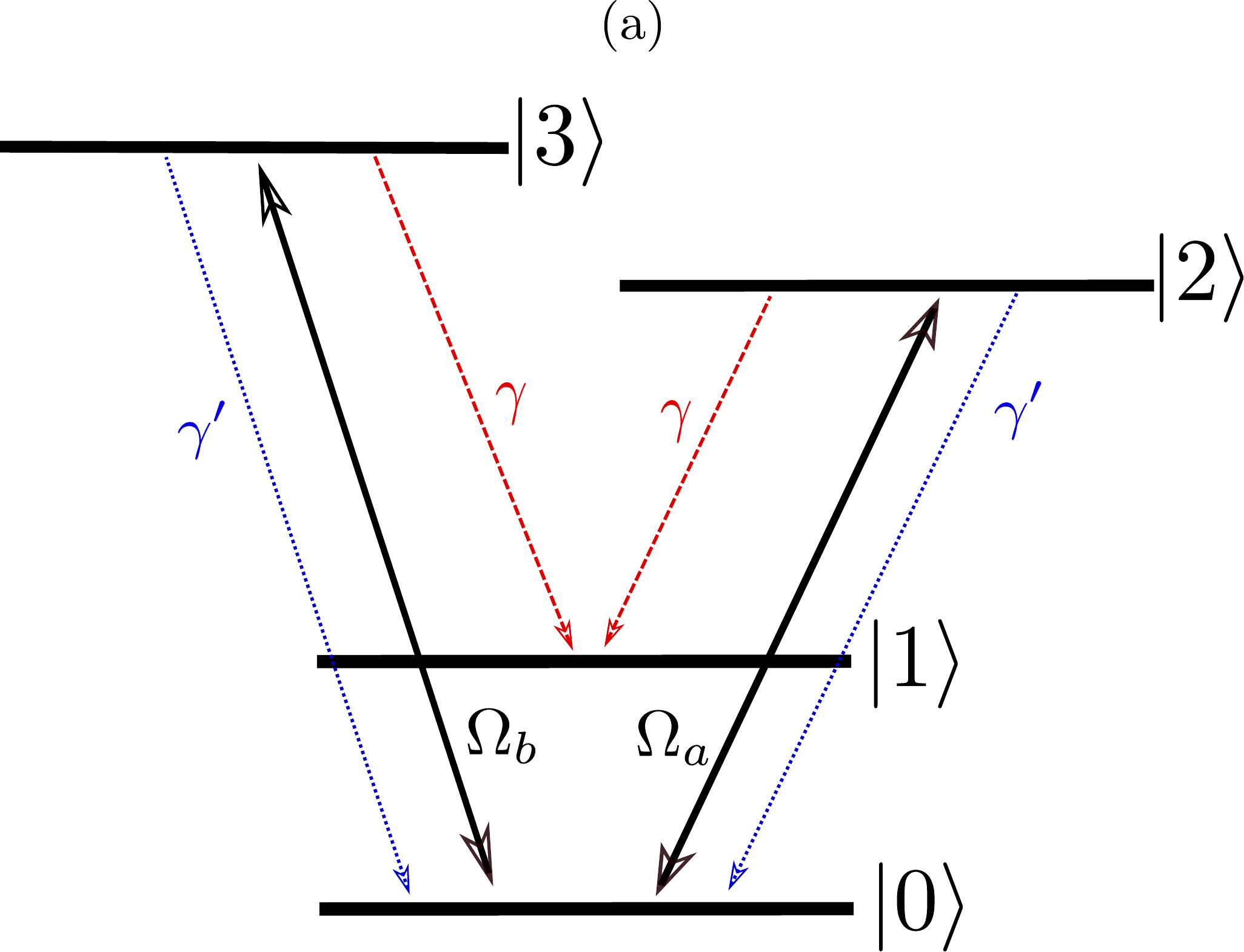}
\includegraphics[width=0.6\columnwidth]{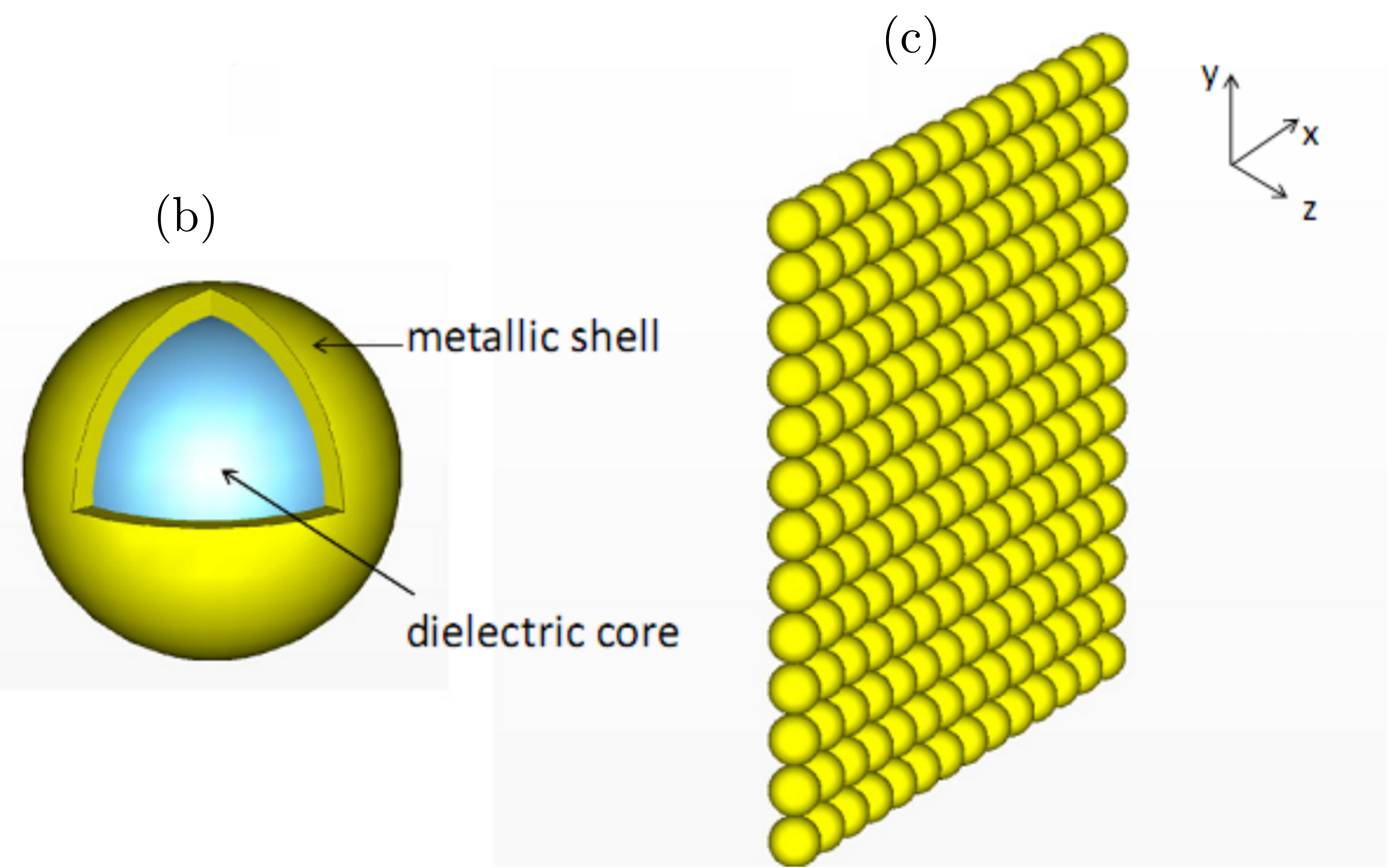}
\caption{Schematic diagram of the four-level double-V-type quantum system (a).
A metal-coated dielectric nanosphere (b) and a 2D array of such spheres
(c).}
\label{fig:scheme}
\end{figure}

The quantum system under study is presented in
Fig.~\ref{fig:scheme}(a): a four-level system containing two
closely lying upper states $|2\rangle$ and $|3\rangle$, and two
lower states $|0\rangle$ and $|1\rangle$, making a four-level
double-V quantum system. The quantum system is in vacuum and at
distance $d$ from the surface of the plasmonic nanostructure. It
is placed right opposite the center of a nanosphere, i.e., at the
center of the 2D unit cell of the (periodic) plasmonic
nanostructure. At this (lateral) placement of the quantum system,
the resulting quantum interference $p$ is maximized. The states
$|2\rangle$ and $|3\rangle$ denote two Zeeman sublevels ($J=1$,
$M_{J}=\pm1$). The two lower states $|0\rangle$ and $|1\rangle$
are corresponding levels with $J=0$. One can define a dipole
moment operator as
\begin{equation}
\overrightarrow{\mu}=\mu^{\prime}(|2\rangle\langle0|\hat{\varepsilon}_{-}+|3\rangle\langle0|\hat{\varepsilon}_{+})+\mu(|2\rangle\langle1|\hat{\varepsilon}_{-}+|3\rangle\langle1|\hat{\varepsilon}_{+}),\label{eq:mu}
\end{equation}
 where $\hat{\varepsilon}_{\pm}=(\mathbf{e}_{z}+i\mathbf{e}_{x})/\sqrt{2}$
stand for the right-rotating ($\hat{\varepsilon}_{+}$) and left-rotating
($\hat{\varepsilon}_{-}$) unit vectors, while $\mu$ and $\mu^{\prime}$
are real.

We assume that the quantum system interacts with two circularly polarized
continuous-wave electromagnetic laser fields with total electric field
\begin{equation}
\overrightarrow{E}(t)=\hat{\varepsilon}_{+}E_{a}\cos(\omega_{a}t+\phi_{a})+\hat{\varepsilon}_{-}E_{b}\cos(\omega_{b}t+\phi_{b}),\label{eq:equatoE}
\end{equation}
where $E_{a}(E_{b})$ characterizes the electric-field amplitude,
$\omega_{a}(\omega_{b})$ denotes the angular frequency, and $\phi_{a}(\phi_{b})$
is the individual phase for the field $a(b)$. The laser field $a$
acts between the lower level $|0\rangle$ and the upper state $|2\rangle$.
The second laser field $b$ couples the lower level $|0\rangle$
to the upper state $|3\rangle$. The transition $|0\rangle\leftrightarrow|1\rangle$
is dipole forbidden. Note that both fields
are taken to have equal frequencies $\omega_{a}=\omega_{b}=\omega_{L}$.

Next, we assume that the upper V-type subsystem containing the states transitions
$|2\rangle$, $|3\rangle$ and $|1\rangle$ lies within the surface-plasmon
bands of the plasmonic nanostructure, whereas the lower V-type subsystem
with states $|2\rangle$, $|3\rangle$ and $|0\rangle$ is spectrally
distant from the surface-plasmon bands, and it is therefore not affected
by the plasmonic nanostructure \cite{Paspalakispra2011}. As a result,
the spontaneous decay in lower V subsystem occurs because of the
interaction of the quantum system with the free-space vacuum electromagnetic
modes. This quantum system can be realized in hyperfine sublevels of D lines in alkali-metal atomic systems, such as  $^{85}$Rb and $^{87}$Rb \cite{PaspalakisPRA2013plasmon,Wang2013SR,Gu2012plasmon}. Similar interactions can also be realized in quantum dots, like in dual CdSe/ZnS/CdSe quantum dots \cite{PaspalakisPRA2013plasmon,Wang2013SR}.

The dynamics of the system is described from the master equation
\begin{equation}
\dot{\rho}_{s}=-\frac{i}{\hbar}[H_{e},\rho_{s}]+\mathcal{L}\rho_{s},\label{eq:masterEq}
\end{equation}
with
\begin{equation}
H_{e}=\hbar\left[(-\delta-\frac{\omega_{32}}{2})|2\rangle\langle2|+(-\delta+\frac{\omega_{32}}{2})|3\rangle\langle3|-\left(\frac{\Omega_{a}e^{i\phi_{a}}}{2}|0\rangle\langle2|+\frac{\Omega_{b}e^{i\phi_{b}}}{2}|0\rangle\langle3|+\mathrm{H.c.}\right)\right], \label{eq:Hinteaction}
\end{equation}
where $\Omega_{a}=\mu^{\prime}E_{a}/\sqrt{2}\hbar$ and $\Omega_{b}=\mu^{\prime}E_{b}/\sqrt{2}\hbar$ are the Rabi frequencies for the two fields. The parameter $\delta=\omega_{L}-\widetilde{\omega}$ is the
detuning from resonance with the average transition energy of states
$|2\rangle$ and $|3\rangle$ from state $|0\rangle$ [$\widetilde{\omega}=(\omega_{2}+\omega_{3})/2-\omega_{0}$]
and $\omega_{32}=(\omega_{3}-\omega_{2})/2$, where $\hbar\omega_{j}=\hbar\omega_{j}$, $j=0-3$ is the energy of state $|j\rangle$. The operator $\mathcal{L}\rho_{s}$
in Eq.~(\ref{eq:masterEq}) represents the dissipation processes
which is given by
\begin{align}
\mathcal{L}\rho_{s} & =\gamma^{\prime}(|0\rangle\langle2|2\rho_{s}|2\rangle\langle0|-|2\rangle\langle2|\rho_{s}-\rho_{s}|2\rangle\langle2|)+\gamma^{\prime}(|0\rangle\langle3|2\rho_{s}|3\rangle\langle0|-|3\rangle\langle3|\rho_{s}-\rho_{s}|3\rangle\langle3|)\nonumber \\
 & +\gamma(|1\rangle\langle2|2\rho_{s}|2\rangle\langle1|-|2\rangle\langle2|\rho_{s}-\rho_{s}|2\rangle\langle2|)+\gamma(|1\rangle\langle3|2\rho_{s}|3\rangle\langle1|-|3\rangle\langle3|\rho_{s}-\rho_{s}|3\rangle\langle3|)\nonumber \\
 & +\kappa(|1\rangle\langle3|2\rho_{s}|2\rangle\langle1|-|2\rangle\langle3|\rho_{s}-\rho_{s}|2\rangle\langle3|)+\kappa(|1\rangle\langle2|2\rho_{s}|3\rangle\langle1|-|3\rangle\langle2|\rho_{s}-\rho_{s}|3\rangle\langle2|)\nonumber \\
 & +\gamma^{\prime\prime}(|0\rangle\langle1|2\rho_{s}|1\rangle\langle0|-|1\rangle\langle1|\rho_{s}-\rho_{s}|1\rangle\langle1|)\,.\label{eq:decay}
\end{align}

The first two terms in Eq.~(\ref{eq:decay}) contain the free-space
spontaneous decay $\gamma^{\prime}=\Gamma_{0}$ \cite{PaspalakisPRB2017plasmon}.
The decay from the two upper states to the lower level is assumed
to be the same. The energy difference of states $|2\rangle$ and $|3\rangle$
is rather small, i.e., $\omega_{32}$ is only a few $\Gamma_{0}$,
where $\Gamma_{0}$ is the decay rate in free space \cite{Paspalakispra2011}.
The term involving $\gamma^{\prime\prime}$ is very small ($\gamma^{\prime\prime}\ll\gamma,\gamma^{\prime}$)
as it arises from a dipole forbidden transition. In this paper we
neglect it by taking $\gamma^{\prime\prime}=0$.

The following equations are obtained for the density
matrix elements by using Eq.~(\ref{eq:masterEq}) which describes
the dynamics of the quantum system
\begin{align}
\dot{\rho}_{20} & =(i\delta+i\frac{\omega_{32}}{2}-\gamma-\gamma^{\prime})\rho_{20}-\kappa\rho_{30}+i\frac{\Omega_{a}}{2}(\rho_{00}-\rho_{22})-i\frac{\Omega_{b}}{2}e^{-i\phi}\rho_{23},\label{eq:d1}\\
\dot{\rho}_{30} & =(i\delta-i\frac{\omega_{32}}{2}-\gamma-\gamma^{\prime})\rho_{30}-\kappa\rho_{20}+i\frac{\Omega_{b}}{2}e^{-i\phi}(\rho_{00}-\rho_{33})-i\frac{\Omega_{a}}{2}\rho_{32},\label{eq:d2}\\
\dot{\rho}_{23} & =(i\omega_{32}-2\gamma-2\gamma^{\prime})\rho_{23}+i\frac{\Omega_{a}}{2}\rho_{03}-i\frac{\Omega_{b}}{2}e^{i\phi}-\kappa(\rho_{22}+\rho_{33}),\label{eq:d3}\\
\dot{\rho}_{00} & =2\gamma^{\prime}(\rho_{22}+\rho_{33})-i\frac{\Omega_{a}}{2}(\rho_{02}-\rho_{20})-i\frac{\Omega_{b}}{2}(\rho_{03}e^{-i\phi}-\rho_{30}e^{i\phi}),\label{eq:d4}\\
\dot{\rho}_{22} & =-2(\gamma+\gamma^{\prime})\rho_{22}+i\frac{\Omega_{a}}{2}(\rho_{02}-\rho_{20})-\kappa(\rho_{23}+\rho_{32}),\label{eq:d5}\\
\dot{\rho}_{33} & =-2(\gamma+\gamma^{\prime})\rho_{33}+i\frac{\Omega_{b}}{2}(\rho_{03}e^{-i\phi}-\rho_{20}e^{i\phi})-\kappa(\rho_{23}+\rho_{32}),\label{eq:d6}
\end{align}
along with the population conservation $\rho_{00}+\rho_{11}+\rho_{22}+\rho_{33}=1$
and $\rho_{ij}=\rho_{ji}^{*}$. The optical coherence corresponding
to the probe transition of $|0\rangle\rightarrow|2\rangle$ ($|0\rangle\rightarrow|3\rangle$)
is $\rho_{20}$($\rho_{30}$), and the relative phase of the applied fields
is denoted by $\phi=\phi_{b}-\phi_{a}$. Note that the probe fields
are assumed to be very weak so that one can treat them as a perturbation.
In the above equations, the parameter $\kappa$ is the coupling coefficient
between states $|2\rangle$ and $|3\rangle$ due to spontaneous emission
in a modified anisotropic vacuum \cite{Agarwal2000} (anisotropic Purcell effect) which is responsible
for the appearance of quantum interference \cite{Kiffner2010}.

The values of $\gamma$ and $\kappa$ are given by \cite{PaspalakisPRL2009,Paspalakispra2011Populationdynamics,Yang2008,Li2009,Jha2015,Hughes2017,Karanikolas2018}
\begin{align}
\gamma & =\frac{\mu_{0}\mu^{2}\bar{\omega}^{2}}{2\hbar}\hat{\varepsilon}_{-}.\,Im\mathbf{G}(\mathbf{r},\mathbf{r};\bar{\omega}).\,\hat{\varepsilon}_{+},\label{eq:4}\\
\kappa & =\frac{\mu_{0}\mu^{2}\bar{\omega}^{2}}{2\hbar}\hat{\varepsilon}_{+}.\,Im\mathbf{G}(\mathbf{r},\mathbf{r};\bar{\omega}).\,\hat{\varepsilon}_{+}.\label{eq:5}
\end{align}
Here, $\mathbf{G}(\mathbf{r},\mathbf{r};\bar{\omega})$ [$\bar{\omega}=(\omega_{3}+\omega_{2})/2-\omega_{1}$]
describes the dyadic electromagnetic Green's tensor,
while $\mathbf{r}$ and $\mu_{0}$ refer to the position of the quantum
emitter and the permeability of vacuum, respectively. One can obtain
the values of $\gamma$ and $\kappa$ from Eqs.~(\ref{eq:4}) and
(\ref{eq:5}) as \cite{PaspalakisPRL2009,Paspalakispra2011Populationdynamics,Yang2008,Li2009,Jha2015,Hughes2017,Karanikolas2018}
\begin{align}
\gamma & =\frac{\mu_{0}\mu^{2}\bar{\omega}^{2}}{2\hbar}\,Im\left[G_{\bot}(\mathbf{r},\mathbf{r};\bar{\omega})+G_{\Vert}(\mathbf{r},\mathbf{r};\bar{\omega})\right]=\frac{1}{2}(\Gamma_{\bot}+\Gamma_{\Vert}),\label{eq:6}\\
\kappa & =\frac{\mu_{0}\mu^{2}\bar{\omega}^{2}}{2\hbar}\,Im\left[G_{\bot}(\mathbf{r},\mathbf{r};\bar{\omega})-G_{\Vert}(\mathbf{r},\mathbf{r};\bar{\omega})\right]=\frac{1}{2}(\Gamma_{\bot}-\Gamma_{\Vert}),\label{eq:7}
\end{align}
where $G_{\bot}(\mathbf{r},\mathbf{r};\bar{\omega})=G_{zz}(\mathbf{r},\mathbf{r};\bar{\omega})$
and $G_{\Vert}(\mathbf{r},\mathbf{r};\bar{\omega})=G_{xx}(\mathbf{r},\mathbf{r};\bar{\omega})$
show components of the electromagnetic Green's tensor,
where the symbol $\bot$($\Vert$) refers to a dipole oriented normal,
along the $z$ axis (parallel, along the $x$ axis) to the surface
of the nanostructure. Let us also define the spontaneous emission
rates normal and parallel to the surface as $\Gamma_{\bot,\Vert}=\mu_{0}\mu^{2}\bar{\omega}^{2}\,Im\left[G_{\bot,\Vert}(\mathbf{r},\mathbf{r};\bar{\omega})\right]/\hbar$. The degree of quantum interference is then given by
\begin{equation}
p=(\Gamma_{\bot}-\Gamma_{\Vert})/(\Gamma_{\bot}+\Gamma_{\Vert}).\label{eq:QI}
\end{equation}
When $p=\pm1$ the maximum quantum interference is obtained in spontaneous
emission \cite{Kiffner2010}. This is achieved by placing the
emitter close to a structure that completely quenches either $\Gamma_{\bot}$ or $\Gamma_{\Vert}$. When the emitter
is placed in vacuum, $\Gamma_{\bot}=\Gamma_{\Vert}$ leading $\kappa=0$,
hence no quantum interference occurs in the system.

The plasmonic nanostructure considered here is a 2D array of touching
metal-coated silica nanospheres {[}see Figs.~\ref{fig:scheme}(b) and
1(c){]}.  The dielectric function of
the shell is provided by a Drude-type electric permittivity
\begin{equation}
\epsilon(\omega)=1-\frac{\omega_{p}^{2}}{\omega(\omega+i/\tau)},\label{eq:8}
\end{equation}
where $\omega_{p}$ is the bulk plasma frequency and $\tau$ the
relaxation time of the conduction-band electrons of the metal. A
typical value of the plasma frequency for gold is
$\hbar\omega_{p}=8.99\,eV$. This also determines the length scale
of the system as $c/\omega_{p}\approx22\,nm$. The dielectric
constant of $SiO_{2}$ is taken to be $\epsilon=2.1$. In the
calculations we have taken $\tau^{-1}=0.05\omega_{p}$. The lattice
constant of the square lattice is $a=2c/\omega_{p}$ and the sphere
radius $S=c/\omega_{p}$ with core radius $S_{c}=0.7c/\omega_{p}$.
Using this particular choice of sphere/ core radius and lattice
constant we achieve maximization of the quantum interference rate
$p$ which prerequisite for the observation of the results present
below.

For the calculation of the spontaneous decay rates next to the plasmonic nanostructure,
we use the layered multiple scattering method \cite{PaspalakisPRL2009,Stefanou1998,Stefanou2000,Sainidou2004}.
We take $\bar{\omega}=0.632\omega_{p}$ while the distance between the
quantum system and the surface of the plasmonic nanostructure, $d$,
varies from $0.5c/\omega_{p}$ to $c/\omega_{p}$. For the results
of $\Gamma_{\bot}$ and $\Gamma_{\Vert}$ that are used here, we refer
to Fig. 3 in Ref. \cite{PaspalakisPRA2012transparenct}. It is found
that $\Gamma_{\Vert}$ gives significant suppression and its
actual value is remarkably lower than the free-space decay rate.
In addition, the value of $\Gamma_{\bot}$ decreases with increasing
distance between the quantum system and the plasmonic nanostructure.
For distances close to the plasmonic nanostructure, $\Gamma_{\bot}$
becomes much larger than the free-space decay rate. The value of $\Gamma_{\bot}$
is larger than the free-space decay rate for distances up to $0.6c/\omega_{p}$,
while for distances between $0.65c/\omega_{p}$ and $c/\omega_{p}$
the value of $\Gamma_{\bot}$ becomes lower than the free-space decay
rate.

\section{Calculation of linear and nonlinear susceptibilities\label{sec:spat-dep-susc}}

In this section we calculate the linear and nonlinear electric susceptibilities
for the laser field $\Omega_{a}$. The probe fields are  weak
enough and are treated as perturbation to the system under steady-state
condition. The method we use extends to third order the method presented in Ref. \cite{WilsonGordon01}, and it is similar to that used in Ref. \cite{Sharma2017}. Under the weak-field approximation, the perturbation approach
is applied to the density-matrix elements, which is expressed
in terms of a perturbative expansion
\begin{equation}
\rho_{ij}=\rho_{ij}^{(0)}+\lambda\rho_{ij}^{(1)}+\lambda^{2}\rho_{ij}^{(2)}+\lambda^{(3)}\rho_{ij}^{(3)}+...,\label{eq:perturbaton}
\end{equation}
where $\lambda$ is a continuously varying parameter ranging from
zero to unity. The constituting terms $\rho_{ij}^{(n)}$ with $n=1,2,3$
are of the n$^{th}$ order in the probe fields. Since the probe fields
are assumed to be weak, the zeroth-order solution is $\rho_{00}^{(0)}=1$,
while the other elements $\rho_{ij}^{(0)}=0$. Replacing
Eq.~(\ref{eq:perturbaton}) into Eqs.~(\ref{eq:d1})-(\ref{eq:d6}),
the equations of motion for the first- and third order density-matrix
elements are given by
\begin{align}
\dot{\rho}_{20}^{(1)}= & (i\delta+i\frac{\omega_{32}}{2}-\gamma-\gamma^{\prime})\rho_{20}^{(1)}-\kappa\rho_{30}^{(1)}+i\frac{\Omega_{a}}{2},\label{eq:l1}\\
\dot{\rho}_{30}^{(1)}= & (i\delta-i\frac{\omega_{32}}{2}-\gamma-\gamma^{\prime})\rho_{30}^{(1)}-\kappa\rho_{20}^{(1)}+i\frac{\Omega_{b}}{2}e^{-i\phi},\label{eq:l2}
\end{align}
and
\begin{align}
\dot{\rho}_{20}^{(3)}= & (i\delta+i\frac{\omega_{32}}{2}-\gamma-\gamma^{\prime})\rho_{20}^{(3)}-\kappa\rho_{30}^{(3)}+i\frac{\Omega_{a}}{2}(\rho_{00}^{(2)}-\rho_{22}^{(2)})-i\frac{\Omega_{b}}{2}e^{-i\phi}\rho_{23}^{(2)},\label{eq:nl1}\\
\dot{\rho}_{30}^{(3)}= & (i\delta-i\frac{\omega_{32}}{2}-\gamma-\gamma^{\prime})\rho_{30}^{(3)}-\kappa\rho_{20}^{(3)}+i\frac{\Omega_{b}}{2}e^{-i\phi}(\rho_{00}^{(2)}-\rho_{33}^{(2)})-i\frac{\Omega_{a}}{2}\rho_{32}^{(2)}.\label{eq:nl2}
\end{align}

After some lengthy but straightforward algebra we get
\begin{align}
\rho_{20}^{(1)} & =i\frac{\Omega_{a}}{2}S_{1}-i\kappa\frac{\Omega_{b}}{2}e^{-i\phi}S_{2},\label{eq:16}\\
\rho_{30}^{(1)} & =i\frac{\Omega_{b}}{2}e^{-i\phi}S_{3}-i\kappa\frac{\Omega_{a}}{2}S_{2}.\label{eq:16-1}
\end{align}
and
\begin{align}
\rho_{20}^{(3)} & =-a_{2}\kappa-a_{1}(i\delta-i\frac{\omega_{32}}{2}-\gamma-\gamma^{\prime}).\label{eq:17}\\
\rho_{30}^{(3)} & =-a_{1}\kappa-a_{2}(i\delta+i\frac{\omega_{32}}{2}-\gamma-\gamma^{\prime}),\label{eq:18-1}
\end{align}
where
\begin{align}
S_{1} & =\frac{(-i\delta+i\frac{\omega_{32}}{2}+\gamma+\gamma^{\prime})}{(-i\delta+i\frac{\omega_{32}}{2}+\gamma+\gamma^{\prime})(-i\delta-i\frac{\omega_{32}}{2}+\gamma+\gamma^{\prime})-\kappa^{2}},\label{eq:s1}\\
S_{2} & =\frac{1}{(-i\delta+i\frac{\omega_{32}}{2}+\gamma+\gamma^{\prime})(-i\delta-i\frac{\omega_{32}}{2}+\gamma+\gamma^{\prime})-\kappa^{2}},\label{eq:s2}\\
S_{3} & =\frac{(-i\delta-i\frac{\omega_{32}}{2}+\gamma+\gamma^{\prime})}{(-i\delta+i\frac{\omega_{32}}{2}+\gamma+\gamma^{\prime})(-i\delta-i\frac{\omega_{32}}{2}+\gamma+\gamma^{\prime})-\kappa^{2}},\label{eq:s3}
\end{align}
and
\begin{align}
a_{1} & =\frac{-i\frac{\Omega_{a}}{2}(\rho_{00}^{(2)}-\rho_{22}^{(2)})-i\frac{\Omega_{b}}{2}e^{-i\phi}\rho_{23}^{(2)}}{(-i\delta+i\frac{\omega_{32}}{2}+\gamma+\gamma^{\prime})(-i\delta-i\frac{\omega_{32}}{2}+\gamma+\gamma^{\prime})-\kappa^{2}},\label{eq:a1}\\
a_{2} & =\frac{-i\frac{\Omega_{b}}{2}e^{-i\phi}(\rho_{00}^{(2)}-\rho_{33}^{(2)})-i\frac{\Omega_{a}}{2}\rho_{32}^{(2)}}{(-i\delta+i\frac{\omega_{32}}{2}+\gamma+\gamma^{\prime})(-i\delta-i\frac{\omega_{32}}{2}+\gamma+\gamma^{\prime})-\kappa^{2}}.\label{eq:a2}
\end{align}

The second-order density matrix elements of Eqs.~(\ref{eq:17}) and (\ref{eq:18-1})
featured in Eqs.~(\ref{eq:a1}) and (\ref{eq:a2}) can be solved
to obtain the steady-state solutions $\rho_{ij}^{(2)}$ (see Appendix~\ref{sec:appendix-A}).
In order to obtain the linear susceptibility $\chi^{(1)}$ and the
third-order nonlinear susceptibility $\chi^{(3)}$, the susceptibility
is assumed to be written as
\begin{equation}
\chi\approx\chi^{(1)}+3\chi^{(3)}E_{a}^{2}/4.\label{eq:susc}
\end{equation}
Then, using
\begin{equation}
\chi(\delta)=\frac{\sqrt{2}N\mu^{\prime}}{\varepsilon_{0}E_{a}}\rho_{20},\label{eq:susc2}
\end{equation}
and expanding $\rho_{20}$ in perturbation series we get
\begin{equation}
\chi^{(1)}(\delta)=\frac{\sqrt{2}N\mu^{\prime}}{\varepsilon_{0}E_{a}}\rho_{20}^{(1)}=\frac{N\mu^{\prime2}}{\varepsilon_{0}\hbar}\frac{\rho_{20}^{(1)}}{\Omega_{a}},\label{eq:linearSuc}
\end{equation}
and
\begin{equation}
\chi^{(3)}(\delta)E_{a}^{2}=\frac{4N\mu^{\prime2}}{3\varepsilon_{0}\hbar}\frac{\rho_{20}^{(3)}}{\Omega_{a}}.\label{eq:NonlinearSusc}
\end{equation}

Substituting Eqs.~(\ref{eq:s1})-(\ref{eq:a2}) [and using Eqs.~(\ref{eq:AA1})-(\ref{eq:AAs})]
into equations~(\ref{eq:16}), (\ref{eq:17}) and defining $x=\frac{\Omega_{b}}{\Omega_{a}}$, Eqs.~(\ref{eq:linearSuc}) and (\ref{eq:NonlinearSusc}) become
\begin{equation}
\chi^{(1)}(\delta)=\frac{N\mu^{\prime2}}{\varepsilon_{0}\hbar}\frac{-i\kappa A+B(\delta-\frac{\omega_{32}}{2}+i\gamma+i\gamma^{\prime})}{(-i\delta+i\frac{\omega_{32}}{2}+\gamma+\gamma^{\prime})(-i\delta-i\frac{\omega_{32}}{2}+\gamma+\gamma^{\prime})-\kappa^{2}},\label{eq:LS}
\end{equation}
and
\begin{equation}
\chi^{(3)}(\delta)=\frac{2N\mu^{\prime4}}{3\varepsilon_{0}\hbar^{3}}\frac{-i\kappa C+D(\delta-\frac{\omega_{32}}{2}+i\gamma+i\gamma^{\prime})}{(-i\delta+i\frac{\omega_{32}}{2}+\gamma+\gamma^{\prime})(-i\delta-i\frac{\omega_{32}}{2}+\gamma+\gamma^{\prime})-\kappa^{2}},\label{eq:NLS}
\end{equation}
where here $\varepsilon_{0}$ is the vacuum permittivity and $N$
is the density of the quantum systems, where $A$, $B$,
$C$ and $D$ are defined in Appendix~\ref{sec:appendix-B}.

The refraction part of the third-order susceptibility $\chi^{(3)}$
corresponds to the Kerr nonlinearity, while its imaginary part determines
the nonlinear absorption. The real and imaginary parts of $\chi^{(1)}$
represent the linear dispersion and absorption, respectively. From Eqs.~(\ref{eq:LS}) and (\ref{eq:NLS}) one can clearly see that the
expressions for the linear and nonlinear susceptibility are very similar
in form with the only difference in their coefficients. So, one may expect
to observe similar variation of the curves for $\chi^{(1)}$ and $\chi^{(3)}$
with the difference in their magnitude. However, this does not happen as the coefficients of the linear susceptibility does not depend on the detuning $\delta$ and the coefficients of the nonlinear susceptibility depends strongly on the detuning $\delta$, so the frequency variation of the two susceptibilities is different. In addition, one can see that
the linear and nonlinear susceptibilities $\chi^{(1)}$ and $\chi^{(3)}$
can be controlled by the system parameters such as the relative phase
of applied fields $\phi$.

\section{Phase dependent nonlinear optical effects }

\begin{figure}
\includegraphics[width=0.4\columnwidth]{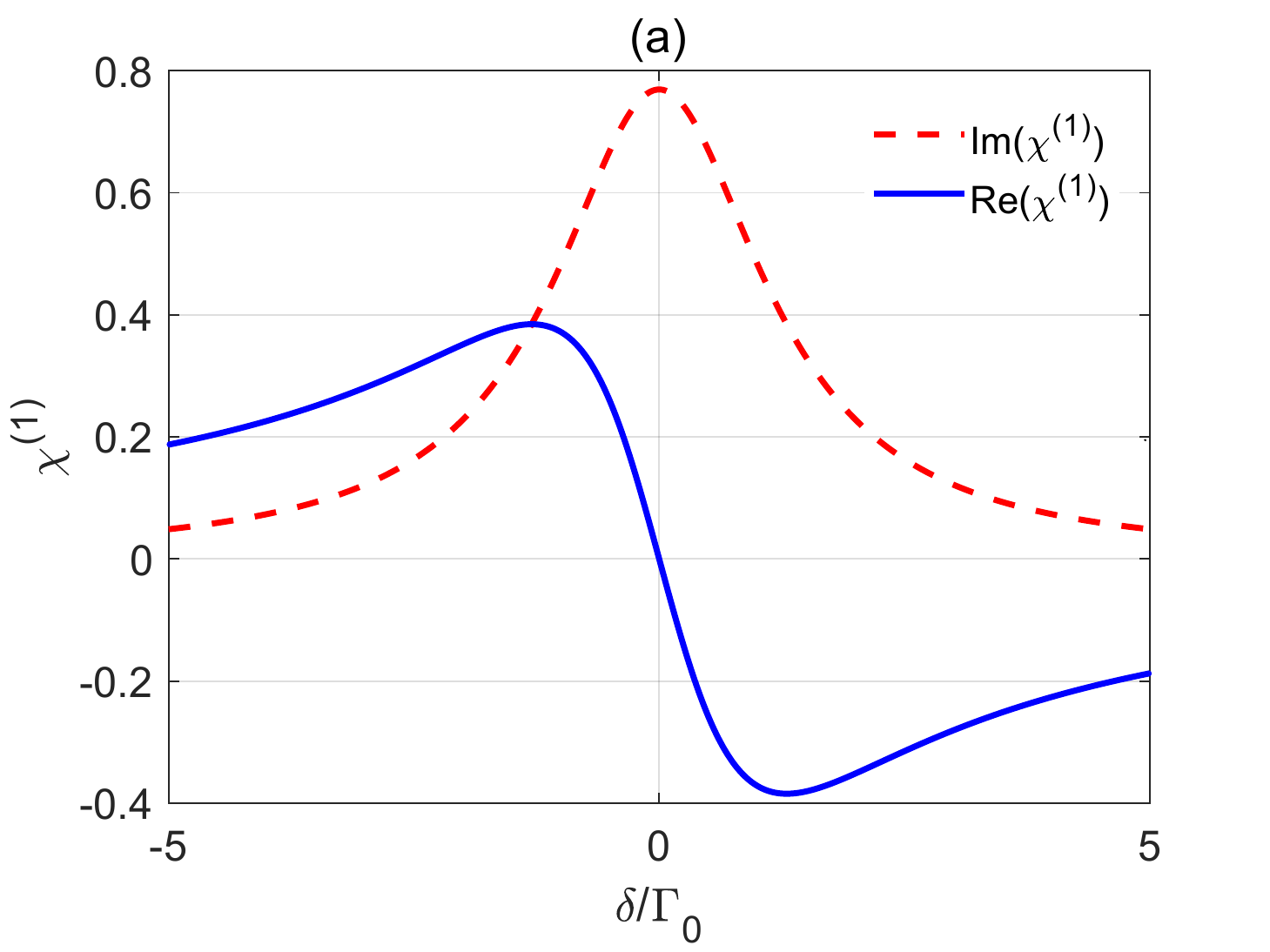} \includegraphics[width=0.4\columnwidth]{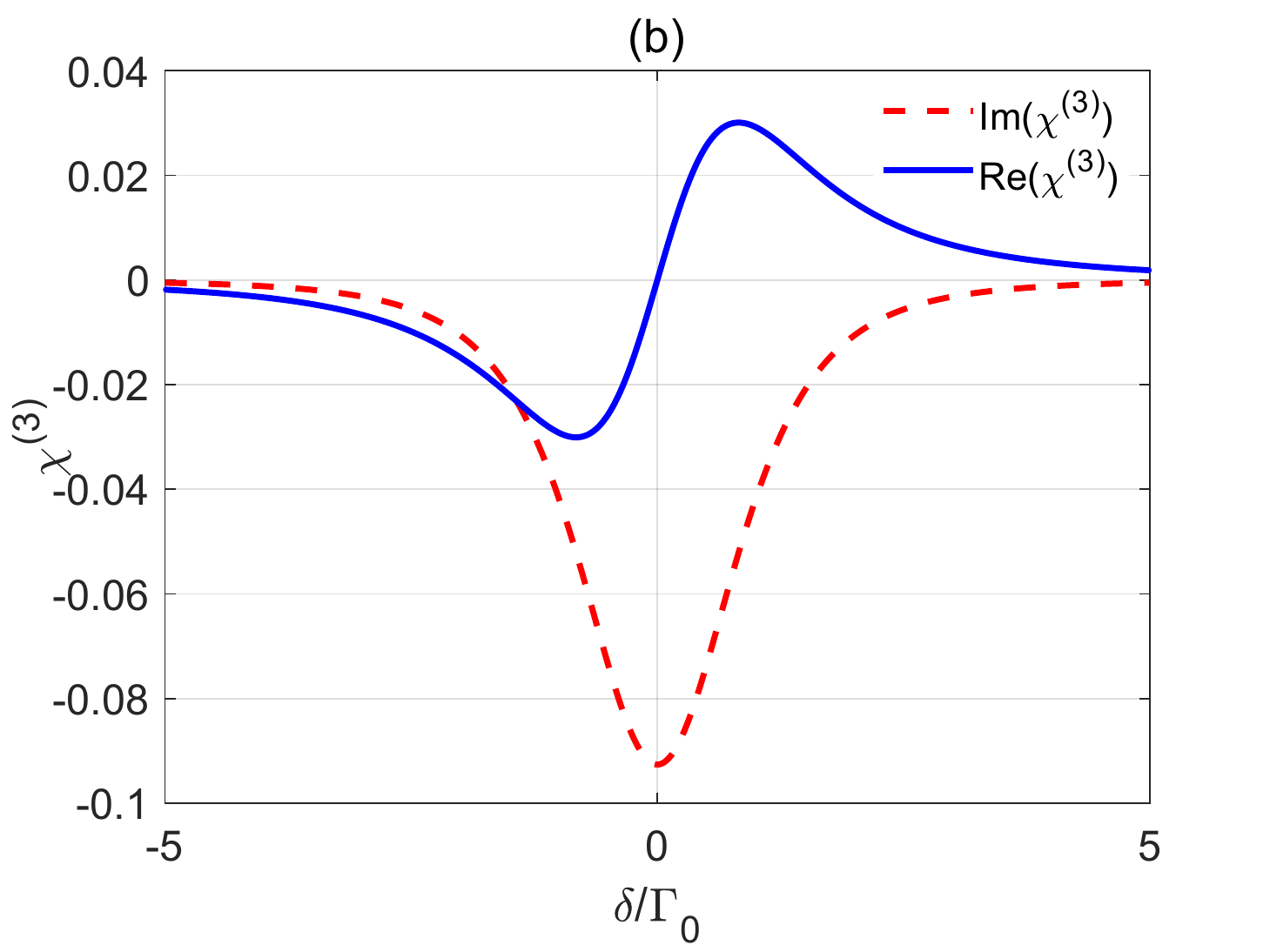}

\caption{(a) Linear susceptibility $\chi^{(1)}$ and (b) nonlinear susceptibility
$\chi^{(3)}$ of the quantum system for the weak probe field $\Omega_{a}$
in arbitrary units as a function of the probe detuning $\delta$ in
the absence of the plasmonic nanostructure. We have assumed that $\omega_{32}=0$,
$\gamma^{\prime}=0.3\Gamma_{0}$ and $\gamma^{\prime\prime}=0$. }
\label{fig:fig2}
\end{figure}

Next we study the nonlinear response of the quantum system to the
probe field $\Omega_{a}$ for weak intensities via numerical
simulation (the linear and nonlinear susceptibilities are plotted
in units of $\frac{N\mu^{\prime2}}{\varepsilon_{0}\hbar}$ and $\frac{2N\mu^{\prime4}}{3\varepsilon_{0}\hbar^{3}}$
, respectively). Figure.~\ref{fig:fig2} shows the real and imaginary
parts of $\chi^{(1)}$ and $\chi^{(3)}$ as a function of the detuning
$\delta$ when the quantum system is in vacuum, i.e., without the
plasmonic nanostructure. We assume that the two upper levels are degenerate
($E_{2}=E_{3}$ leading to $\omega_{32}=0$). This assumption significantly
simplifies Eqs.~(\ref{eq:LS}) and (\ref{eq:NLS}) giving (for $\delta=0$)
analytical expressions for the linear as well as nonlinear absorption
and dispersion coefficients (see Appendix~\ref{sec:appendix-C}).
The typical linear [Fig.~\ref{fig:fig2}(a)] and nonlinear [Fig.~\ref{fig:fig2}(b)]
susceptibility spectra for this case are such that
that the medium experiences strong linear and nonlinear absorption
at $\delta=0$. This is already expected from Eqs.~(\ref{eq:LS})
and (\ref{eq:NLS}) when the quantum system is not near the plasmonic
nanostructure ($\kappa=0$ and $\gamma=\Gamma_{0}$). Setting $\kappa=0$
and $\gamma=\Gamma_{0}$ into Eqs.~(\ref{eq:AC1})- (\ref{eq:AC4}),
one can simplify these equations giving the resonant linear and nonlinear
absorption and dispersion coefficients

\begin{equation}
Im(\chi^{(1)}(\delta=0))=\frac{N\mu^{\prime2}}{\varepsilon_{0}\hbar}\frac{1}{(\Gamma_{0}+\gamma^{\prime})},\label{eq:imL-1}
\end{equation}

\begin{equation}
Re(\chi^{(1)}(\delta=0))=0,\label{eq:reL-1}
\end{equation}

\begin{equation}
Im(\chi^{(3)}(\delta=0))=\frac{2N\mu^{\prime4}}{3\varepsilon_{0}\hbar^{3}}\frac{1-x^{2}}{8(\Gamma_{0}+\gamma^{\prime})^{2}},\label{eq:ImN-1}
\end{equation}

\begin{equation}
Re(\chi^{(3)}(\delta=0))=0.\label{eq:ReN-1}
\end{equation}

On exact resonance, the Kerr nonlinearity is zero for the quantum
system [see Fig.~\ref{fig:fig2}(b) and (Eq.~(\ref{eq:ReN-1}))],
while its magnitude is very weak around the resonance accompanied
by the linear (Eq.~(\ref{eq:imL-1})) and nonlinear (Eq.~(\ref{eq:ImN-1}))
absorption. The slope of linear dispersion is negative around zero
probe detuning suggesting superluminal light propagation [Fig.~\ref{fig:fig2}(a)].
We note that no phase dependence is obtained in this case.

\begin{figure}
\includegraphics[width=0.3\columnwidth]{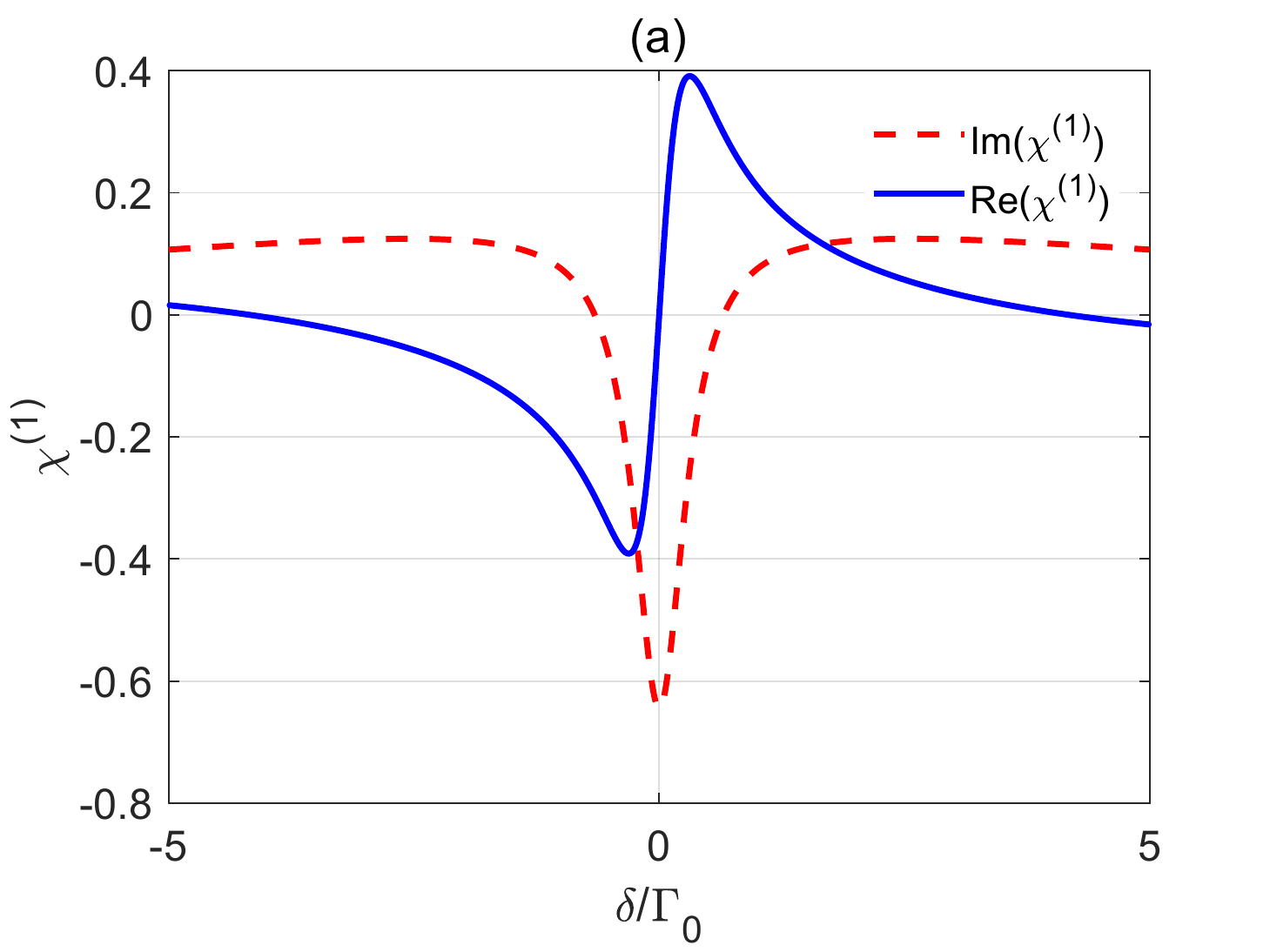} \includegraphics[width=0.3\columnwidth]{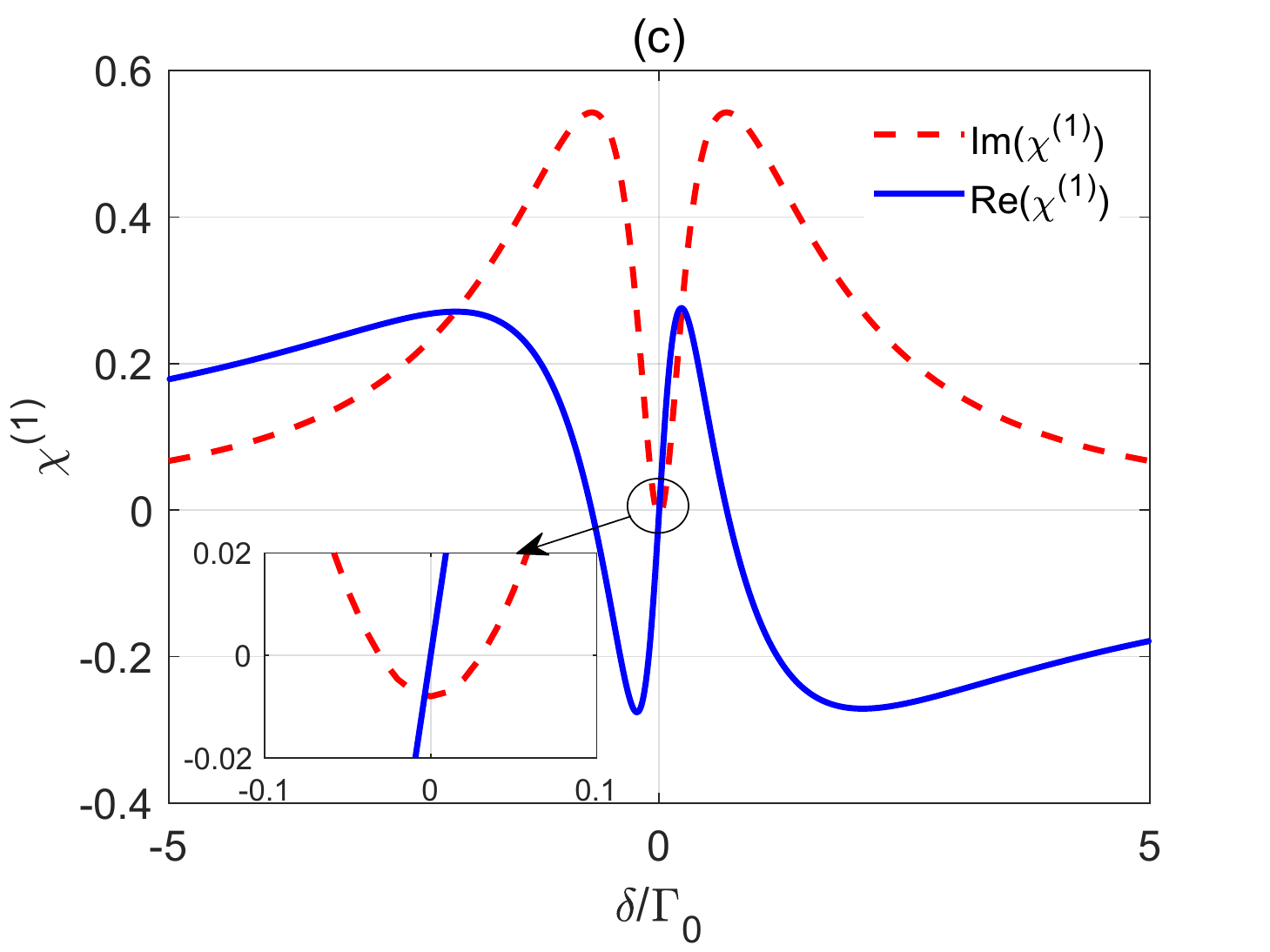}

\includegraphics[width=0.3\columnwidth]{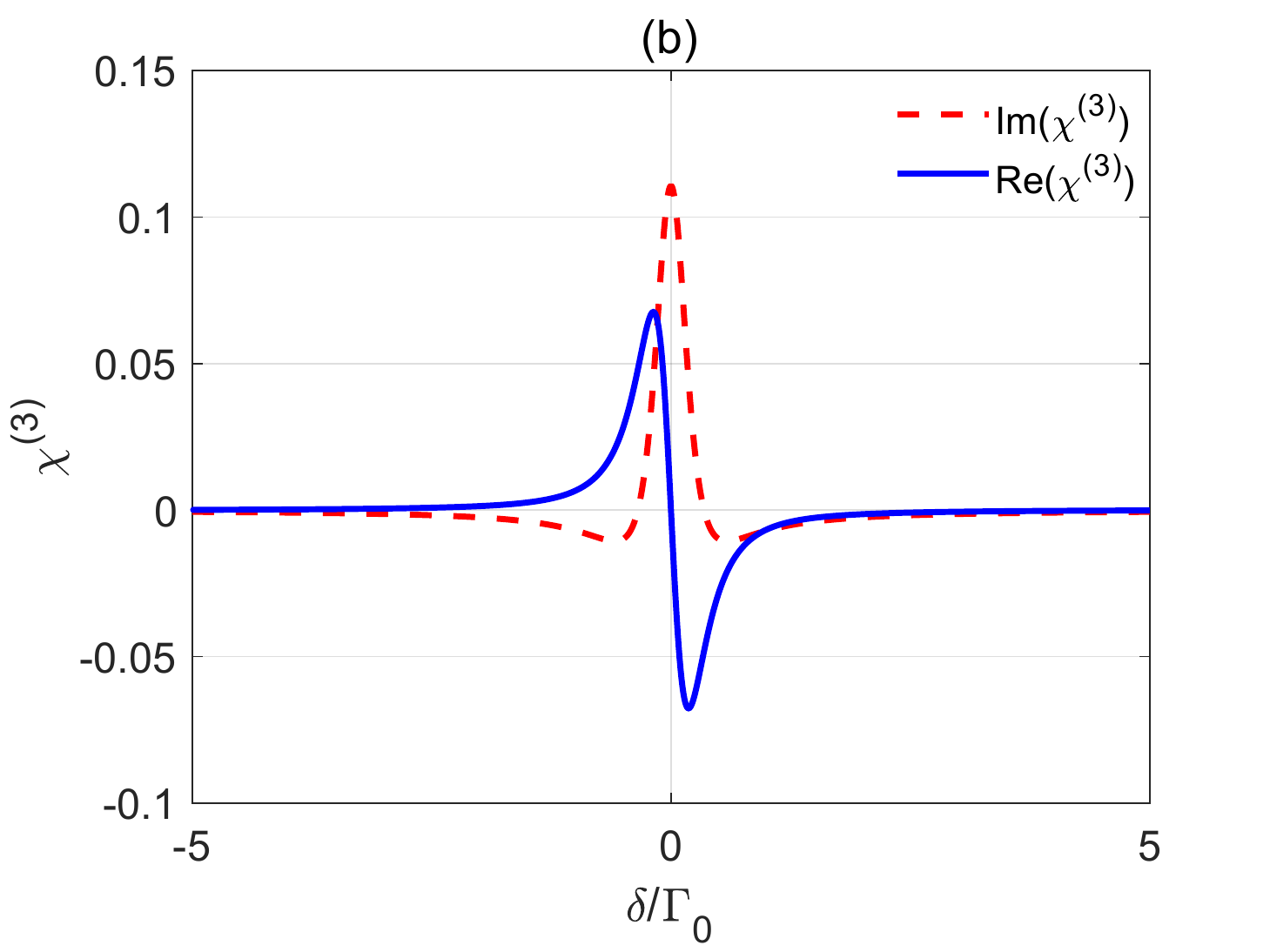}\includegraphics[width=0.3\columnwidth]{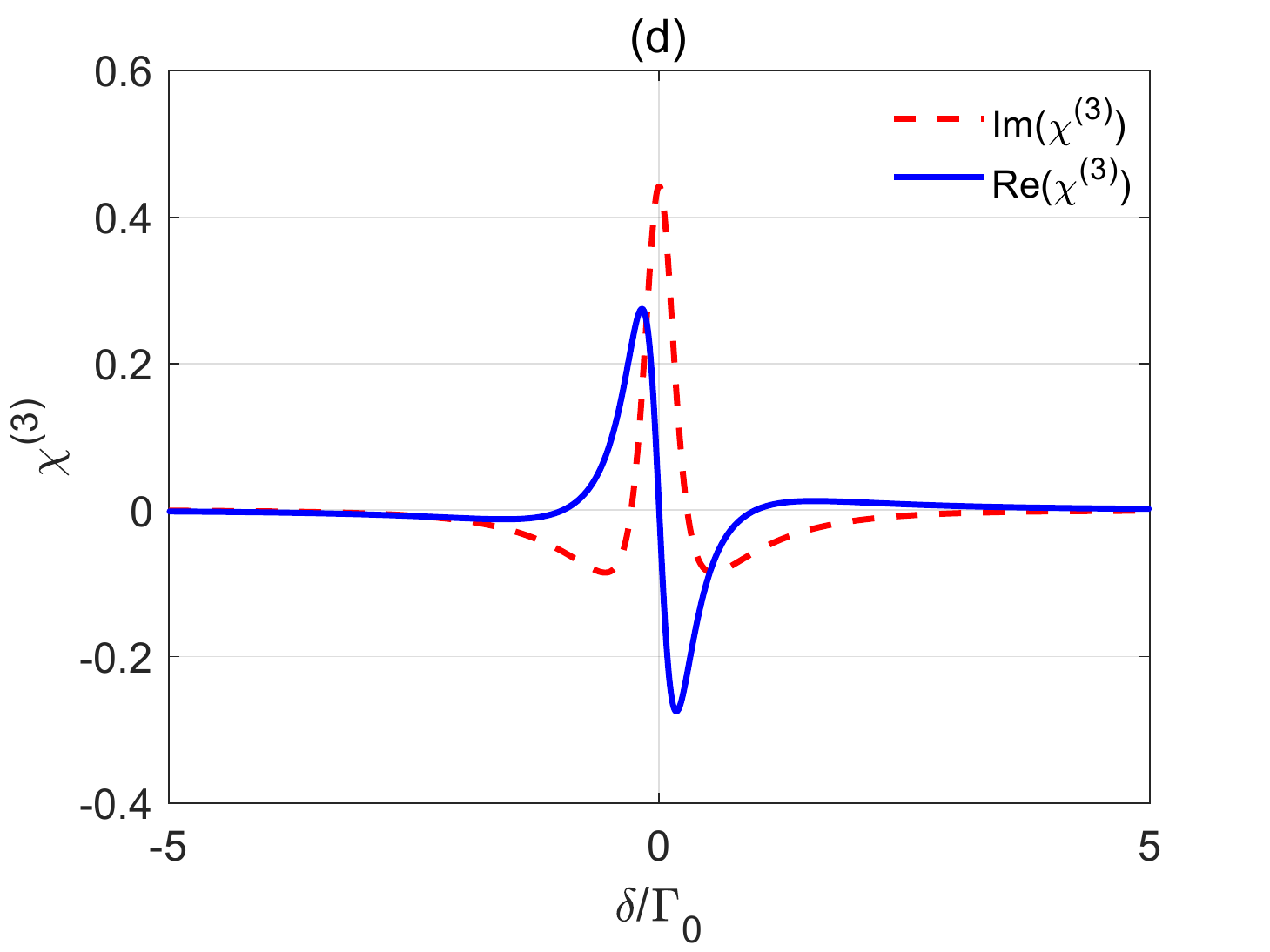}

\caption{(a,c) Linear susceptibility $\chi^{(1)}$ and (b,d) nonlinear susceptibility
$\chi^{(3)}$ of the quantum system for the weak probe field $\Omega_{a}$
in arbitrary units as a function of the probe detuning $\delta$ in
the presence of the plasmonic nanostructure. We take here $\omega_{32}=0$,
$\gamma^{\prime}=0.3\Gamma_{0}$, $\gamma^{\prime\prime}=0$, $x=1.5$,
$\phi=0$, $\bar{\omega}=0.632\omega_{p}$, and $d=0.3c/\omega_{p}$
(a,b), $d=0.6c/\omega_{p}$ (c,d). }
\label{fig:figs3}
\end{figure}

The linear and nonlinear optical properties of the quantum system
are very different when the quantum system is placed near the plasmonic
nanostructure. In Figs.~ \ref{fig:figs3}(a) and (c) where the quantum
system is near the plasmonic nanostructure , we obtain a gain dip
in the linear absorption profile at $\delta=0$. The slope of linear dispersion
becomes positive, indicating slow light condition. As shown in Figs.~\ref{fig:figs3}(b)
and (d), the enhanced Kerr nonlinearity appears inside the linear
gain regions. The maximal Kerr nonlinearity around resonance is enhanced by almost four times when the distance between
the quantum emitter and the nanostructure increases from $d=0.3c/\omega_{p}$
[Figs.~\ref{fig:figs3}(b)] to $d=0.6c/\omega_{p}$ [Figs.~\ref{fig:figs3}(d)].

\begin{figure}
\includegraphics[width=0.3\columnwidth]{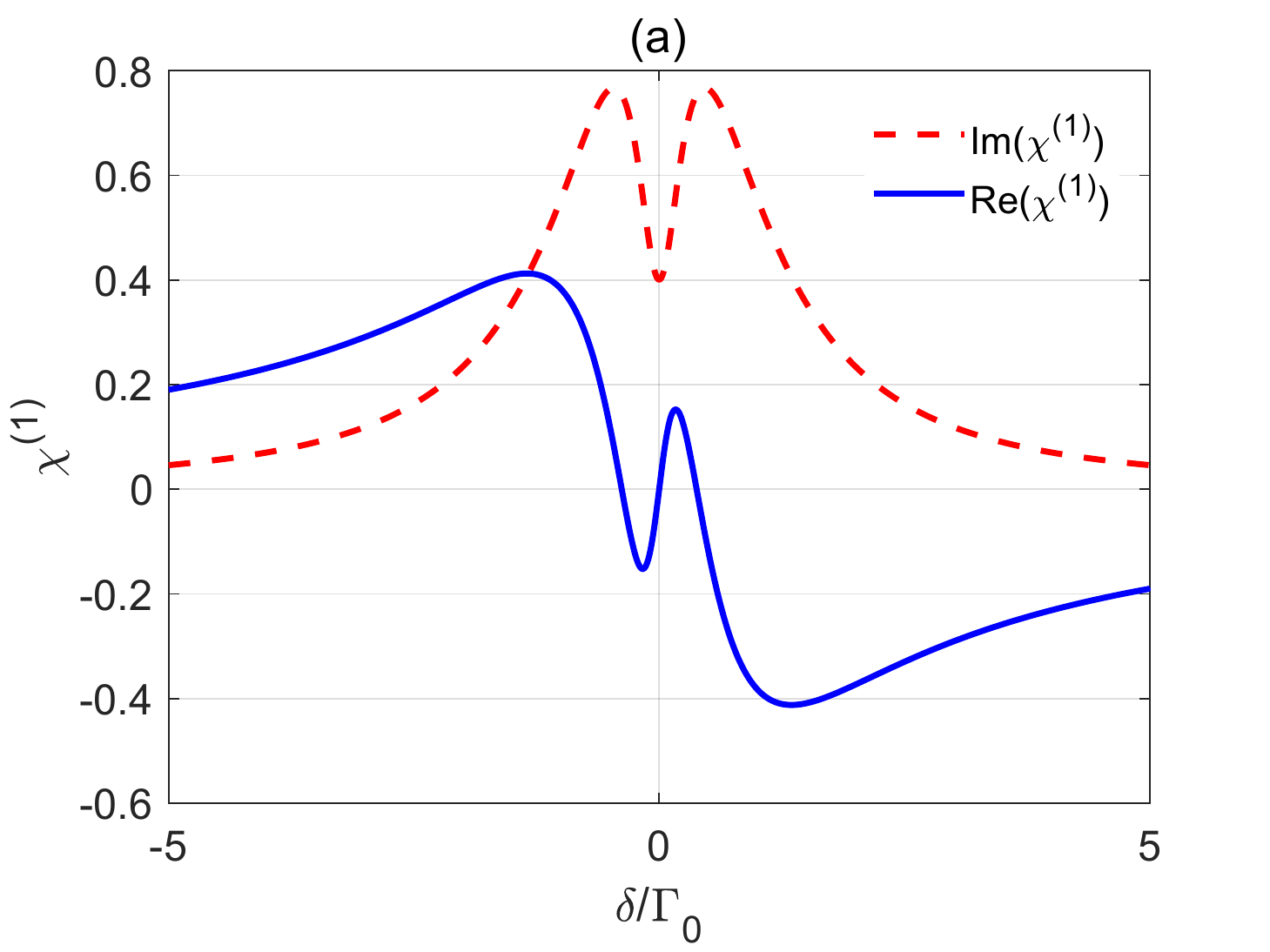} \includegraphics[width=0.3\columnwidth]{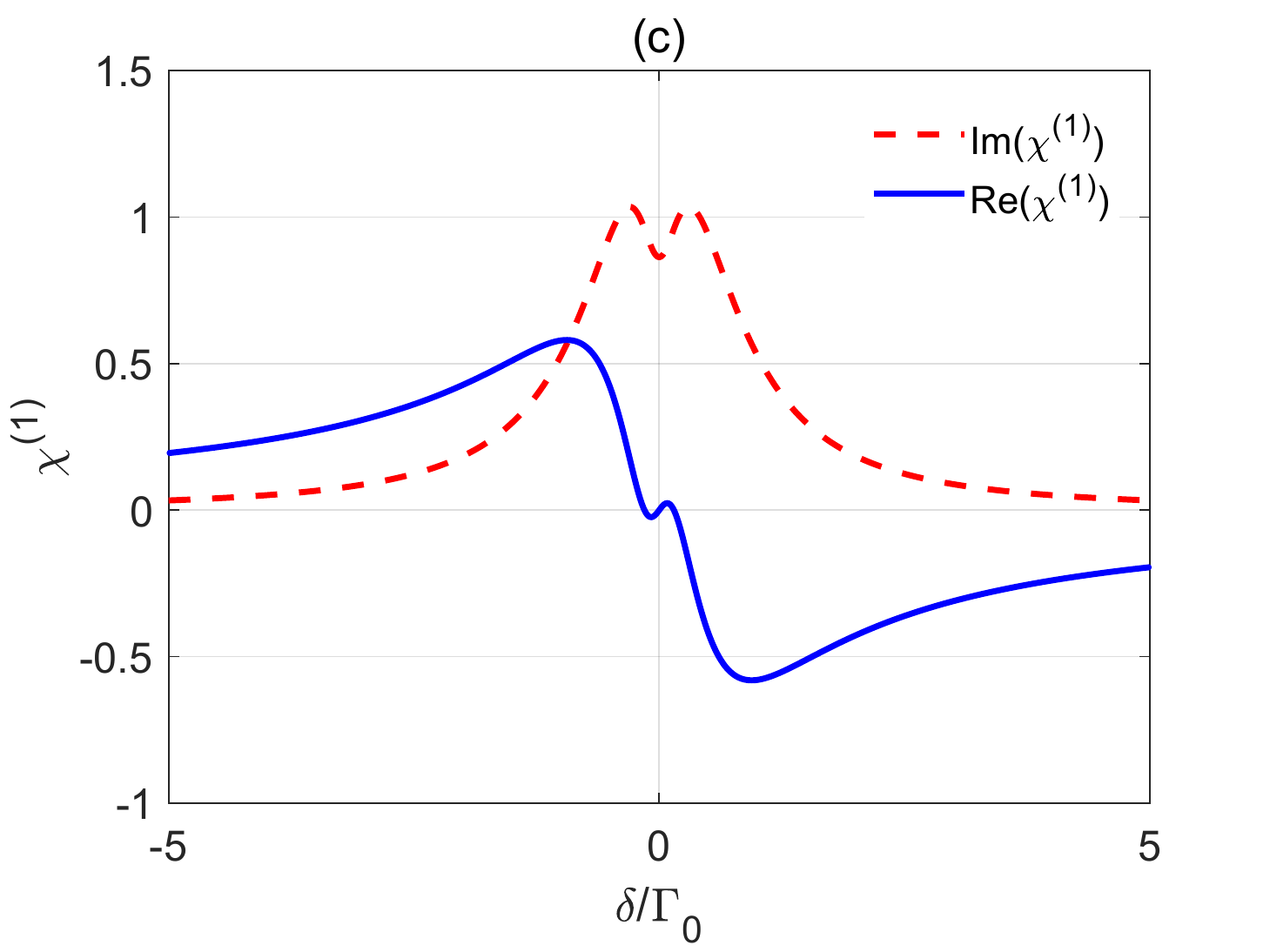}

\includegraphics[width=0.3\columnwidth]{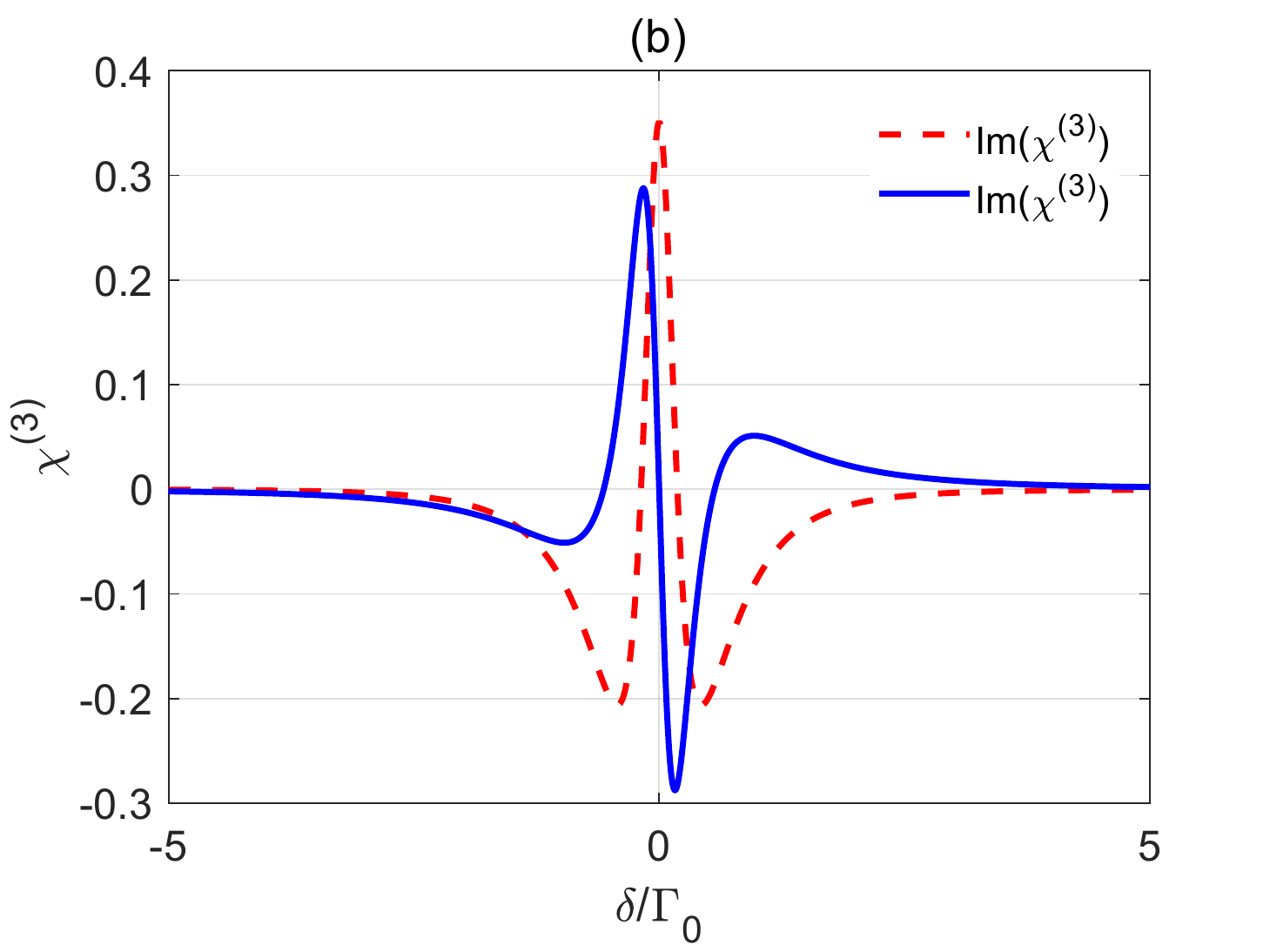}~\includegraphics[width=0.3\columnwidth]{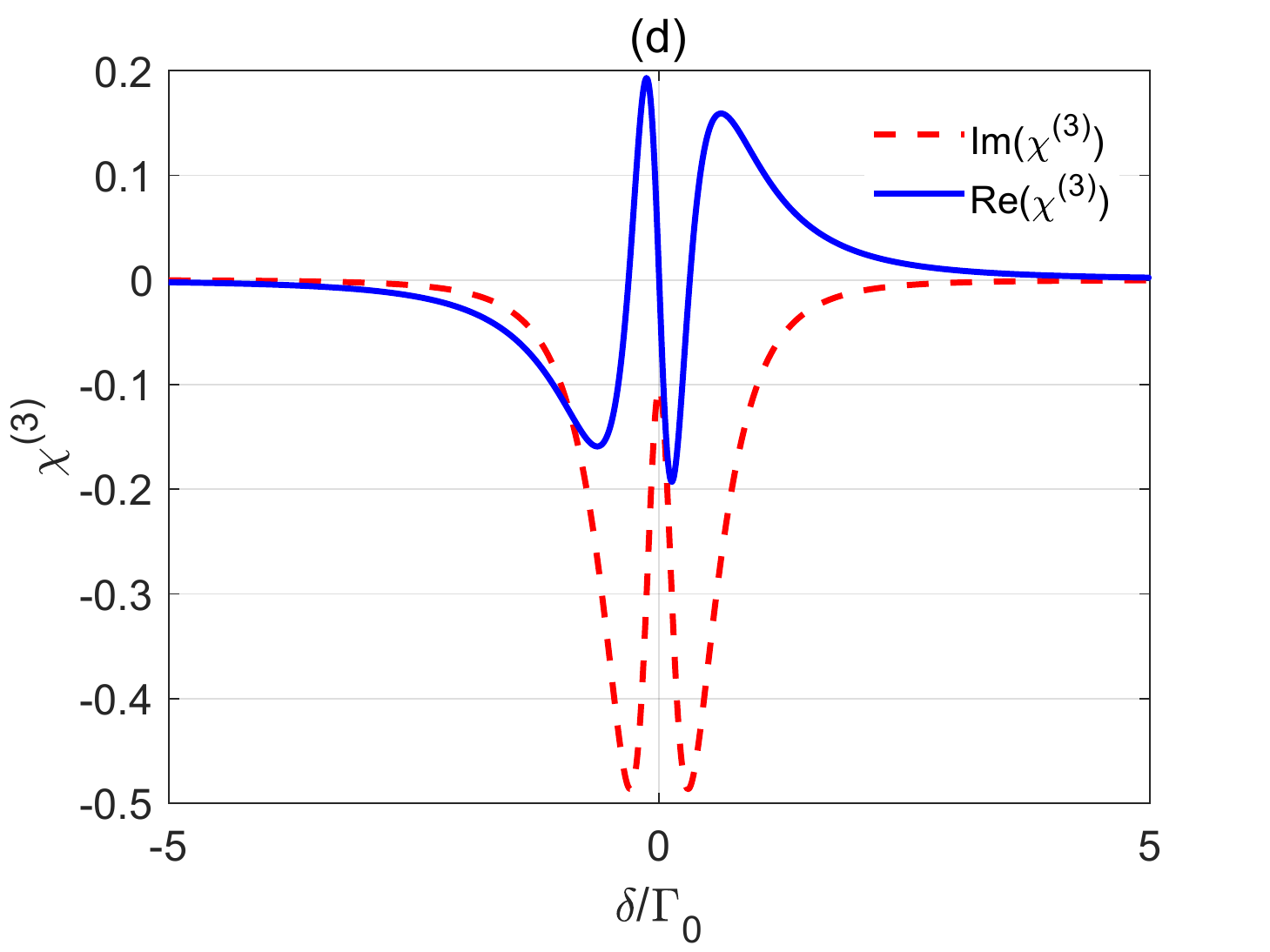}

\caption{(a,c) Linear susceptibility $\chi^{(1)}$ and (b,d) nonlinear susceptibility
$\chi^{(3)}$ of the quantum system for the weak probe field $\Omega_{a}$
in arbitrary units as a function of the probe detuning $\delta$ in
the presence of the plasmonic nanostructure. We take here $\omega_{32}=0$,
$\gamma^{\prime}=0.3\Gamma_{0}$, $\gamma^{\prime\prime}=0$, $x=1.5$,
$\phi=0$, $\bar{\omega}=0.632\omega_{p}$, and $d=0.7c/\omega_{p}$
(a,b), $d=0.8c/\omega_{p}$ (c,d). }
\label{fig:figs4}
\end{figure}

\begin{figure}
\includegraphics[width=0.3\columnwidth]{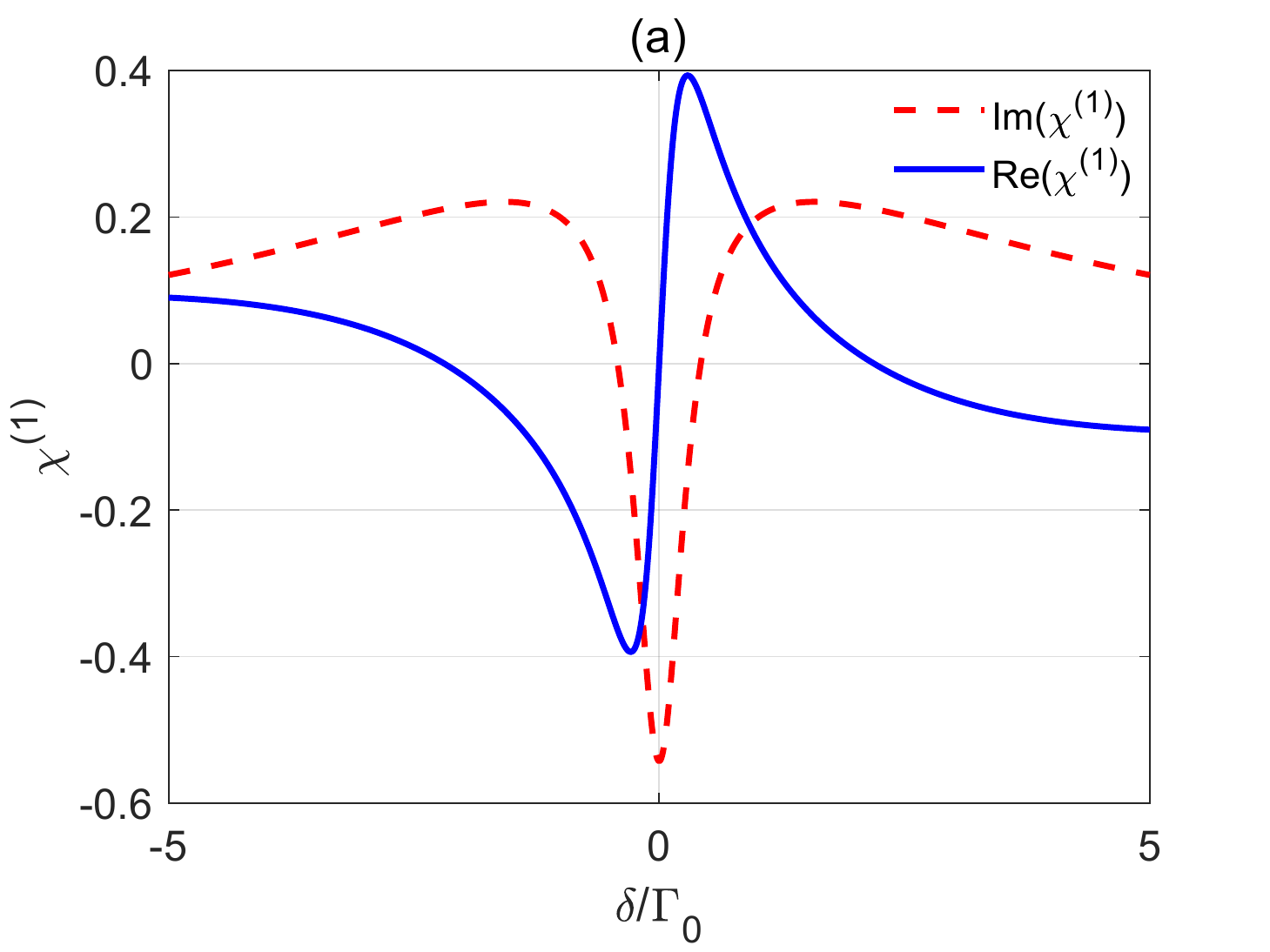} \includegraphics[width=0.3\columnwidth]{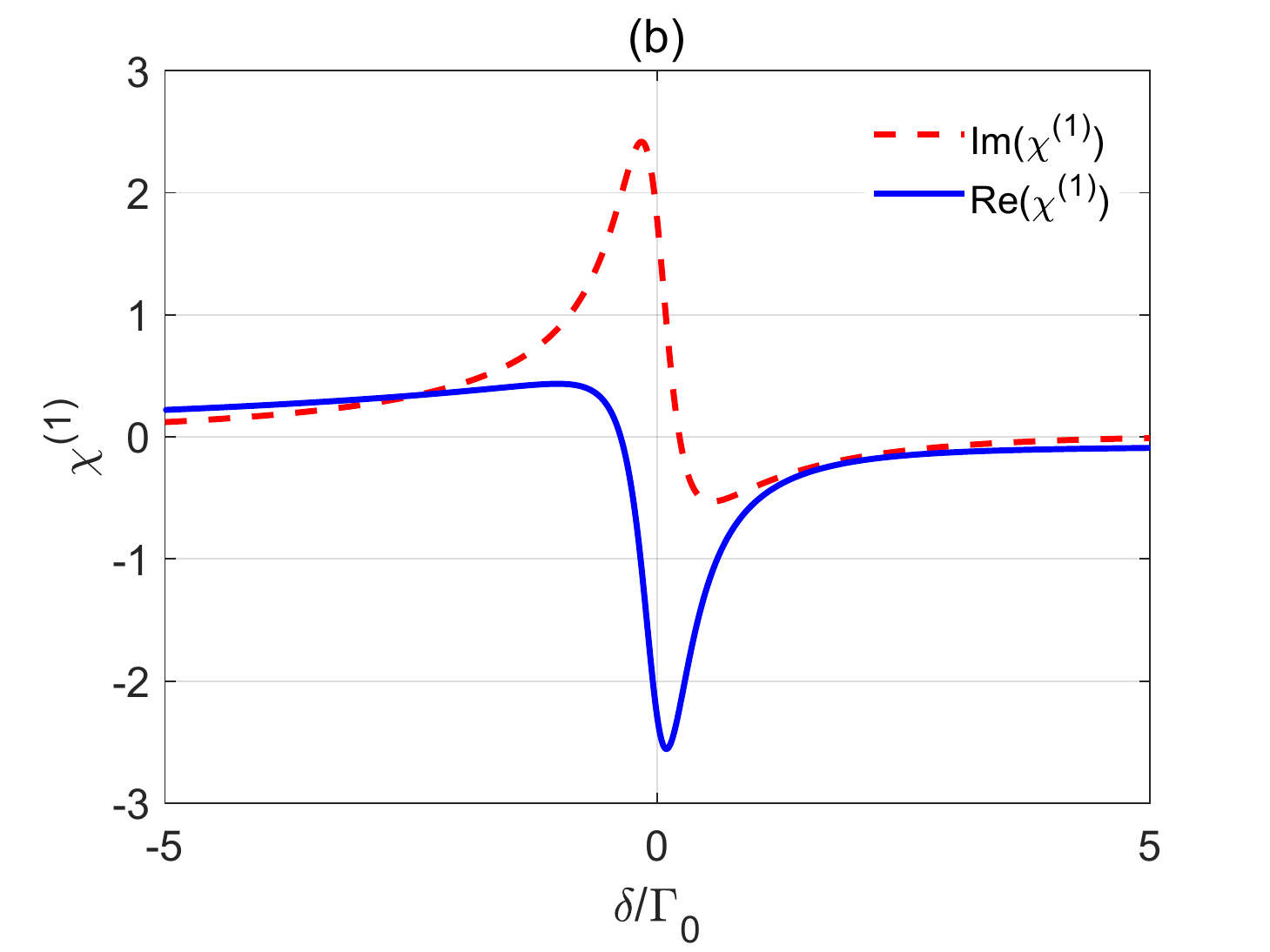}

\includegraphics[width=0.3\columnwidth]{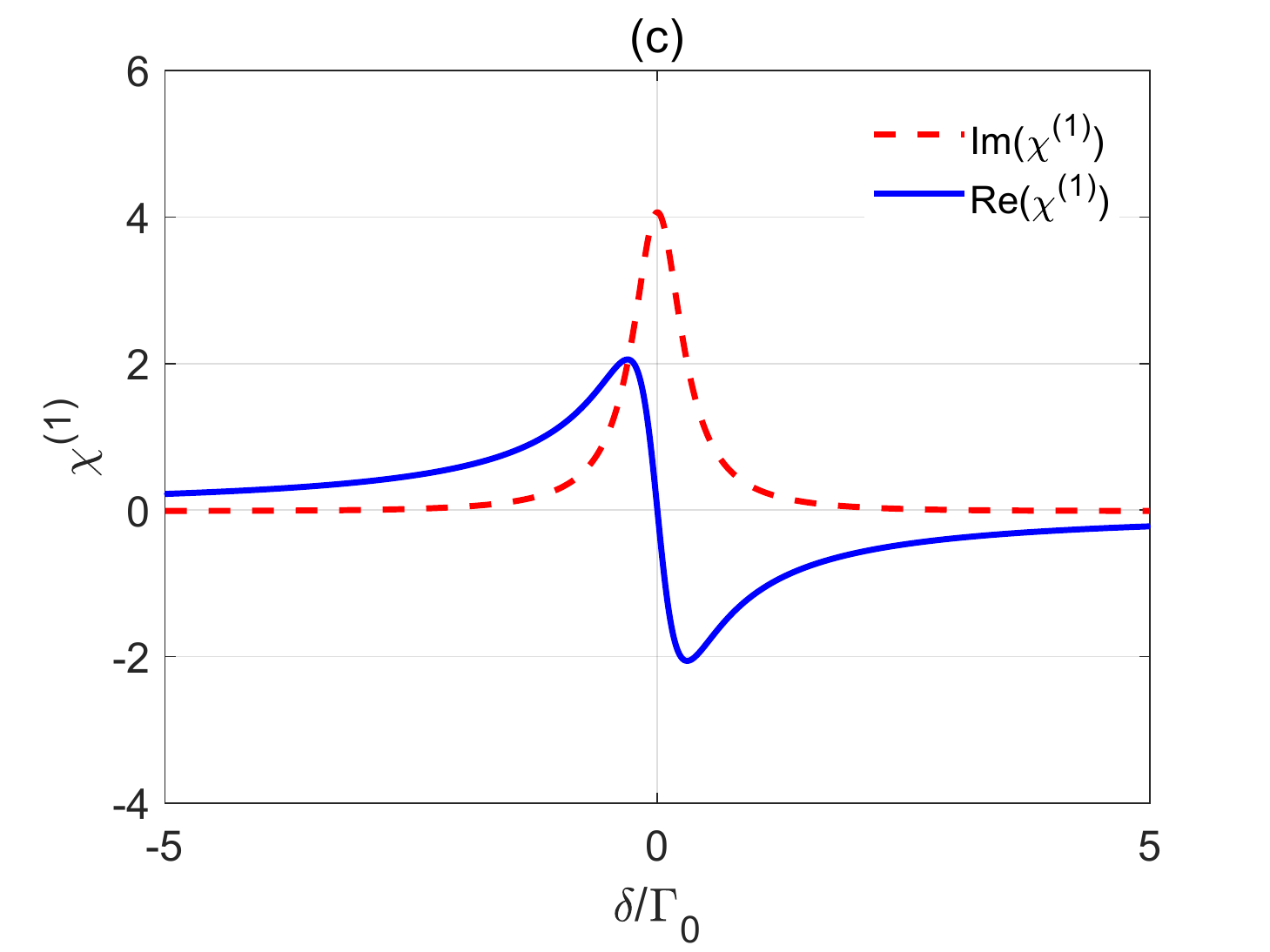}~\includegraphics[width=0.3\columnwidth]{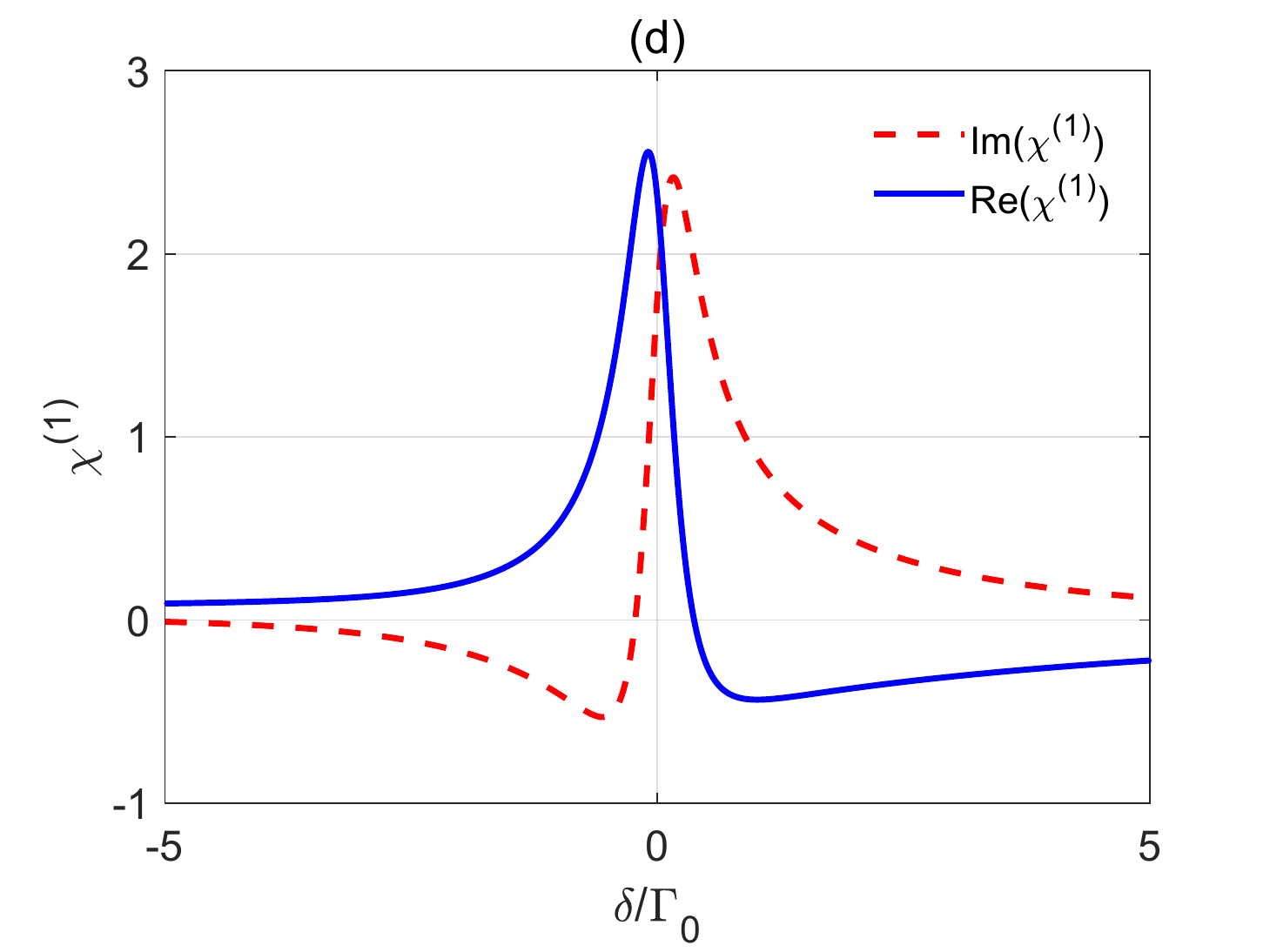}

\caption{Linear susceptibility $\chi^{(1)}$ of the quantum system for the
weak probe field $\Omega_{a}$ in arbitrary units as a function of
the probe detuning $\delta$ in the presence of the plasmonic nanostructure.
We have assumed that $\omega_{32}=0$, $\gamma^{\prime}=0.3\Gamma_{0}$, $\gamma^{\prime\prime}=0$,
$x=1.5$, $\phi=0$, $\bar{\omega}=0.632\omega_{p}$, $d=0.4c/\omega_{p}$
and (a), $\phi=0$, (b), $\phi=\pi/2$, (c) $\phi=\pi$, and (d) $\phi=3\pi/2$. }
\label{fig:figs5}
\end{figure}

\begin{figure}
\includegraphics[width=0.3\columnwidth]{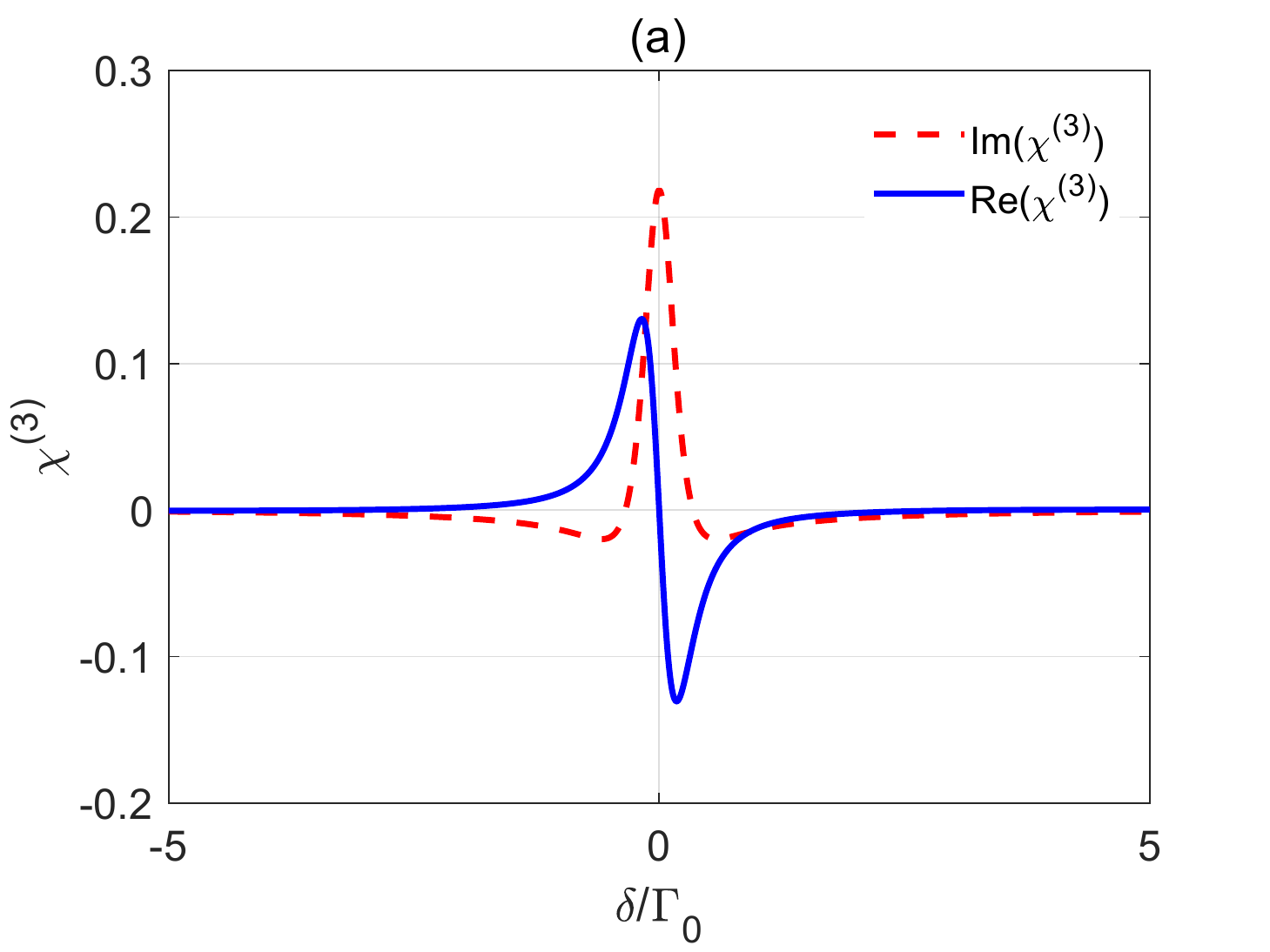} \includegraphics[width=0.3\columnwidth]{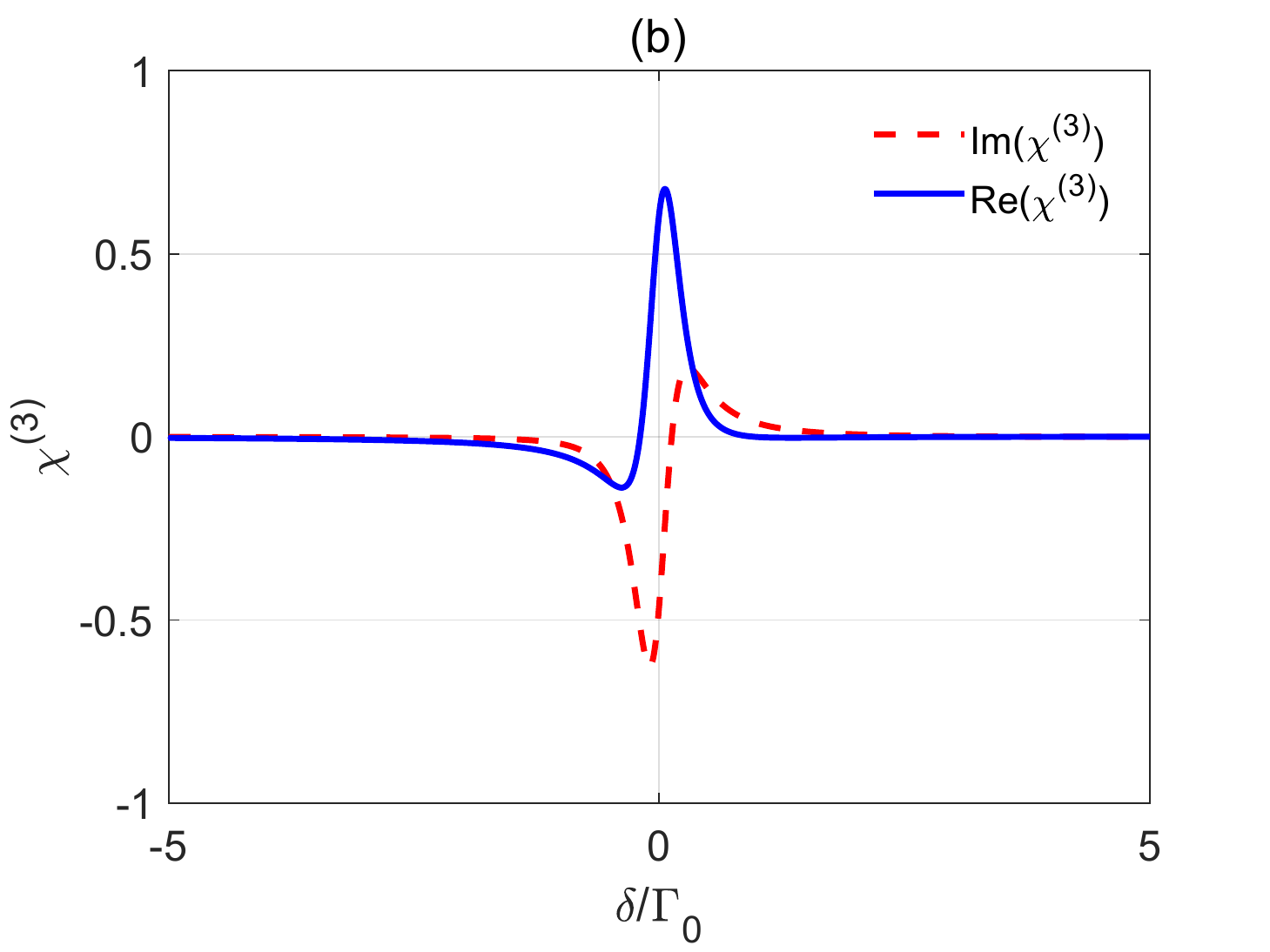}

\includegraphics[width=0.3\columnwidth]{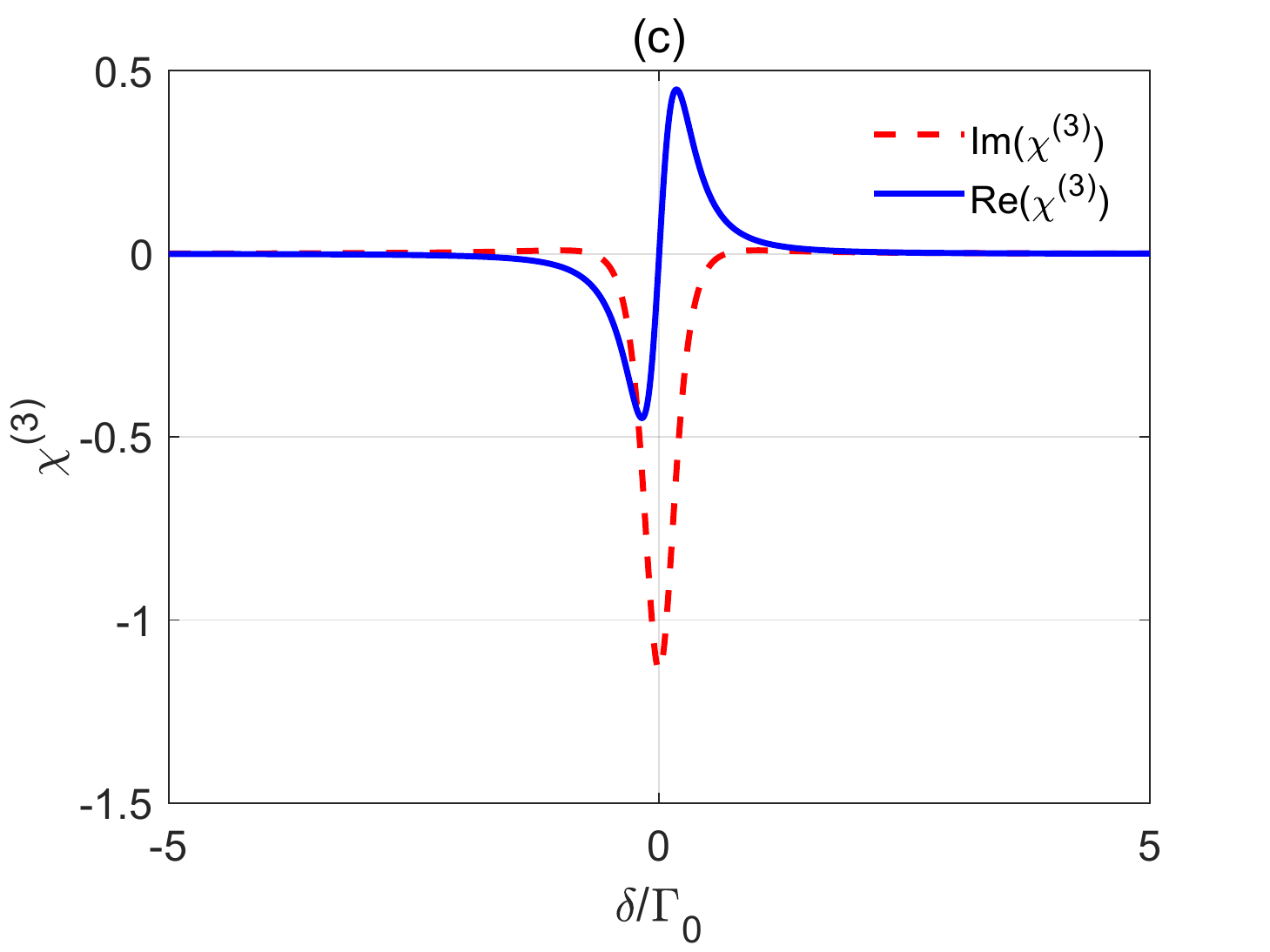}~\includegraphics[width=0.3\columnwidth]{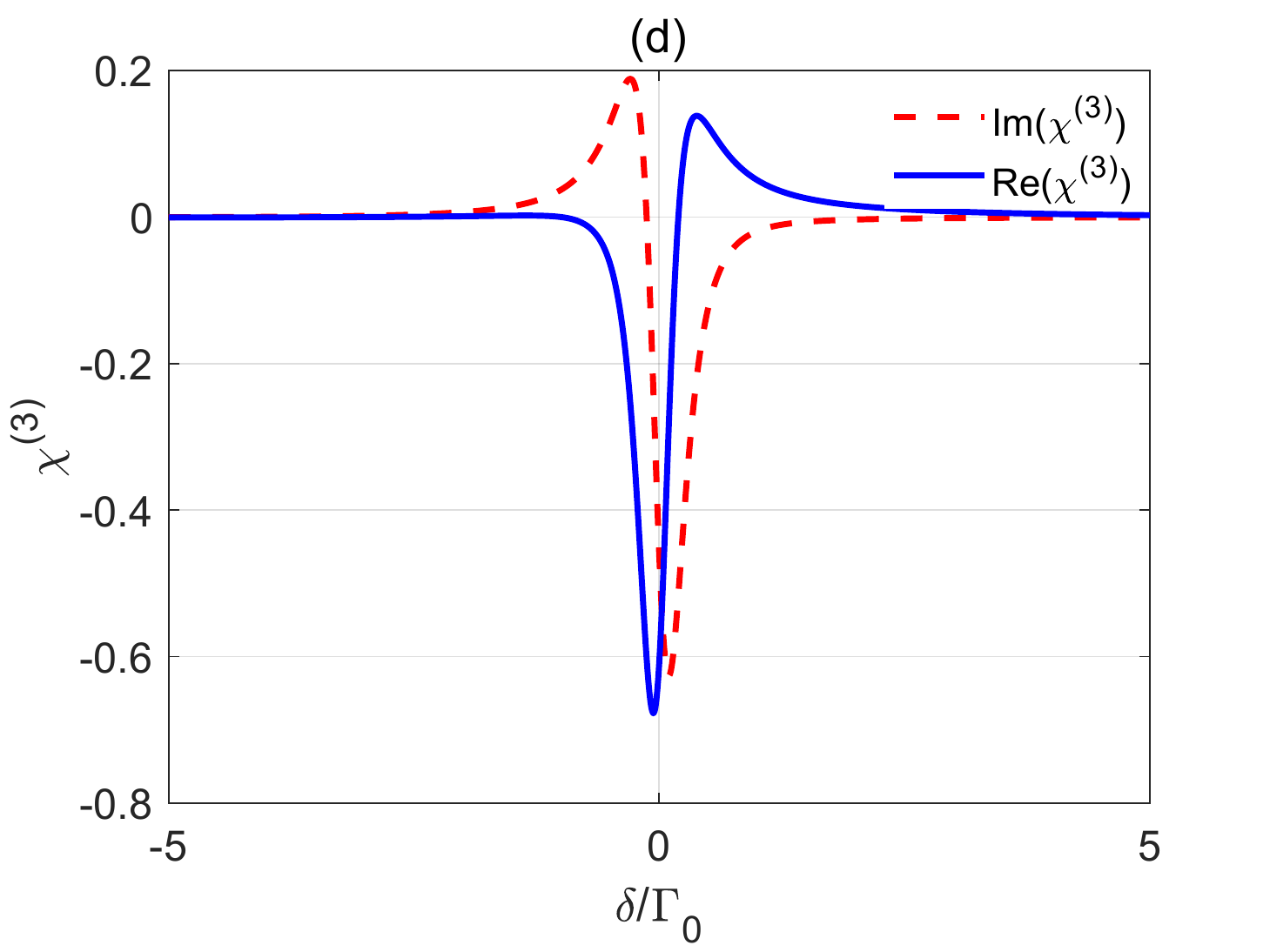}

\caption{Nonlinear susceptibility $\chi^{(3)}$ of the quantum system for the
weak probe field $\Omega_{a}$ in arbitrary units as a function of
the probe detuning $\delta$ in the presence of the plasmonic nanostructure.
We have assumed that $\omega_{32}=0$, $\gamma^{\prime}=0.3\Gamma_{0}$, $\gamma^{\prime\prime}=0$,
$x=1.5$, $\phi=0$, $\bar{\omega}=0.632\omega_{p}$, $d=0.4c/\omega_{p}$
and (a), $\phi=0$, (b), $\phi=\pi/2$, (c) $\phi=\pi$, and (d) $\phi=3\pi/2$. }
\label{fig:figs6}
\end{figure}

As illustrated in Figs.~\ref{fig:figs4}, both $\chi^{(1)}$ and
$\chi^{(3)}$are observed to behave differently for large distances
of the quantum system from the plasmonic nanostructure. We observe
that linear gain changes to a double-peaked absorption spectrum for
$d=0.7c/\omega_{c}$ (Fig.~\ref{fig:figs4}(a)). The Kerr nonlinearity
find its maximal value around the zero probe field detuning [Fig.~\ref{fig:figs4}(b)].
In Fig.~\ref{fig:figs4}(d), nonlinear gain takes also place in the
medium by altering the distance $d$ (which
leads to the change in values of $\Gamma_{\bot}$ and $\Gamma_{\Vert}$).
We obtain different behaviors of $\chi^{(1)}$ and $\chi^{(3)}$.
This is the main reason for appearance of gain or absorption in the
quantum system as demonstrated in Figs.~\ref{fig:figs3} and Figs.~\ref{fig:figs4}.
The minima of linear (and nonlinear) absorption or gain in Figs.~\ref{fig:figs3}
and \ref{fig:figs4} are given by Eqs.~(\ref{eq:AC1}) (and (\ref{eq:AC3})).
One can show from Eqs.~(\ref{eq:AC1}), that the gain is present at $\delta=0$
when (Figs.~\ref{fig:figs3})
\begin{equation}
\Omega_{b}>\frac{2\gamma^{\prime}+\Gamma_{\bot}+\Gamma_{\Vert}}{(\Gamma_{\bot}-\Gamma_{\Vert})\cos\phi}\Omega_{a},\label{eq:gain}
\end{equation}
while absorption takes place when when (Figs.~\ref{fig:figs4})
\begin{equation}
\Omega_{b}<\frac{2\gamma^{\prime}+\Gamma_{\bot}+\Gamma_{\Vert}}{(\Gamma_{\bot}-\Gamma_{\Vert})\cos\phi}\Omega_{a}.\label{eq:absorption}
\end{equation}
Equations.~(\ref{eq:LS}) and (\ref{eq:NLS}) and their corresponding
coefficients in Appendix~\ref{sec:appendix-B} prove that in the presence
of the plasmonic nanostructure, the linear and nonlinear susceptibilities
are sensitive to the relative phase of the weak probe fields. Figures.~\ref{fig:figs5}
and \ref{fig:figs6} illustrate the dependence of $\chi^{(1)}$ and
$\chi^{(3)}$ on $\phi$ when the quantum system is placed at a distance
$d=0.4c/\omega_{p}$ from the surface of the plasmonic nanostructure.
The strong variation of linear and nonlinear absorption and dispersion
profiles for different values of $\phi$ is obvious. In particular,
for $\phi=0$ the maximal of Kerr nonlinearity is placed in a region
of linear gain around $\delta=0$. Subluminal response takes place in
this situation on resonance [see Figs.~\ref{fig:figs5}(a)
and \ref{fig:figs6}(a)]. When $\phi$ becomes $\pi$, a strong
absorption instead of gain appears at line center for the $\chi^{(1)}$ profile, as can be seen in Fig.~\ref{fig:figs5}(c). Such a phase
sensitive gain and absorption is well understood through Eqs.~(\ref{eq:gain})
and (\ref{eq:absorption}). In both cases, the value of the Kerr index
at exact resonance is zero. According to Eq.~(\ref{eq:AC4}), for $\phi=0$ and $\phi=\pi$, both sine terms in Eq.~(\ref{eq:AC4})
vanish leading to zero Kerr nonlinearity on resonance. It should be
mentioned that a nonzero resonant Kerr nonlinearity can be obtained
for $\phi=\pi/2$ [Fig.~\ref{fig:figs6}(b)] and $\phi=3\pi/2$ [Fig.~\ref{fig:figs6}(d)].

\begin{figure}
\includegraphics[width=0.5\columnwidth]{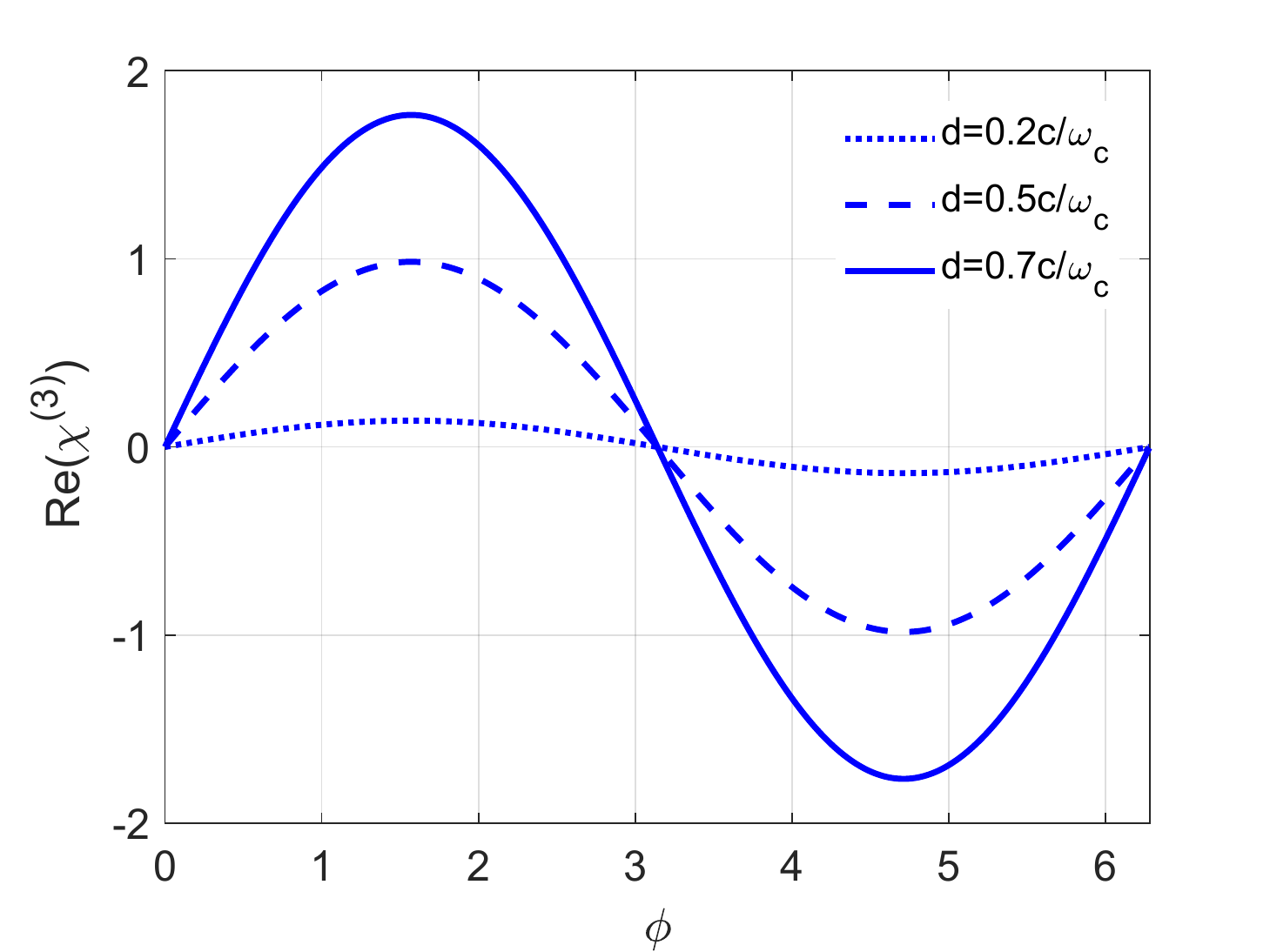}

\caption{The refractive part of third-order nonlinear susceptibility $Re(\chi^{(3)})$
(Kerr nonlinearity) of the quantum system for the weak probe field
$\Omega_{a}$ in arbitrary units as a function of the relative phase
$\phi$ for different distances from the plasmonic nanostructure $d=0.2c/\omega_{p}$
(dot line), $d=0.5c/\omega_{p}$ (dash line), $d=0.7c/\omega_{p}$
(solid line). We take here $\omega_{32}=0$, $\gamma^{\prime}=0.3\Gamma_{0}$,
$\gamma^{\prime\prime}=0$, $x=1.5$, $\phi=0$, $\bar{\omega}=0.632\omega_{p}$. }
\label{fig:figs7}
\end{figure}

The results obtained here may suggest a tunable control over the Kerr
nonlinearity of the quantum system near the plasmonic nanostructure
by using the relative phase of the applied fields. In Fig.~\ref{fig:figs7}
we present an example of the variation of the Kerr nonlinearity spectra
for different distances of the quantum system from the plasmonic
nanostructure, $d=0.2c/\omega_{p}$ (dot line), $d=0.5c/\omega_{p}$
(dash line), $d=0.7c/\omega_{p}$ (solid line). A wide range of tunability
can be observed over the refractive part of third-order nonlinear
susceptibility $\chi^{(3)}$ spectra just by adjusting the relative
phase parameter. We find that the whole profile for Kerr nonlinearity
is enhanced for larger distances due to the reduction of both $\Gamma_{\bot}$ and $\Gamma_{\Vert}$
reduces by distance. The nonlinear dispersion becomes zero at $\phi=n\pi$,
while it obtains its maximal amplitude for $\phi=n\frac{\pi}{2}$.
It also changes from negative to positive and back to positive
as the relative phase changes from $0$ to $2\pi$.

\section{Spatially structured optical effects}

Up to now no assumption has been made about the spatial profile of
laser fields. Next, we will consider the case where the incident
field possesses a nontrivial structural profile which, however, is
almost unaffected by the plasmonic nanostructure (marginal
reflection and absorption) as it is tuned to the resonant
frequencies of the lower V-type subsystem which lie well beyond,
say well above, the surface-plasmon bands of the nanostructure in
which case the nanostructure is almost transparent to the
impinging structured laser field. Since, the position of the
quantum is kept fixed, i.e. right opposite the center of the
nanosphere, we will study the role of the position of the quantum
system within the structured-field landscape, which, given the
fixed position of the quantum system, is translated into the
dependence of the applied structured field relative to the
nanostructure.

We assume that the probe field $\Omega_{b}$ has an orbital angular
momentum $\hbar l$ along the propagation axis $z$
\cite{Allen1999OAM}. In this case, the vortex probe field
$\Omega_{b}$ is characterized by the Rabi frequency
\begin{equation}
\Omega_{b}=A_{b}\exp(il\Phi).\label{eq:12}
\end{equation}
For a Laguerre-Gaussian (LG) doughnut beam we may write the amplitude
of a vortex beam $A_{b}$ as
\begin{equation}
A_{b}(\varrho)=|\Omega_{b}|(\frac{\varrho}{w})^{|l|}\exp(-\frac{\varrho^{2}}{w^{2}}),\label{eq:13}
\end{equation}
where $\Phi=\tan^{-1}(y/x)$ is the azimuthal angle, $x$ and $y$
are transverse directions, $\varrho=\sqrt{x^{2}+y^{2}}$ represents
the distance from the vortex core (cylindrical radius), $w$ denotes
the beam waist parameter, and $|\Omega_{b}|$ is the strength of the
vortex beam. The Rabi frequency of the other probe field does not
have a vortex and is given by
\begin{equation}
\Omega_{a}=|\Omega_{a}|.\label{eq:14}
\end{equation}

In this case, Eqs.~(\ref{eq:l1})-(\ref{eq:a2}) for the evolution
of the system and their corresponding coefficients featured in Appendix~\ref{sec:appendix-A}
remain the same, with the only difference that $\phi$ changes to
$l\Phi$. In addition, Eqs.~(\ref{eq:LS}) and (\ref{eq:NLS}) will
describe the azimuthally varying linear and nonlinear susceptibilities
under the transformation $\phi\rightarrow l\Phi$, yet one needs to perform
also the transformation $x\rightarrow X(\frac{\varrho}{w})^{|l|}\exp(-\frac{\varrho^{2}}{w^{2}})$,
where $X=\frac{|\Omega_{b}|}{|\Omega_{a}|}$ in the corresponding
coefficients given in Appendix~\ref{sec:appendix-B}. This allows
to study the azimuthal modulation of the linear and nonlinear response
of a weak non-vortex probe field $\Omega_{a}$ at weak intensity regime.

\begin{figure}
\includegraphics[width=0.2\columnwidth]{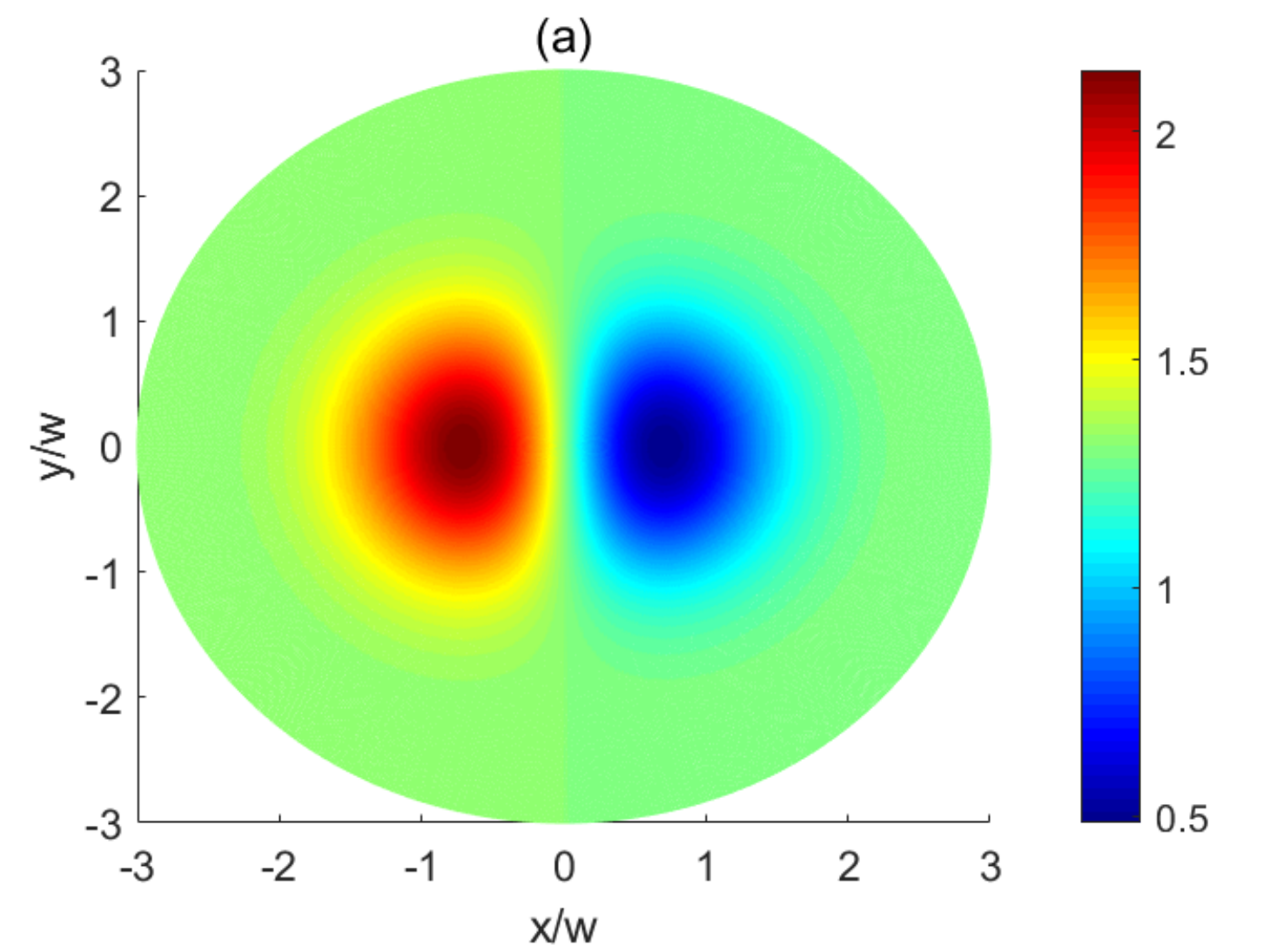} \includegraphics[width=0.2\columnwidth]{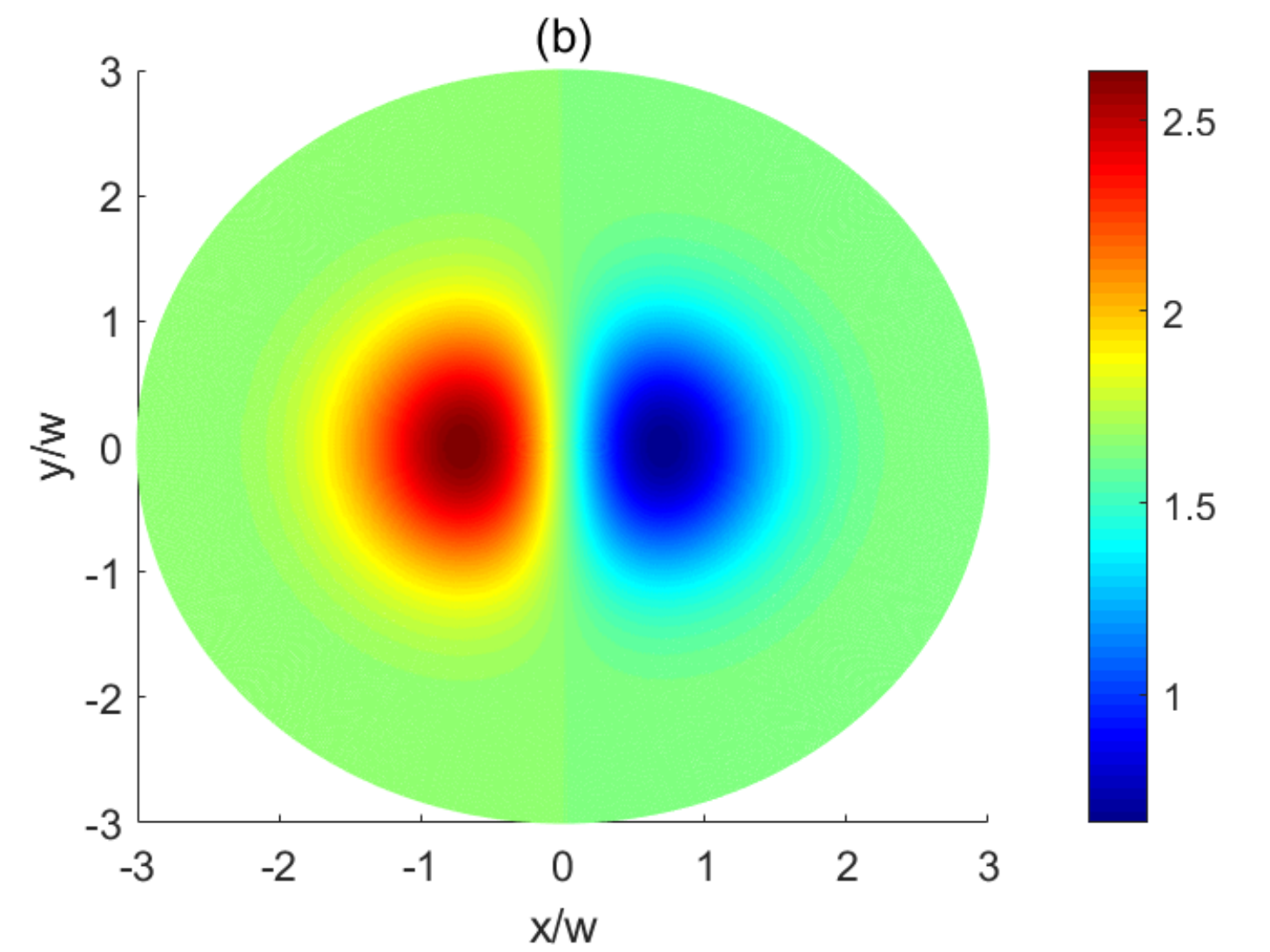}
\includegraphics[width=0.2\columnwidth]{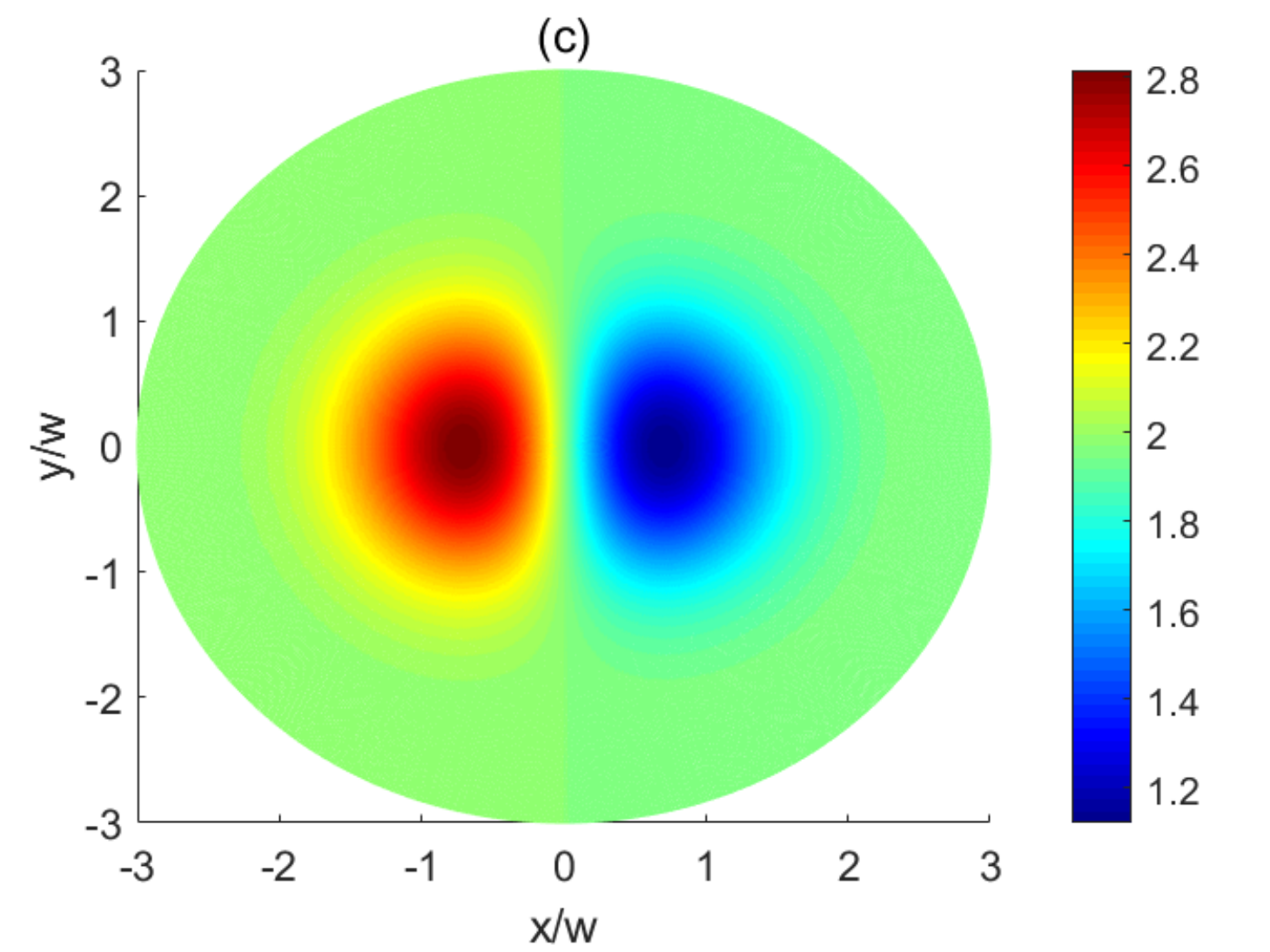} \includegraphics[width=0.2\columnwidth]{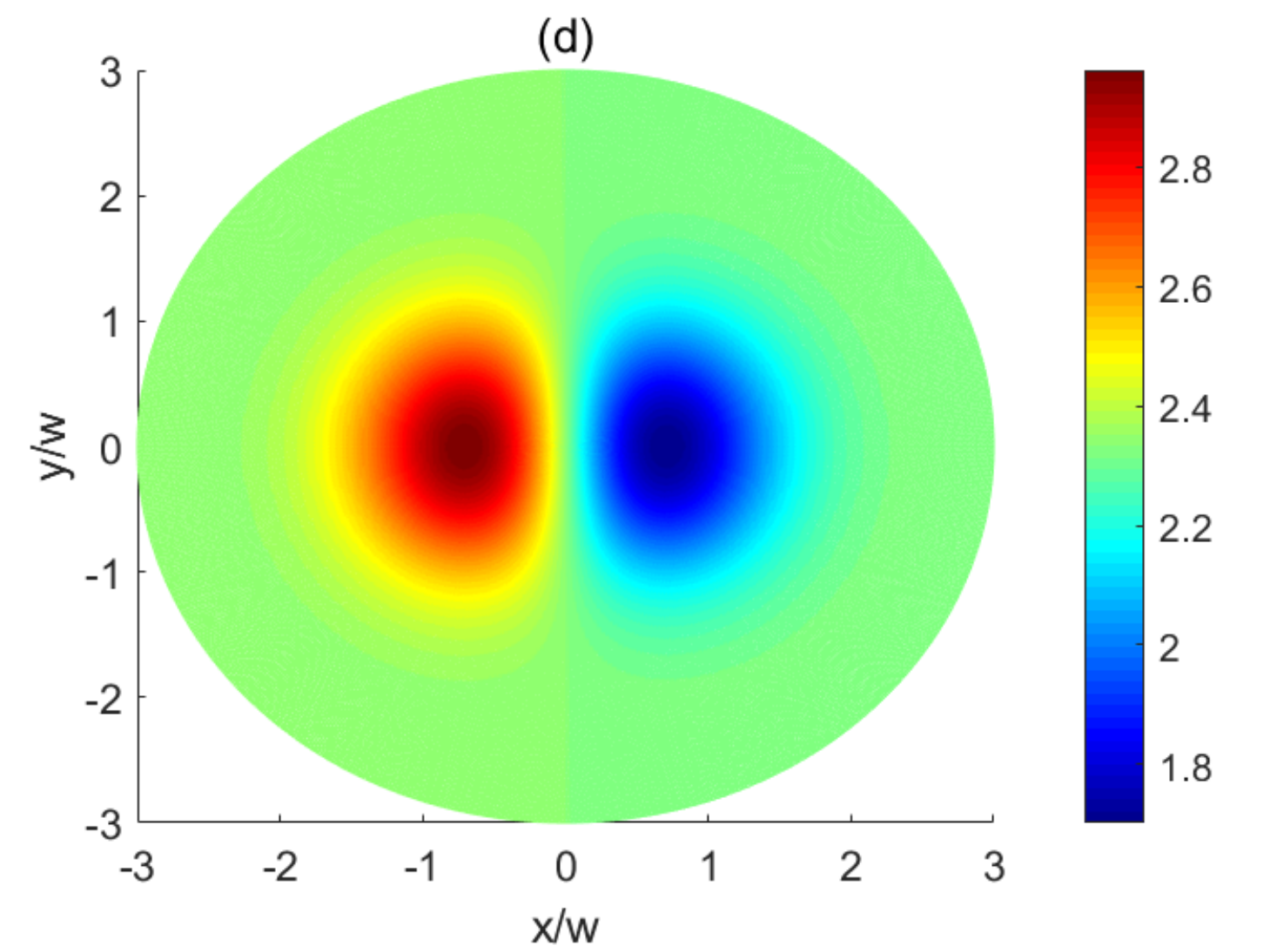}

\includegraphics[width=0.2\columnwidth]{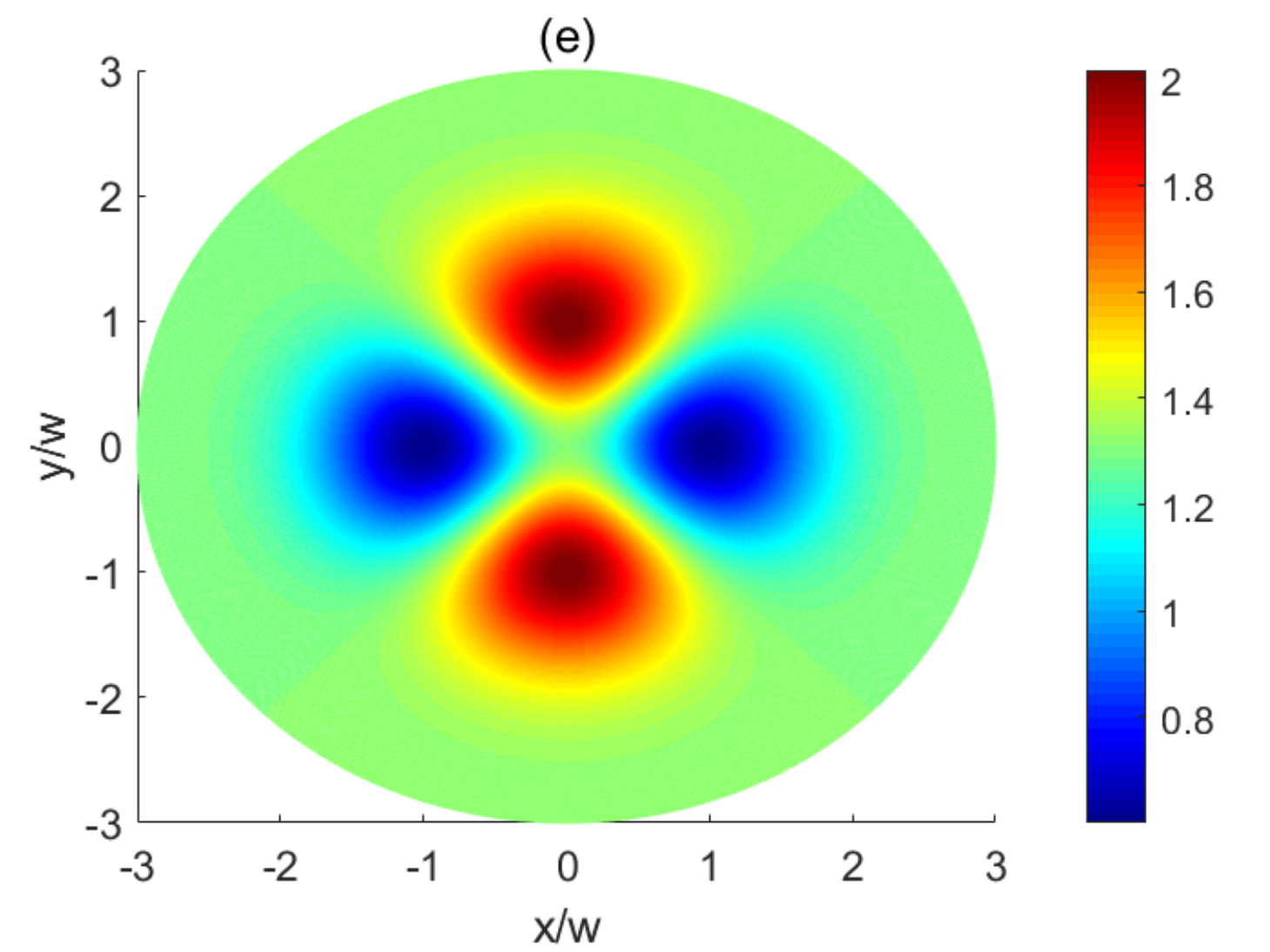} \includegraphics[width=0.2\columnwidth]{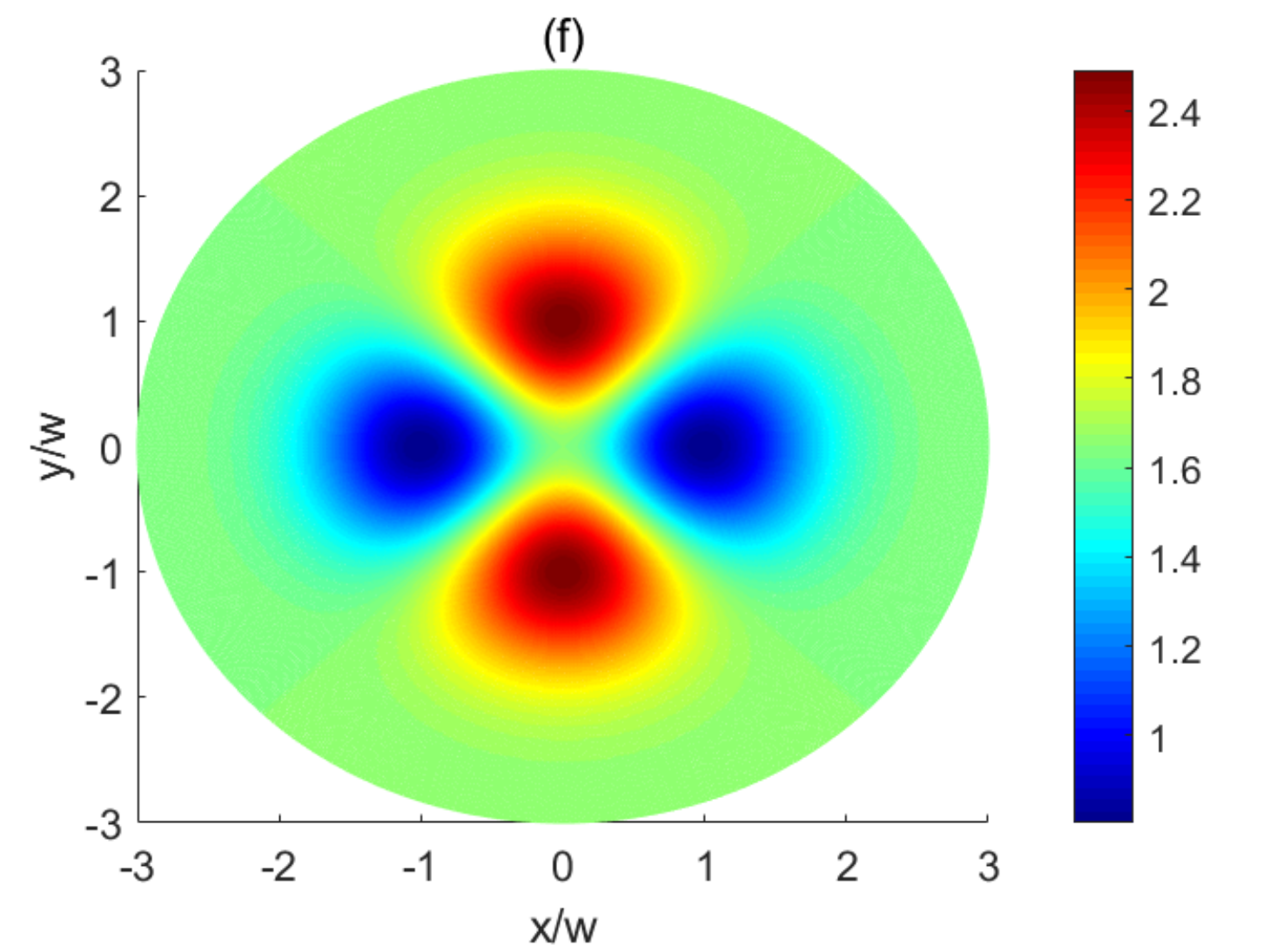}
\includegraphics[width=0.2\columnwidth]{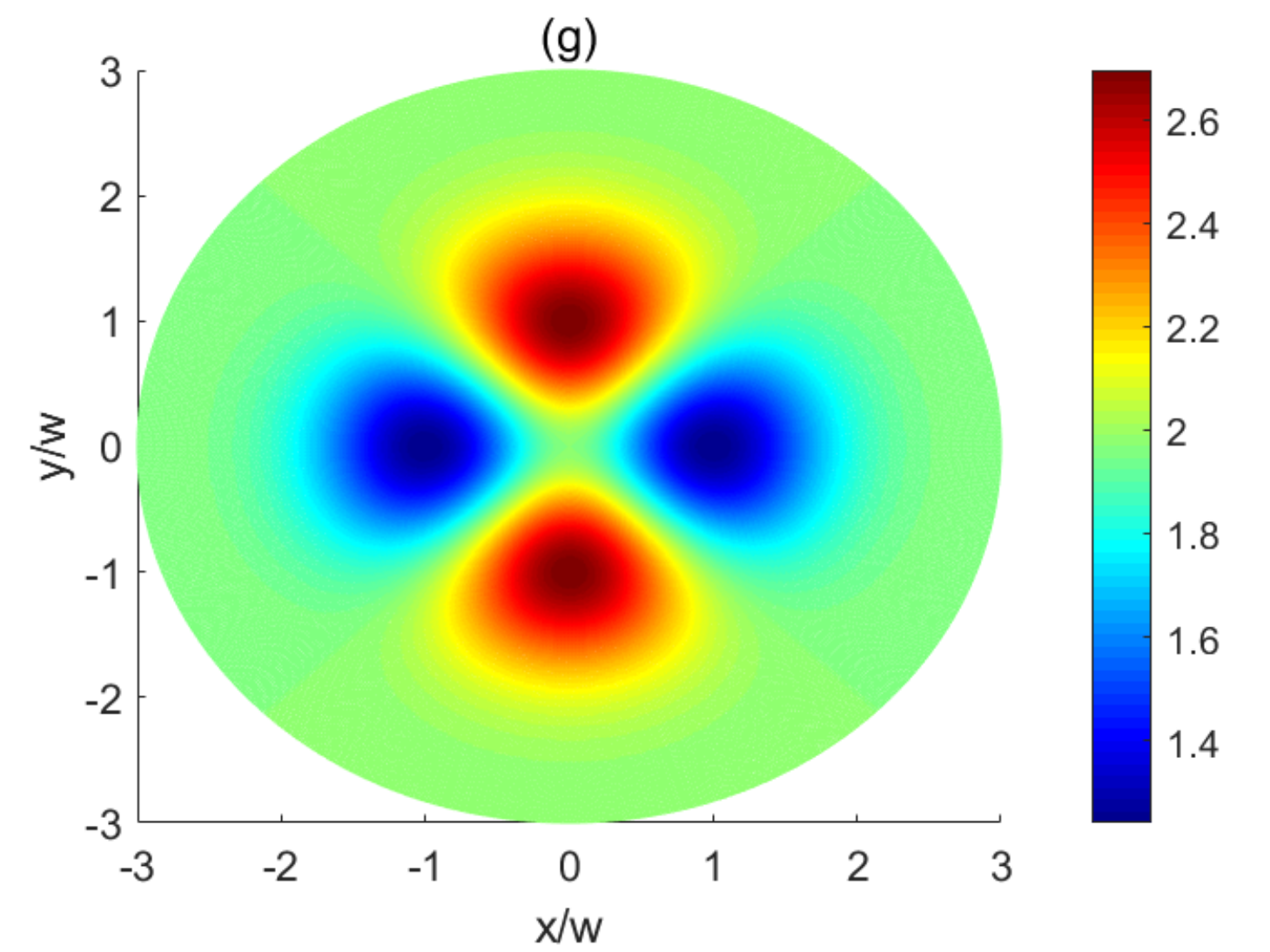} \includegraphics[width=0.2\columnwidth]{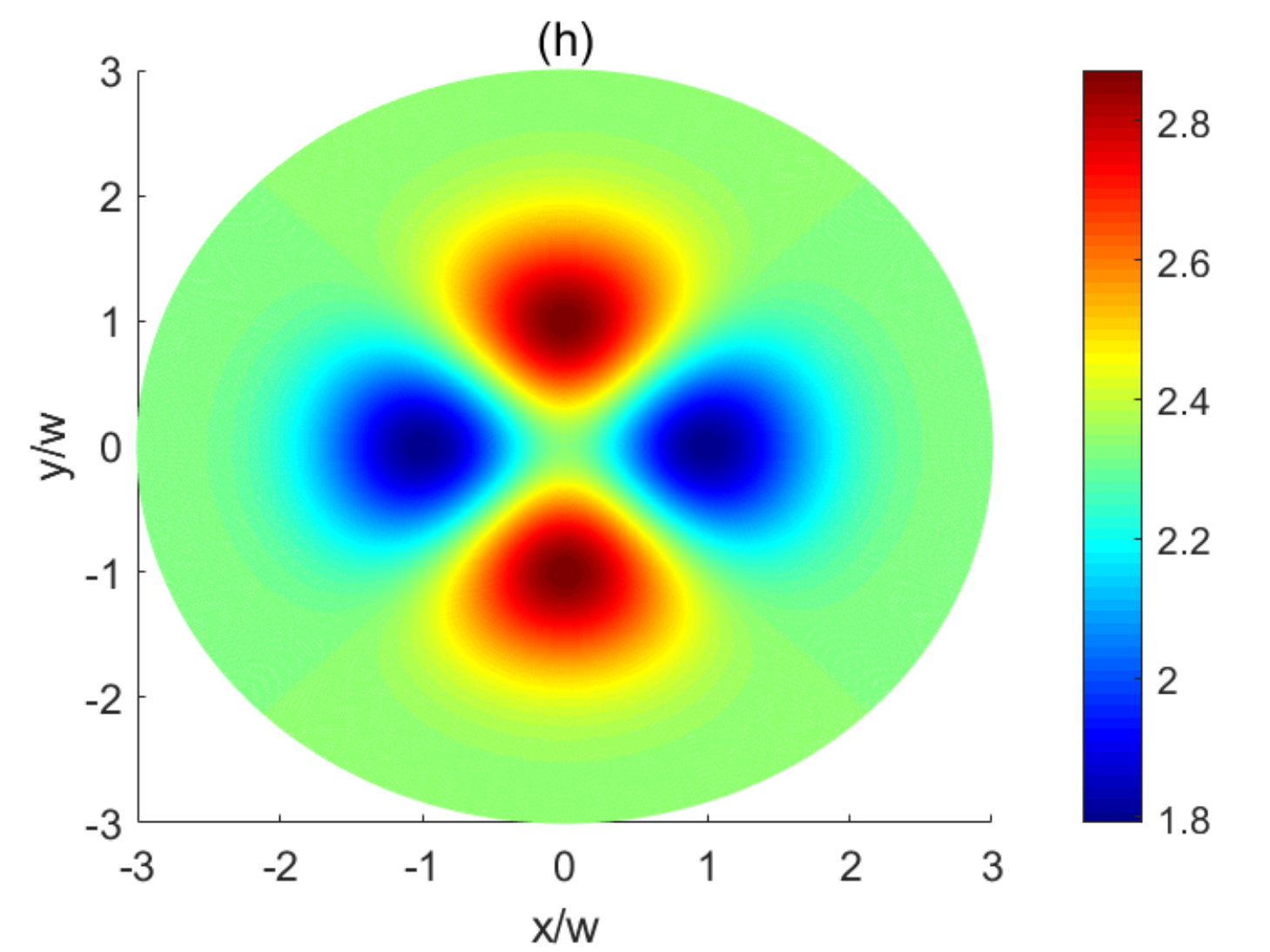}

\caption{Spatially structured linear absorption $Im(\chi^{(1)})$ profiles
of the probe beam $\Omega_{a}$ in arbitrary units, in the presence
of the plasmonic nanostructure and for different values of the distance
$d$ of the quantum system from the plasmonic nanostructure: $d=0.1c/\omega_{c}$
(a,e), $d=0.3c/\omega_{p}$ (b,f), $d=0.6c/\omega_{p}$ (c,g) and
$d=0.8c/\omega_{p}$ (d,h). Here, the winding number is $l=1$ for
(a,b,c,d) and $l=2$ (e,f,g,h), while the other parameters are $\delta=0$,
$\omega_{32}=0$, $\frac{|\Omega_{b}|}{|\Omega_{a}|}=1.5$, $\bar{\omega}=0.632\omega_{p}$,
$\gamma^{\prime}=0.3\Gamma_{0}$ and $\gamma^{\prime\prime}=0$. }
\label{fig:figs8}
\end{figure}

We will consider a situation where the laser fields are at exact resonance
with the corresponding transitions ($\delta=0$). We also assume that
the quantum system is degenerate ($\omega_{32}=0$). In this case,
the imaginary part of Eq.~(\ref{eq:LS}) for the linear absorption
of probe field $\Omega_{a}$ simplifies to
\begin{equation}
Im(\chi^{(1)}(\delta=0))=\frac{N\mu^{\prime2}}{\varepsilon_{0}\hbar}\frac{(\gamma+\gamma^{\prime})-\kappa A\cos(l\Phi)}{(\gamma+\gamma^{\prime})^{2}-\kappa^{2}}.\label{eq:absorptionAzimuth}
\end{equation}
Eq.~(\ref{eq:absorptionAzimuth})  implies that the linear
absorption of the probe field $\Omega_{a}$ can be influenced by the
vortex probe beam $\Omega_{b}$ through the term $\kappa A\cos(l\Phi)$.
This term contains a phase factor $l\Phi$ accounting for the spatial
variation of the probe absorption. It is indeed the existence
of the quantum interference term $\kappa$, which makes the quantum
system sensitive to the azimuthal phase, resulting in the spatially-dependent linear absorption when the quantum system is near the plasmonic
nanostructure ($d\neq0$), as shown in Fig.~\ref{fig:figs8}.

Fig.~\ref{fig:figs8} demonstrates the resulting absorption spectra
for different values of the distance $d$. The results are presented in Figs.~\ref{fig:figs8}
for two different vorticities $l=1$ [Figs.~\ref{fig:figs8} (a,b,c,d)]
and $l=2$ [Figs.~\ref{fig:figs8} (e,f,g,h)]. From Fig.~\ref{fig:figs8},
we observe that the linear absorption increases with the distance $d$
for the whole region of transverse spatial profile. The spatially
structured absorption profiles oscillate sinusoidally in the presence
of the plasmonic nanostructure [see also Eq.~(\ref{eq:absorptionAzimuth})].

\begin{figure}
\includegraphics[width=0.2\columnwidth]{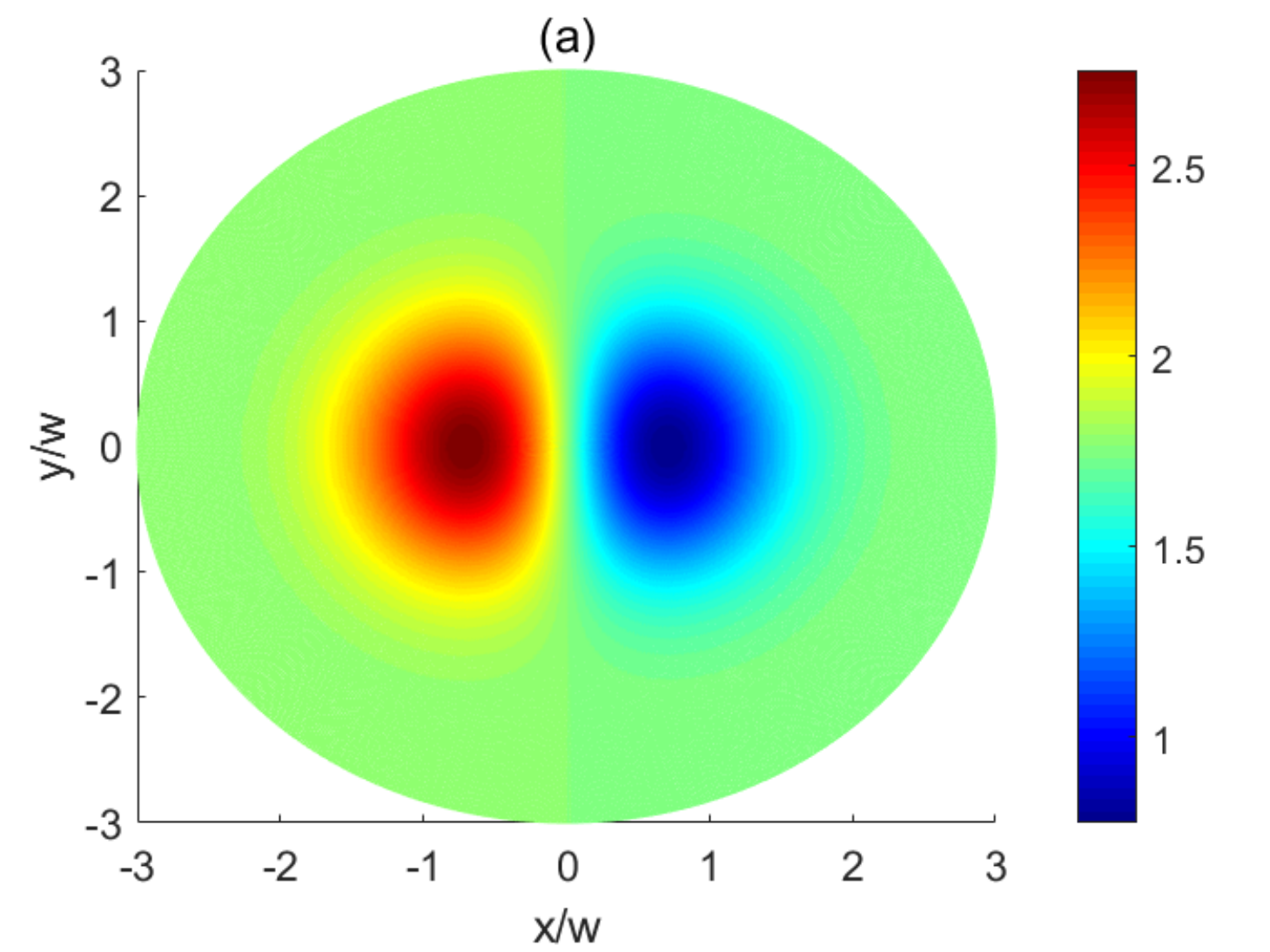} \includegraphics[width=0.2\columnwidth]{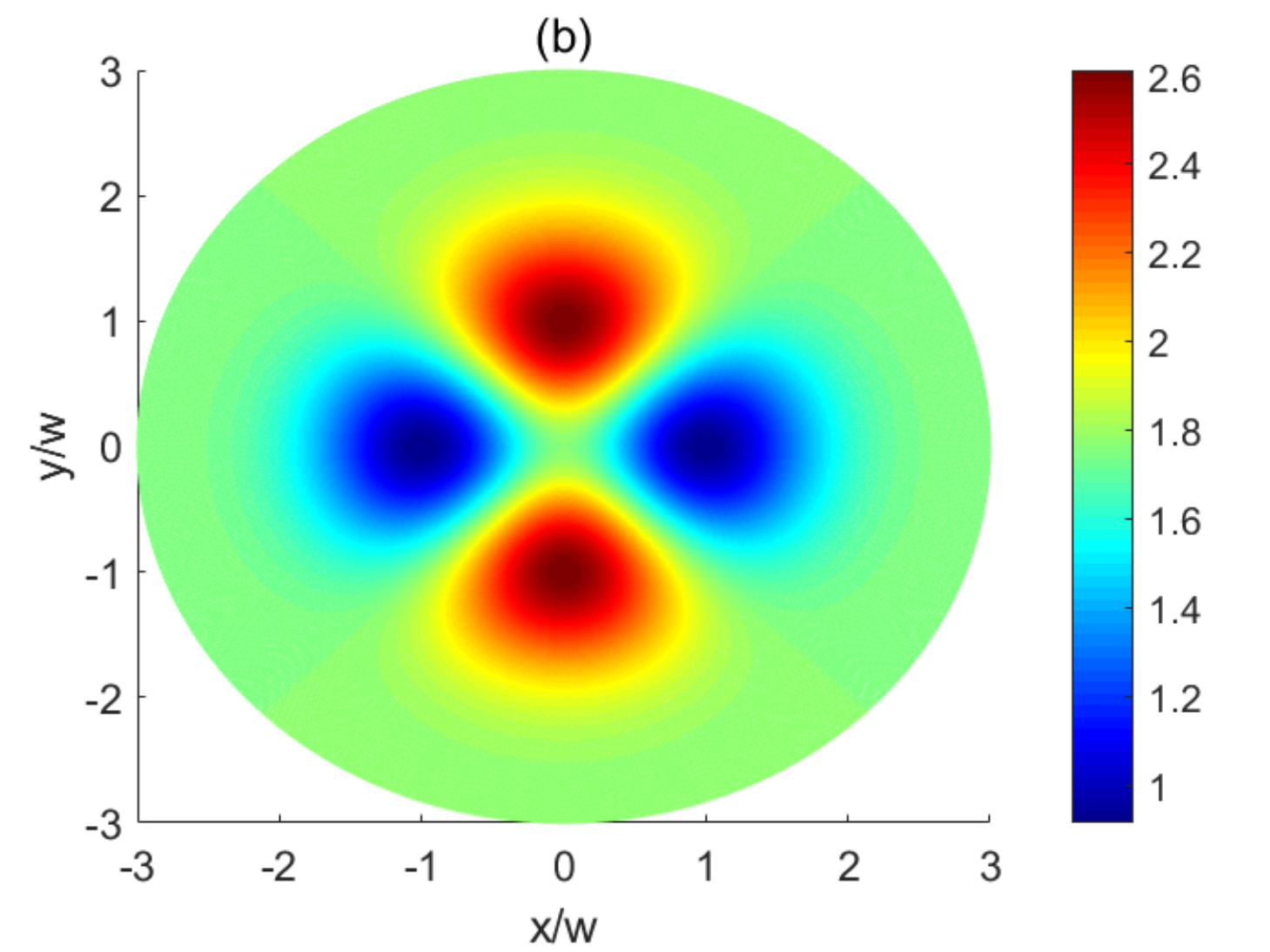}
\includegraphics[width=0.2\columnwidth]{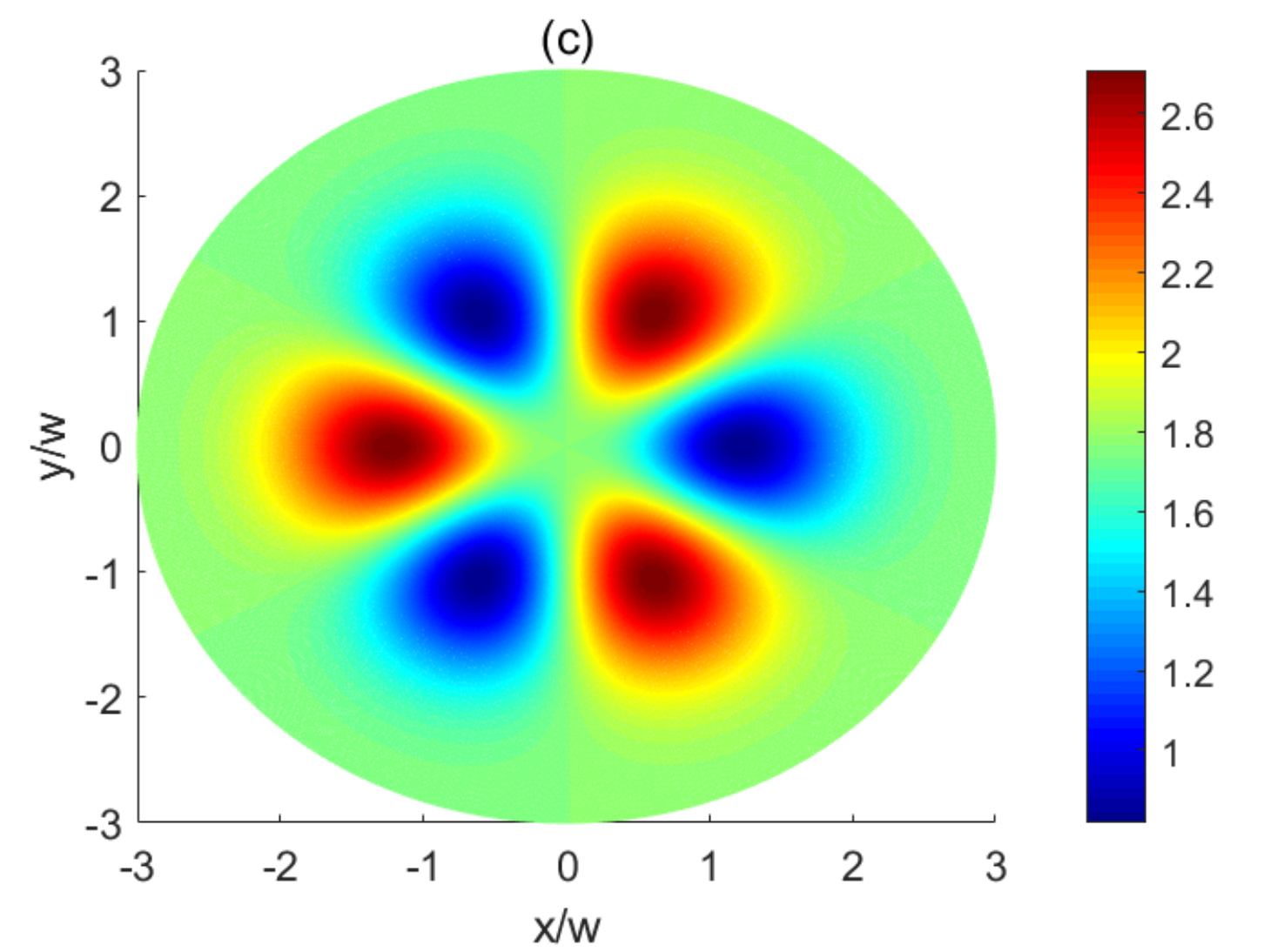}

\includegraphics[width=0.2\columnwidth]{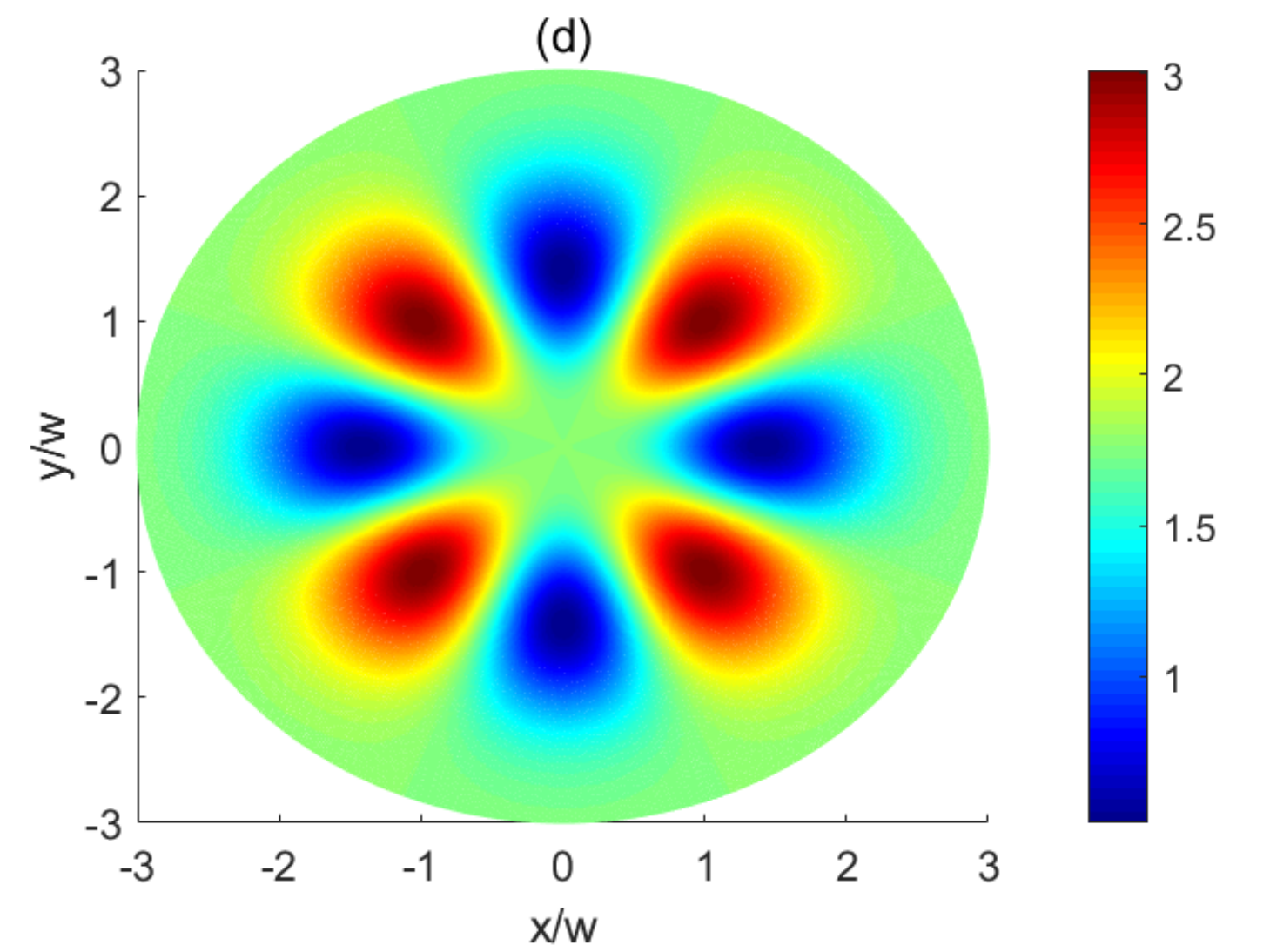} \includegraphics[width=0.2\columnwidth]{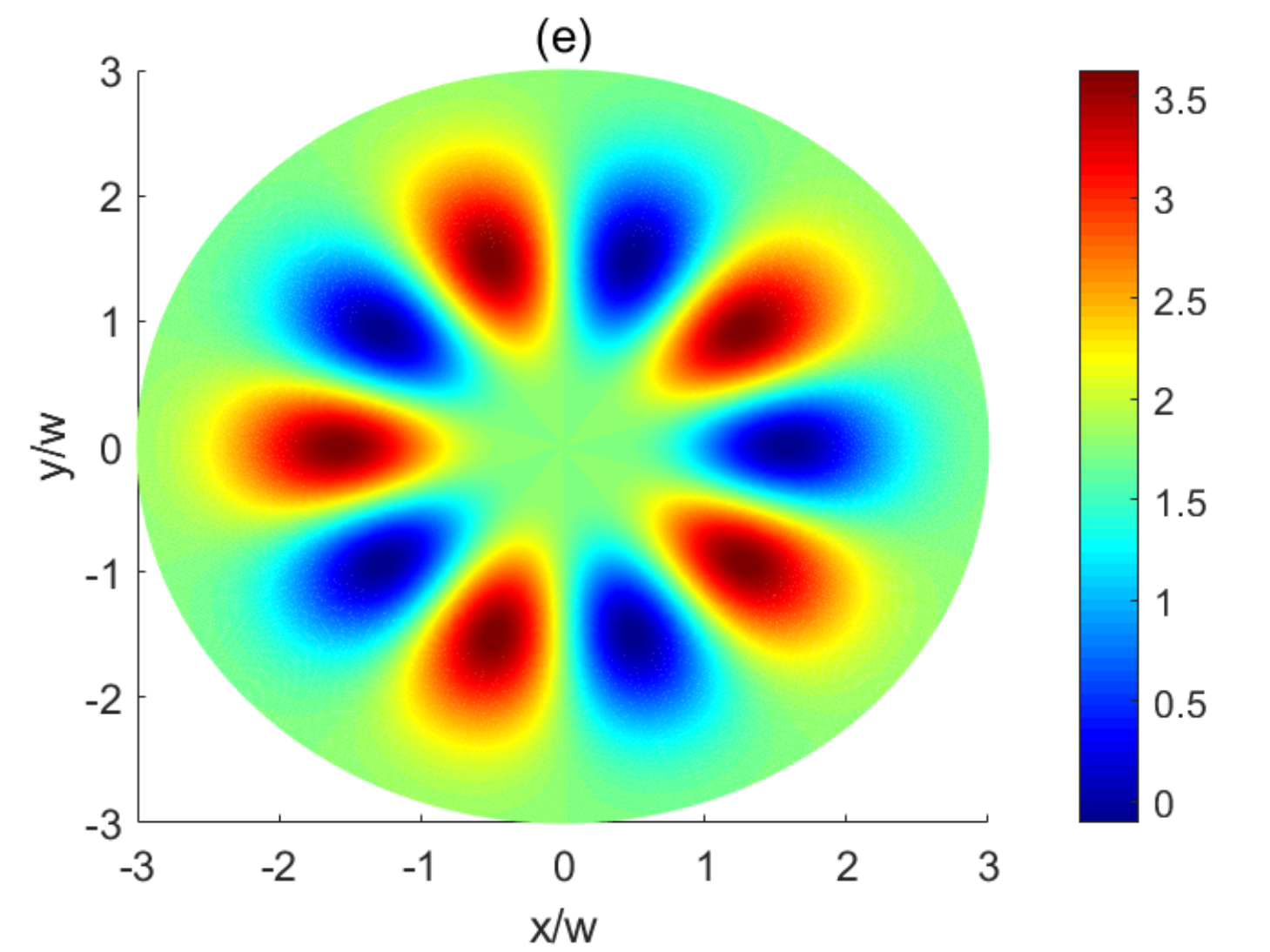}
\includegraphics[width=0.2\columnwidth]{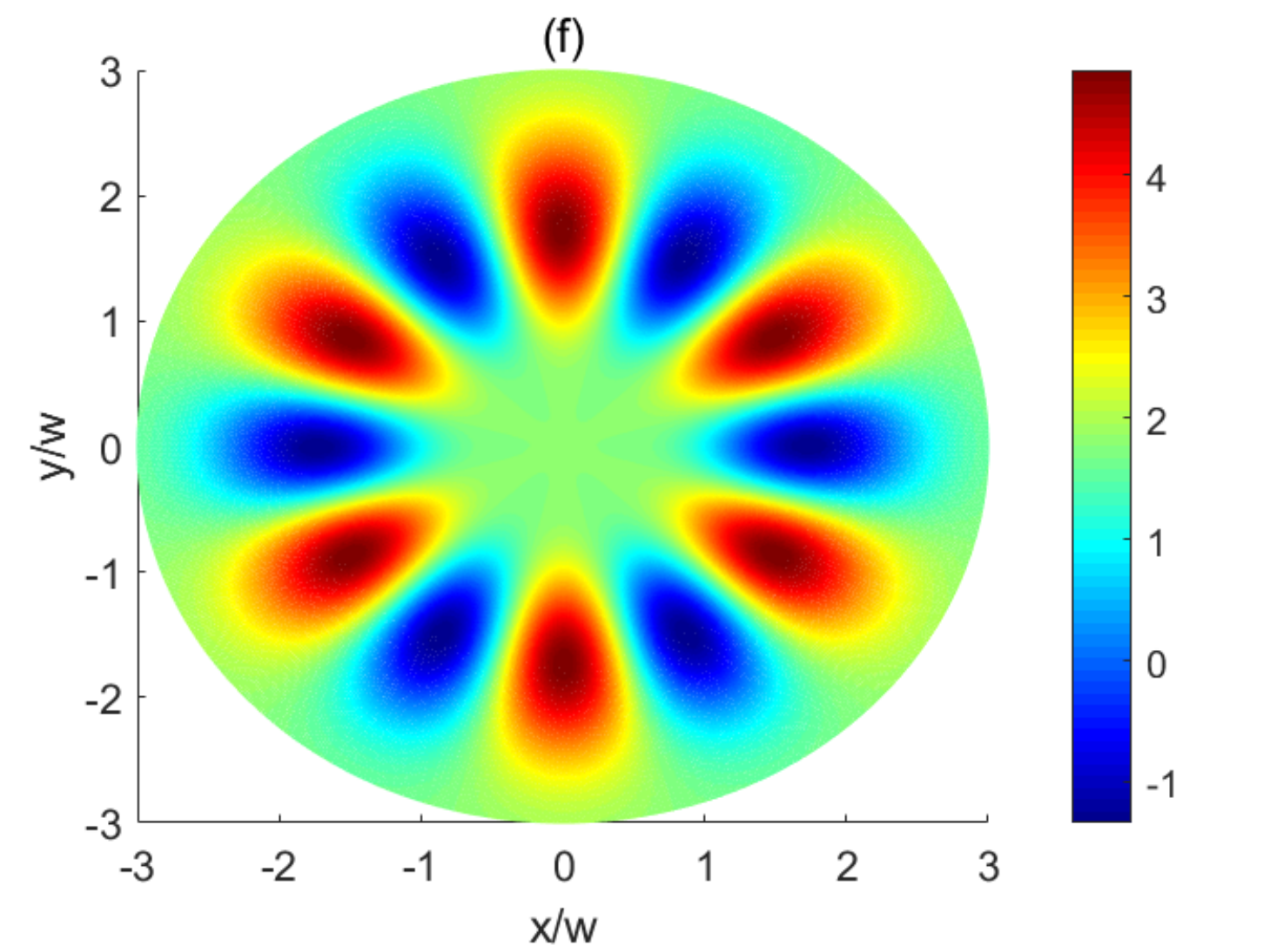}

\caption{Spatially structured linear absorption $Im(\chi^{(1)})$ profiles
of the probe beam $\Omega_{a}$ in arbitrary units, in the presence
of the plasmonic nanostructure and for different winding $l=1$(a)$-$$l=6$
(f). Here, $d=0.4c/\omega_{p}$ and the other parameters are the same
as Fig.~\ref{fig:figs8}.}
\label{fig:figs9}
\end{figure}

Eq.~(\ref{eq:absorptionAzimuth}) implies that the linear absorption
of the system for the transition $|0\rangle\leftrightarrow|2\rangle$
of the quantum system near the plasmonic nanostruture can be manipulated
through the winding number $l$ (OAM number). The $l$ factor in the cosine
term of Eq.~(\ref{eq:absorptionAzimuth}) governs the number of absorption
peaks (or dips) in the transverse ($x-y$) plane. The periodic oscillatory
behavior of the absorption profile in the transverse plane for
a given value of distance $d=0.4c/\omega_{c}$ but different OAM numbers
$l=1-6$ is observed in Fig.~\ref{fig:figs9}. Because of the
angular dependence, the spatially structured absorption profile displays
a $l$-fold symmetry. The number of absorption peaks (or dips) increases with larger winding number $l$. As a result, one can
easily distinguish an unknown vorticity of a vortex probe beam $\Omega_{b}$
solely by counting the bright spots appearing in the absorption
profile of the probe field $\Omega_{a}$. Furthermore, the maximum
of the linear absorption curve is enhanced in some regions of the transverse
plane by increasing the winding number, while gain appears
in some other regions.

\begin{figure}
\includegraphics[width=0.2\columnwidth]{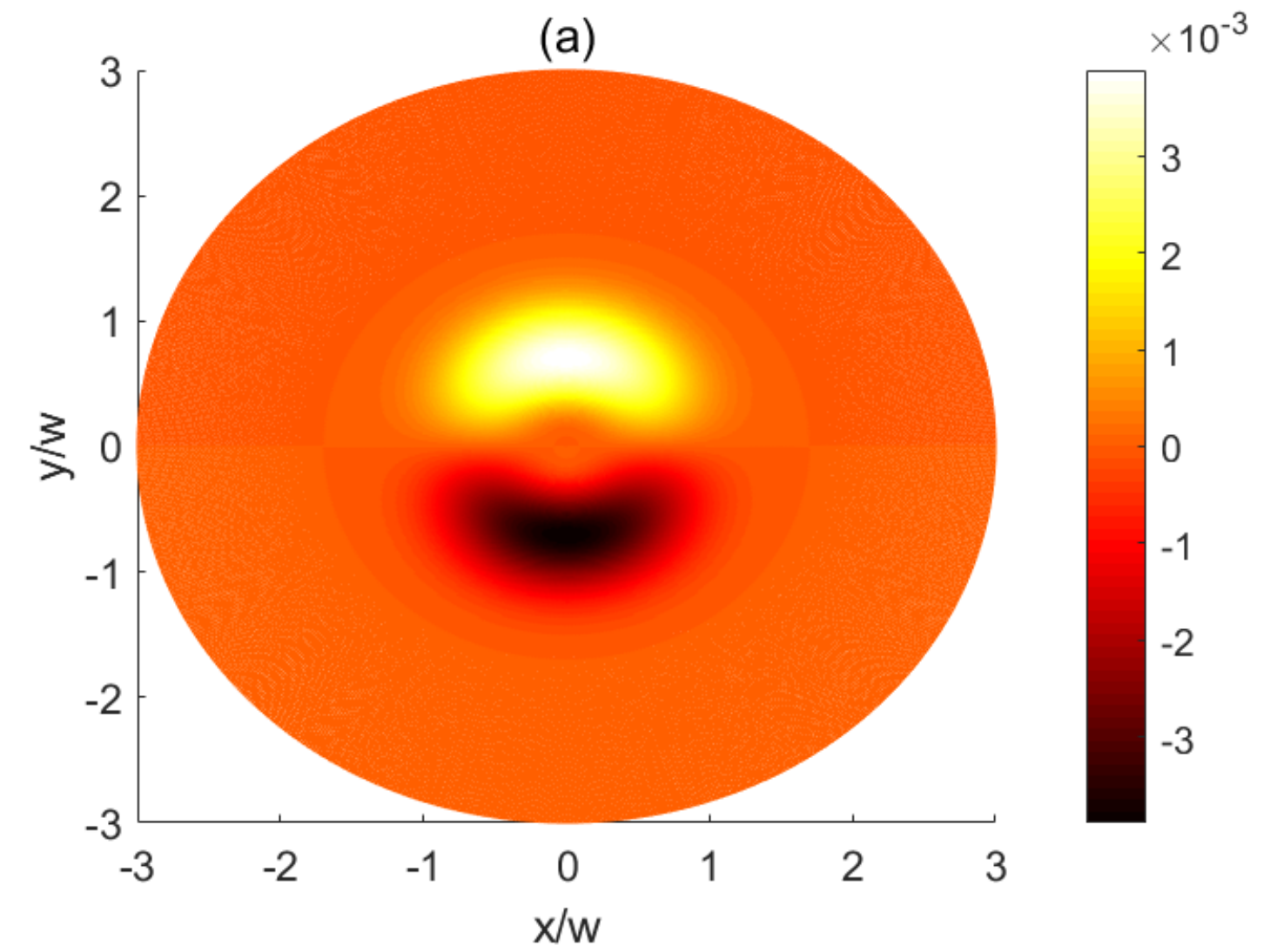} \includegraphics[width=0.2\columnwidth]{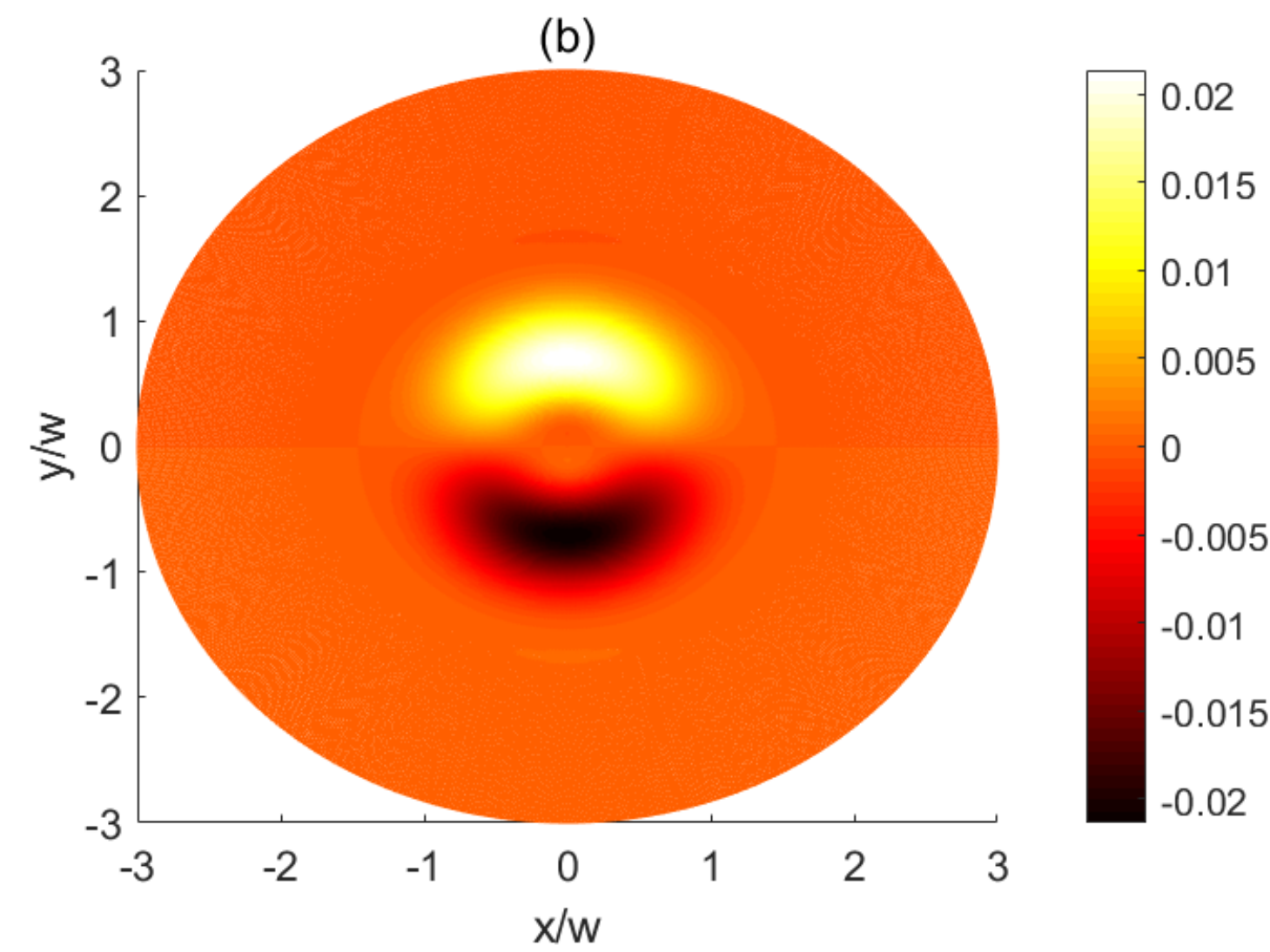}
\includegraphics[width=0.2\columnwidth]{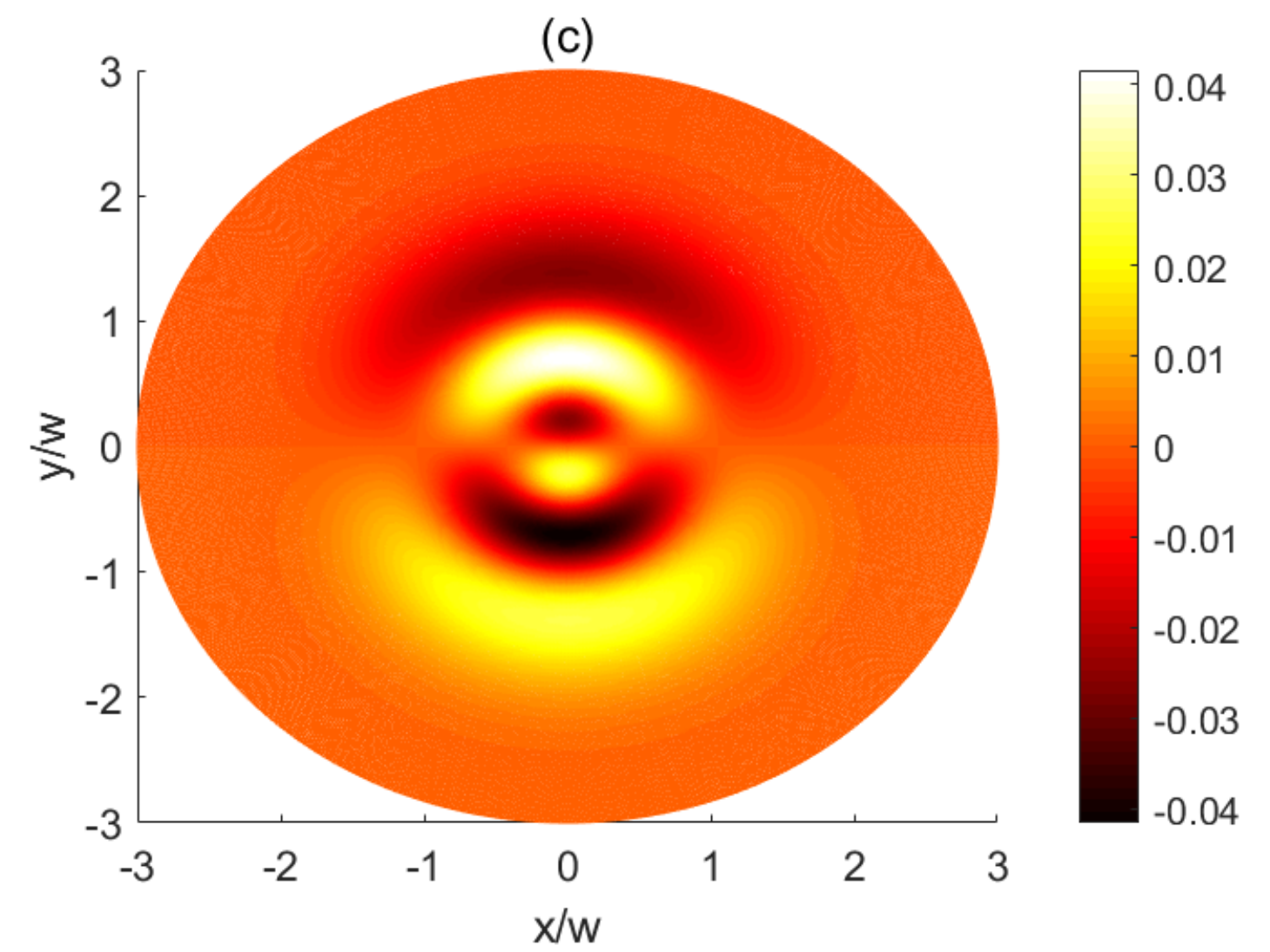} \includegraphics[width=0.2\columnwidth]{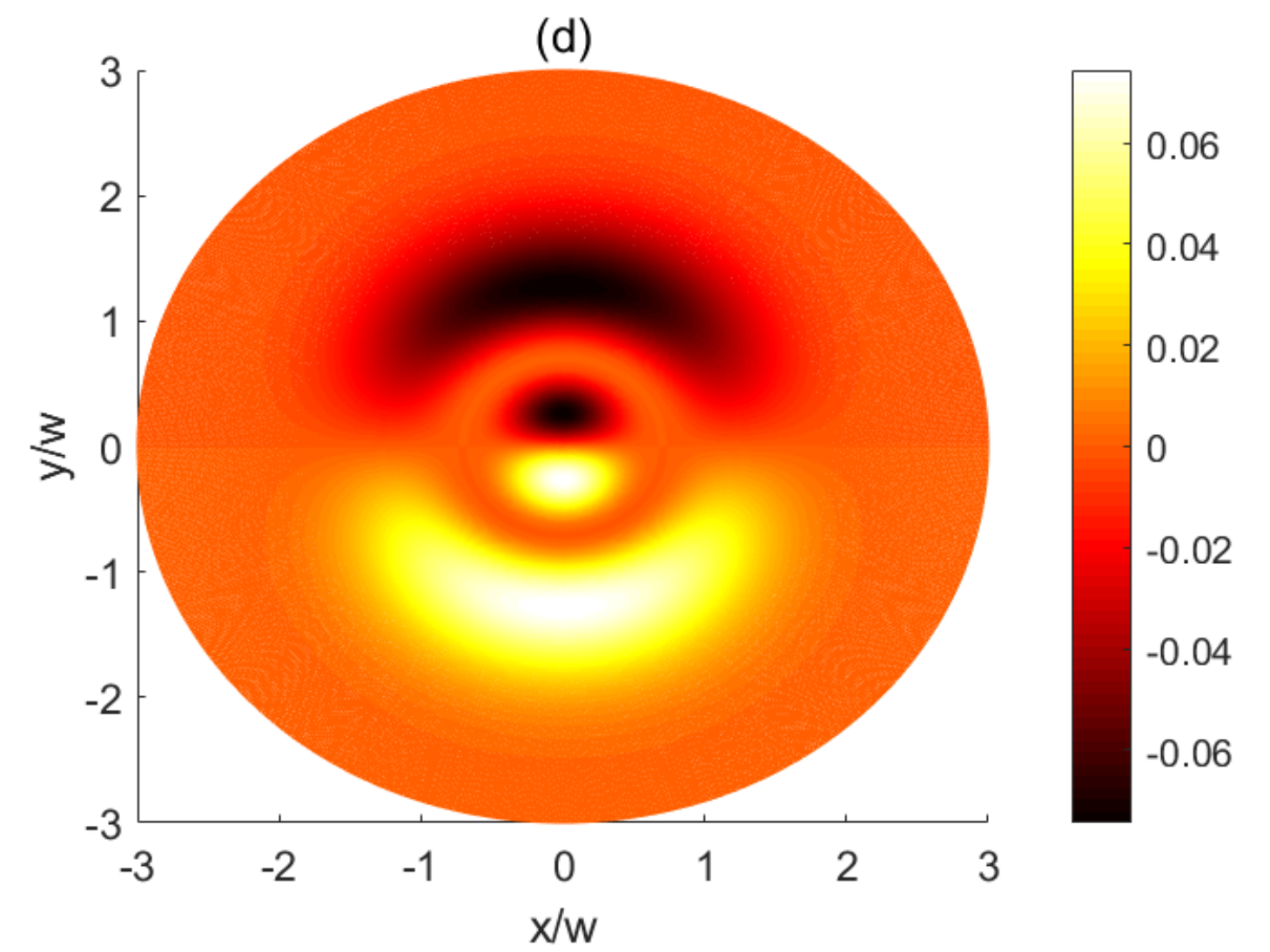}

\includegraphics[width=0.2\columnwidth]{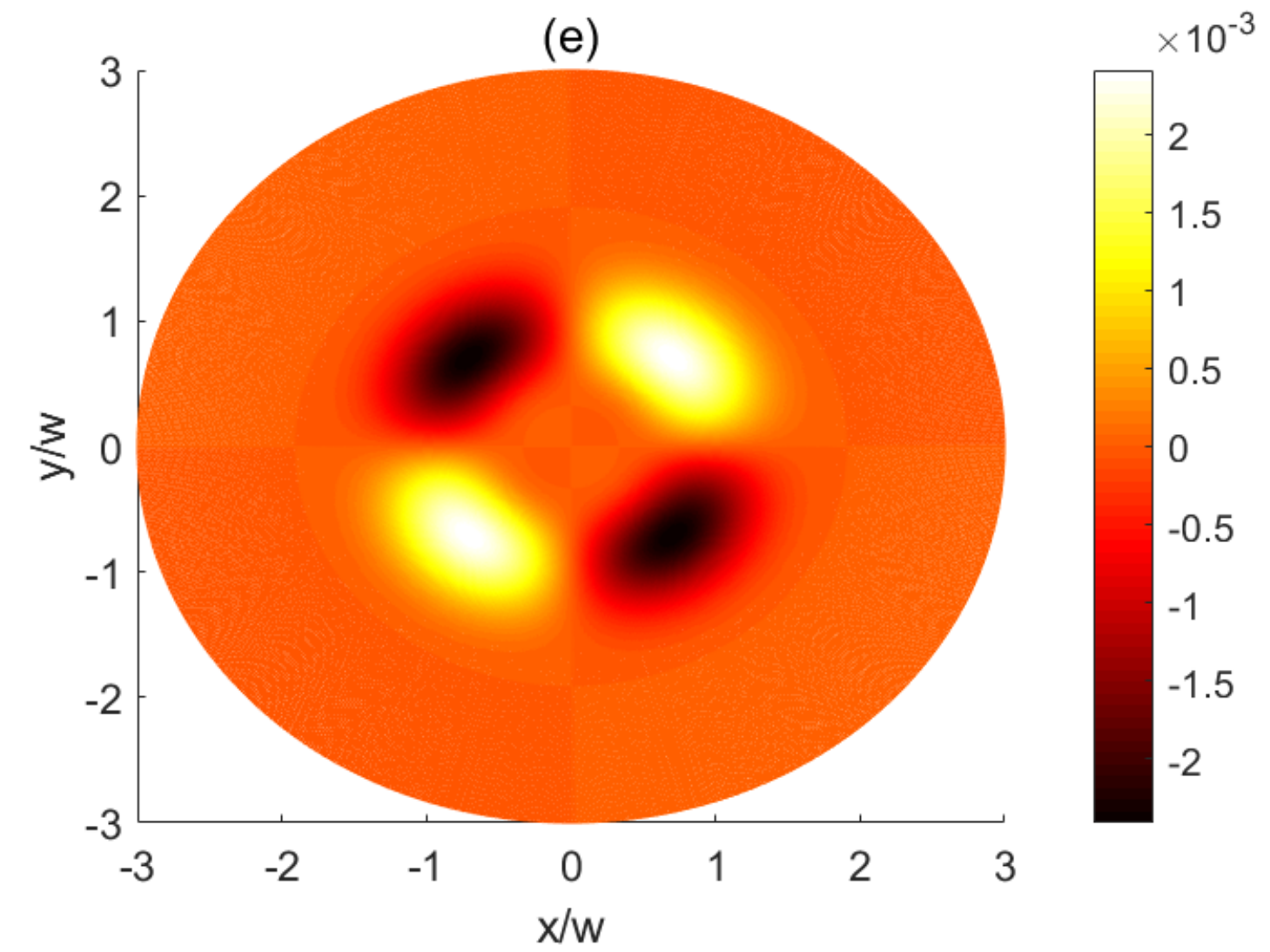} \includegraphics[width=0.2\columnwidth]{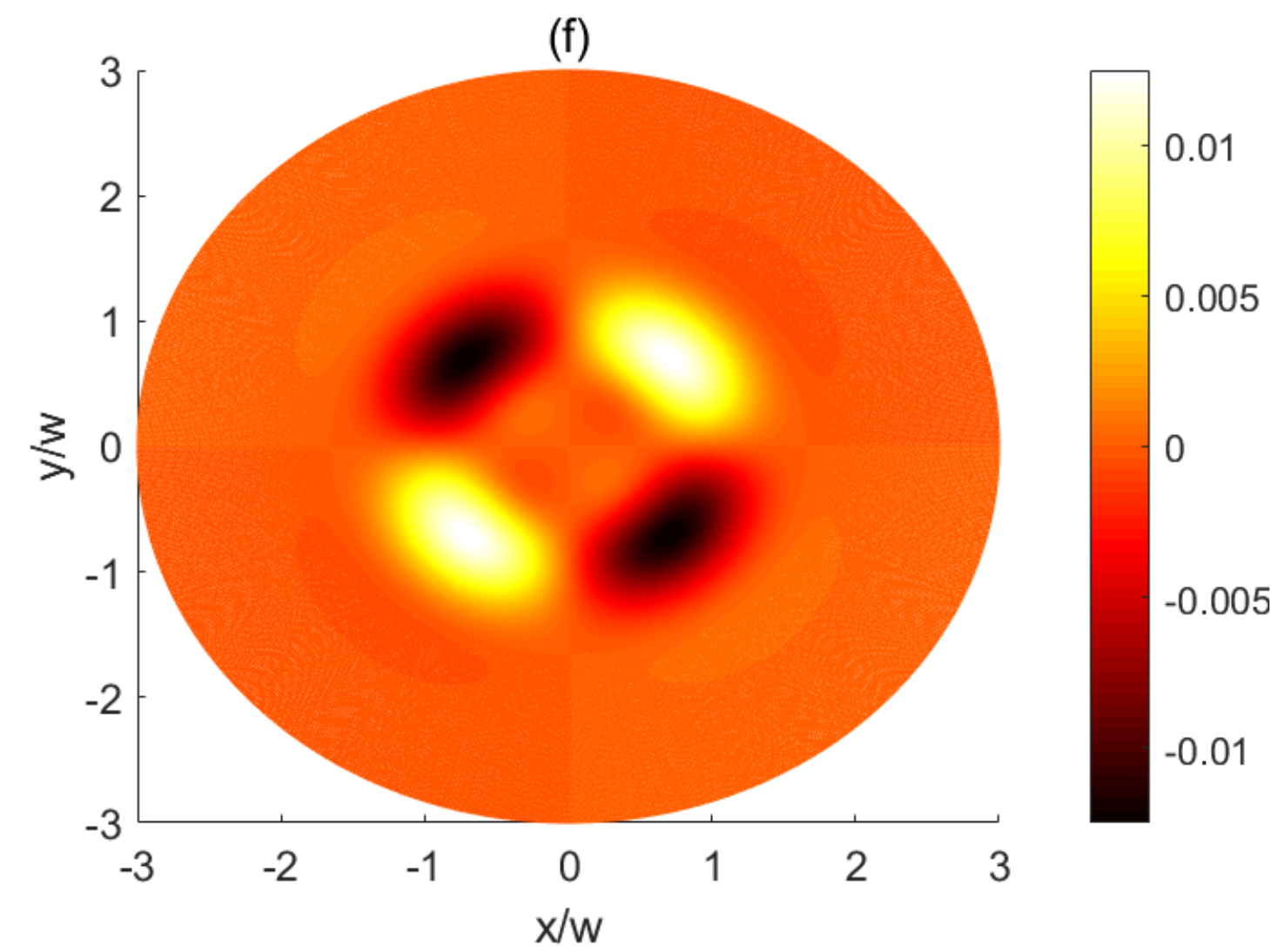}
\includegraphics[width=0.2\columnwidth]{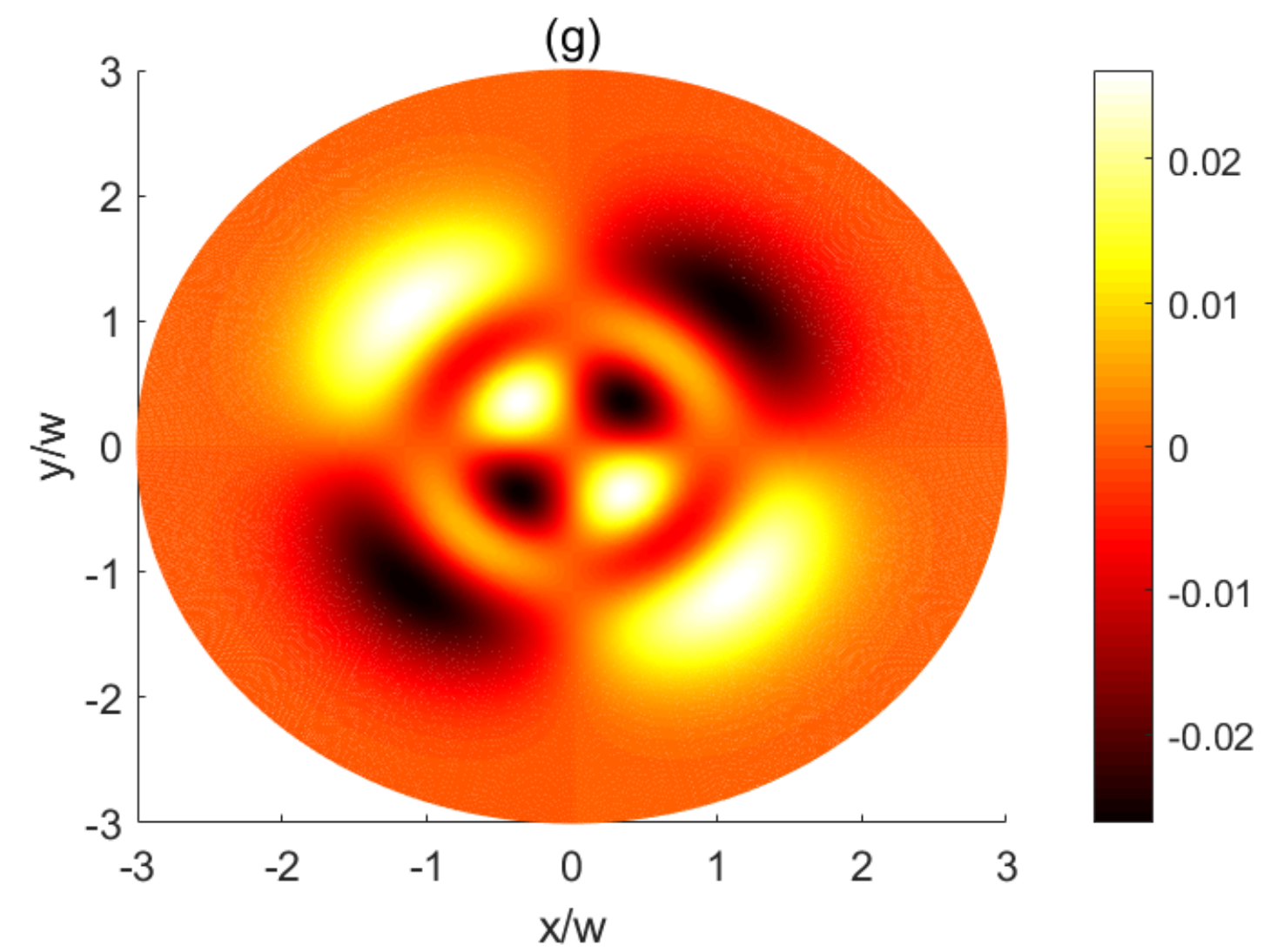} \includegraphics[width=0.2\columnwidth]{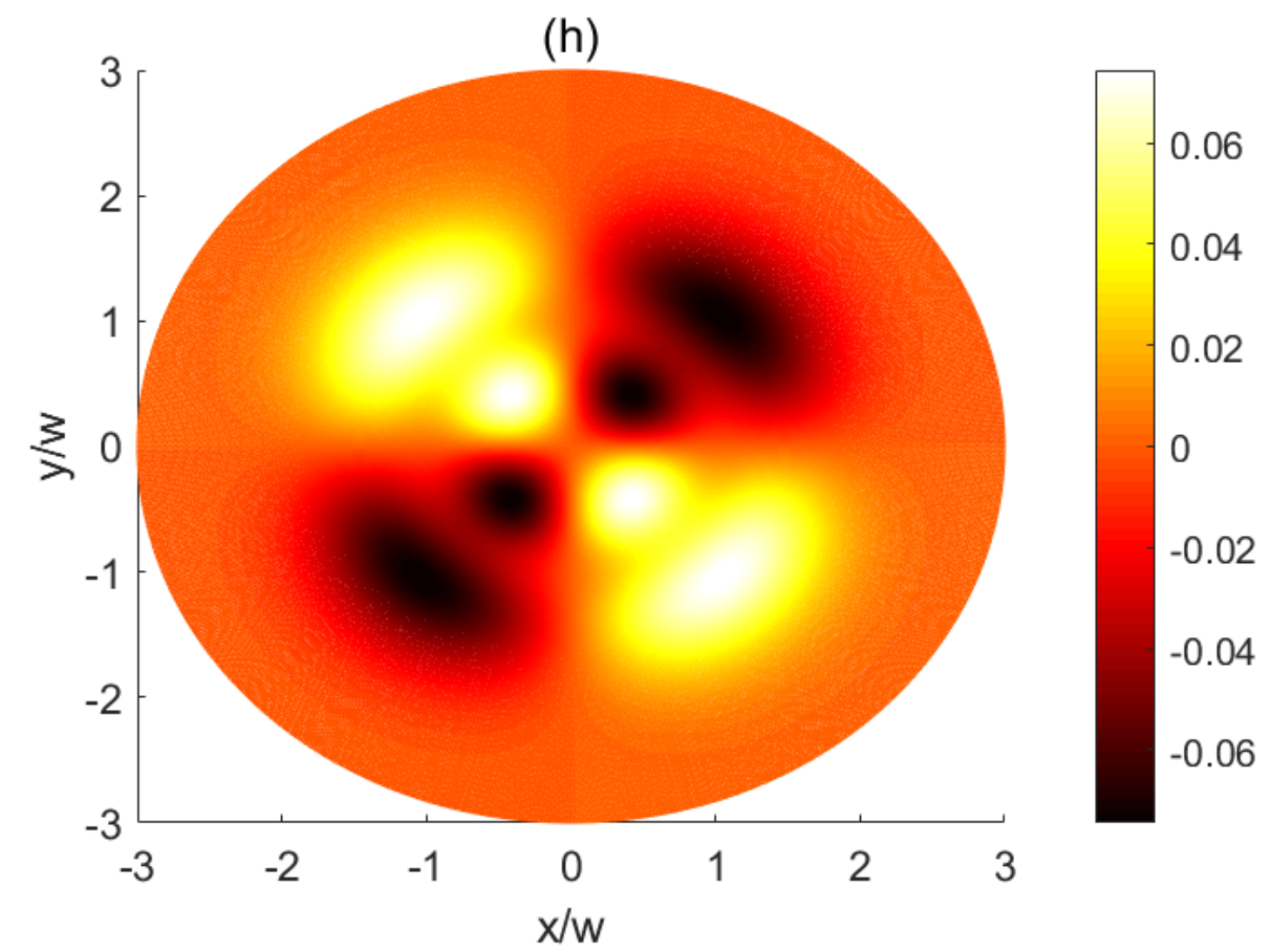}

\caption{Spatially structured Kerr nonlinearity $Re(\chi^{(3)})$
profiles of the probe beam $\Omega_{a}$ in arbitrary units, in the
presence of the plasmonic nanostructure and for different values of
distance $d$ of the quantum system from the plasmonic nanostructure:
$d=0.1c/\omega_{p}$ (a,e), $d=0.3c/\omega_{p}$ (b,f), $d=0.6c/\omega_{p}$
(c,g) and $d=0.8c/\omega_{p}$ (d,h). Here, the winding number is
$l=1$ for (a,b,c,d) and $l=2$ (e,f,g,h), and the other parameters
are the same as Fig.~\ref{fig:figs8}.}
\label{fig:figs10}
\end{figure}

\begin{figure}
\includegraphics[width=0.2\columnwidth]{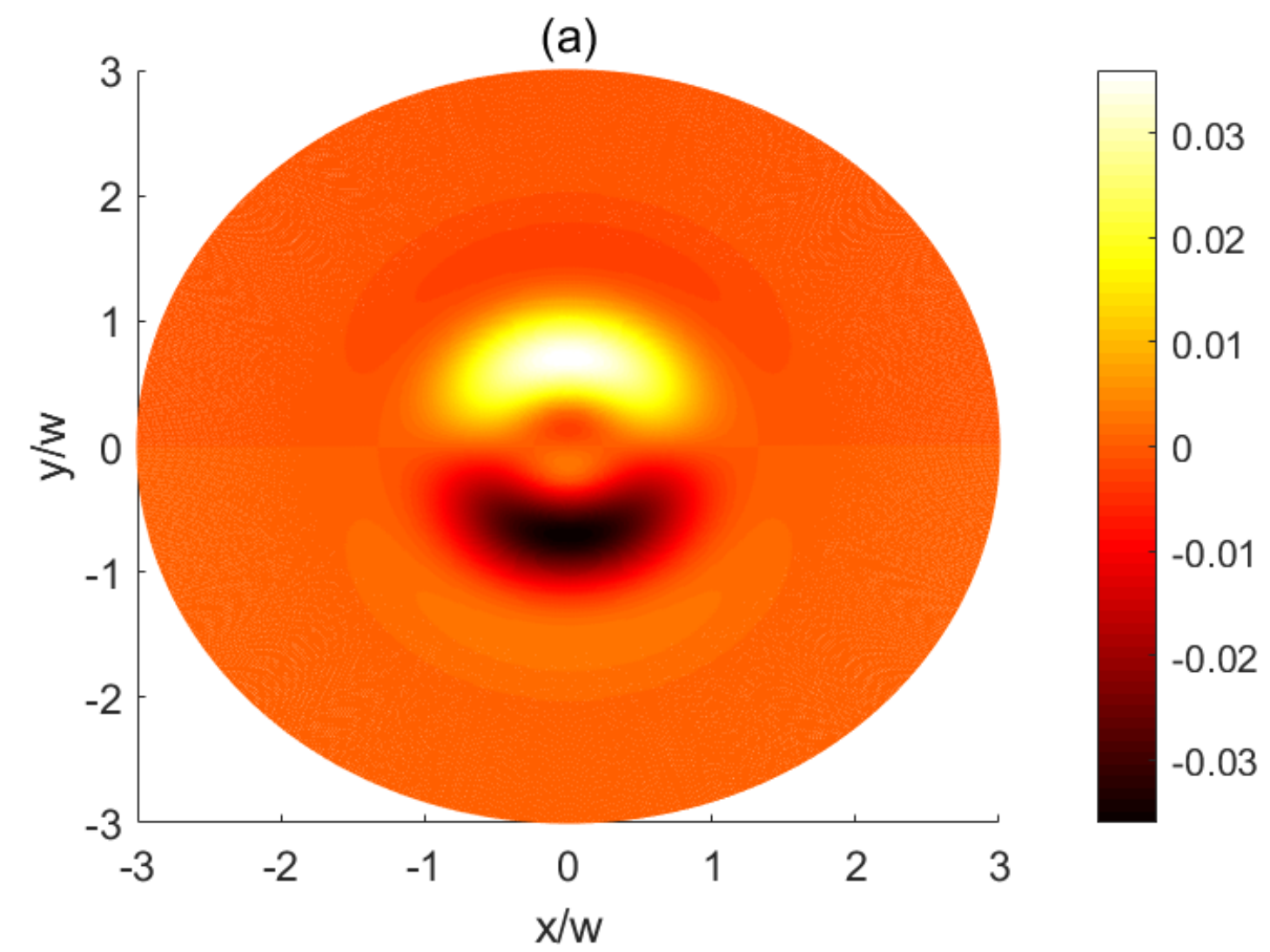} \includegraphics[width=0.2\columnwidth]{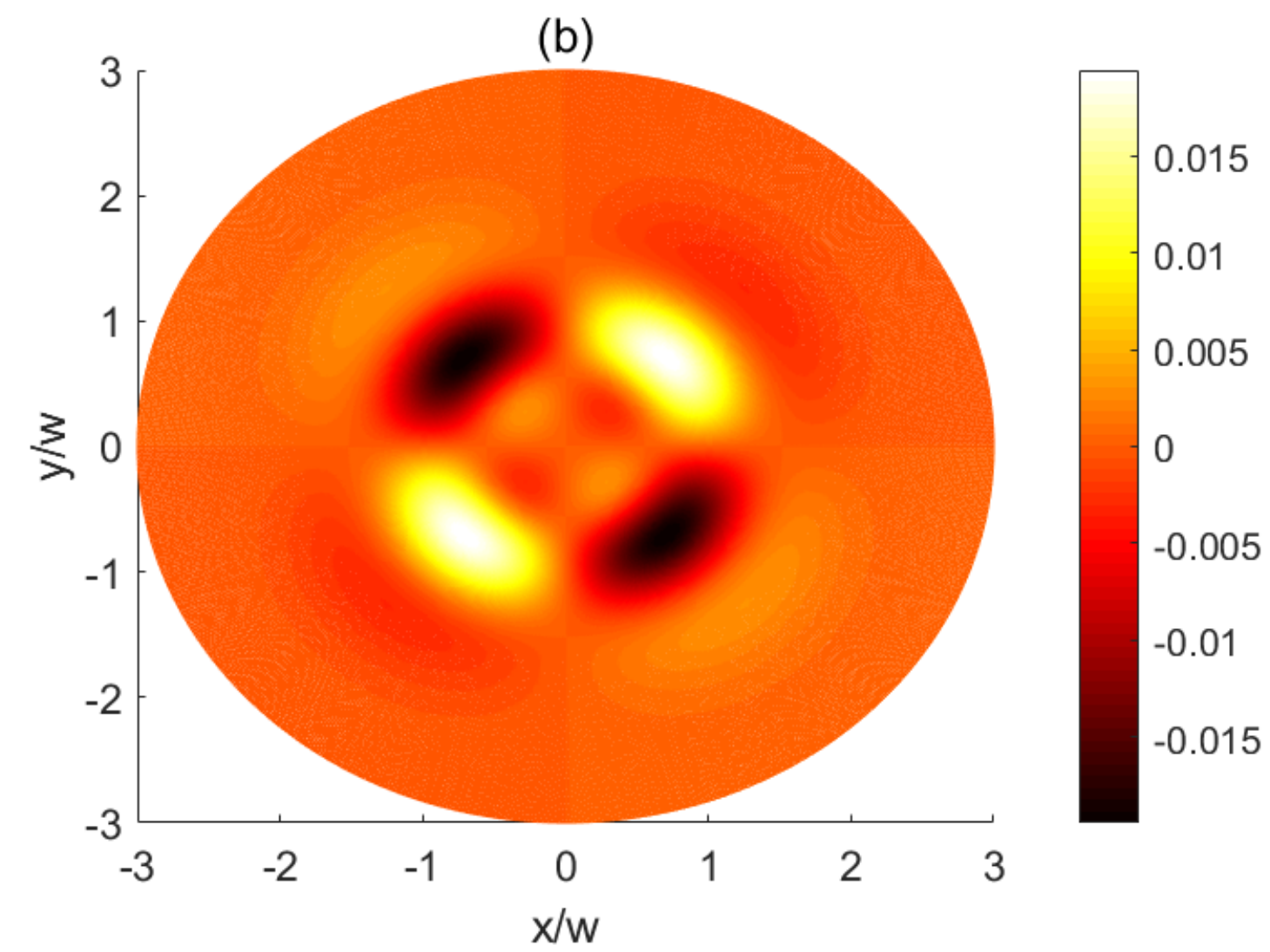}
\includegraphics[width=0.2\columnwidth]{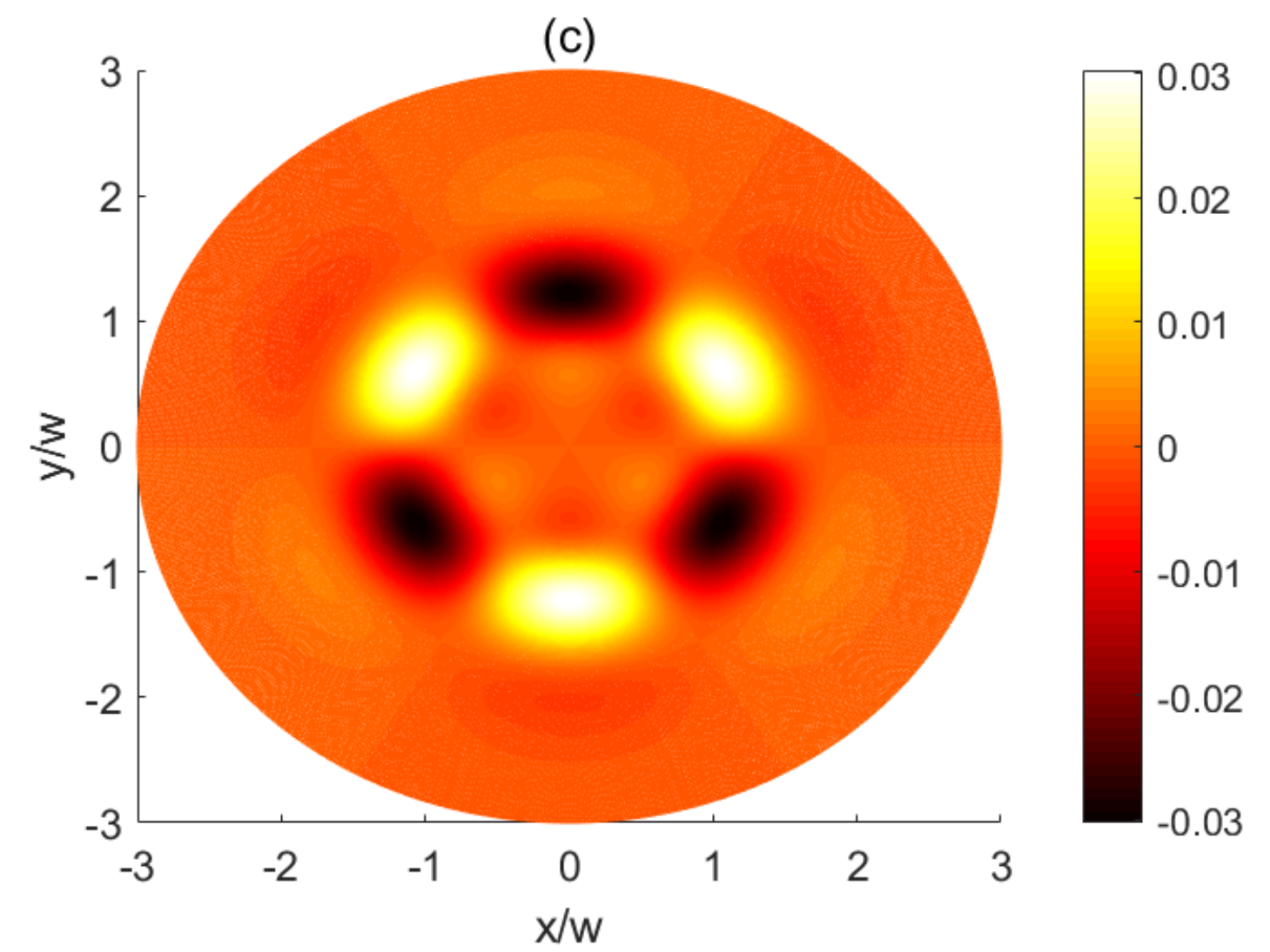} \includegraphics[width=0.2\columnwidth]{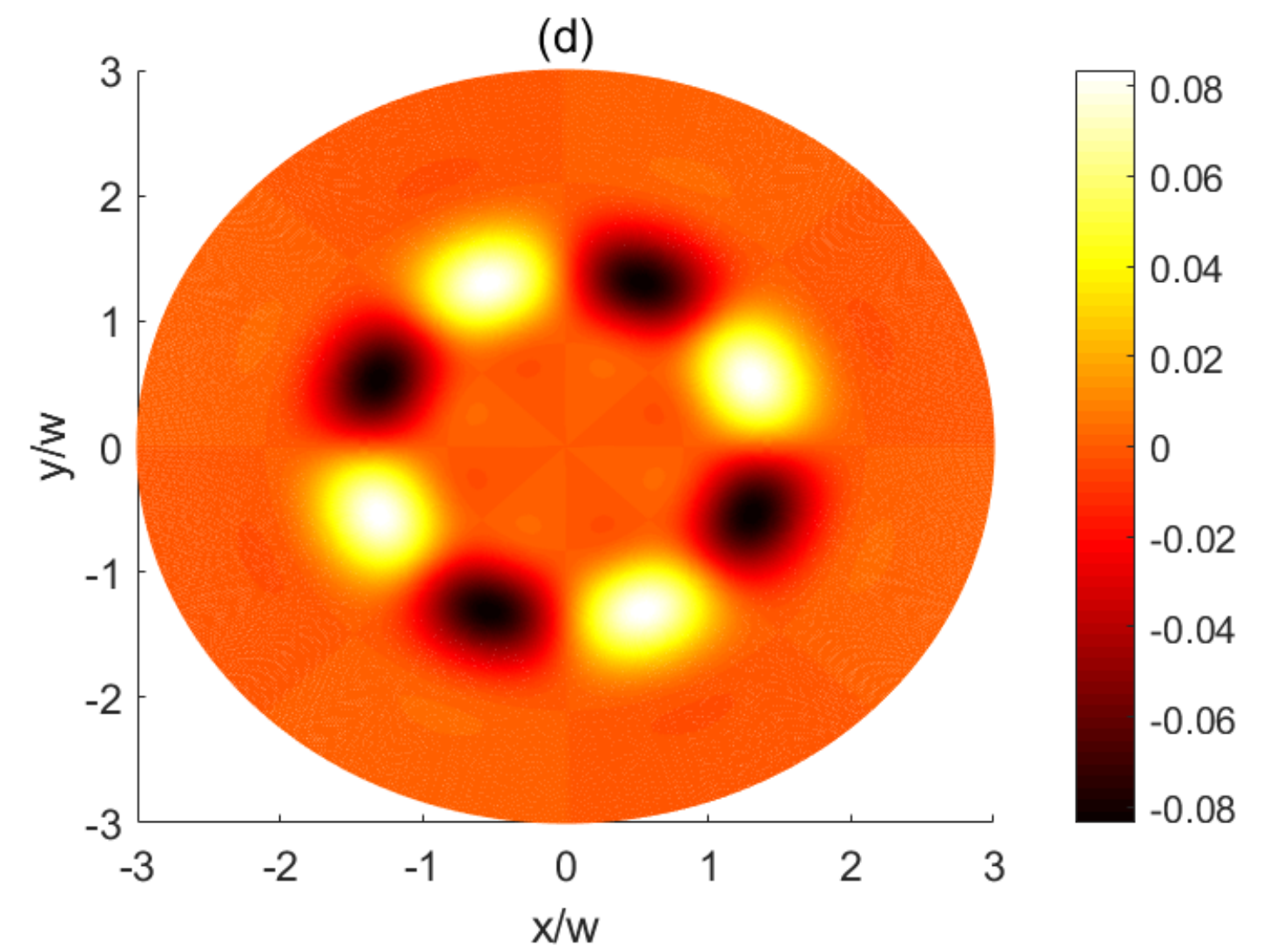}

\includegraphics[width=0.2\columnwidth]{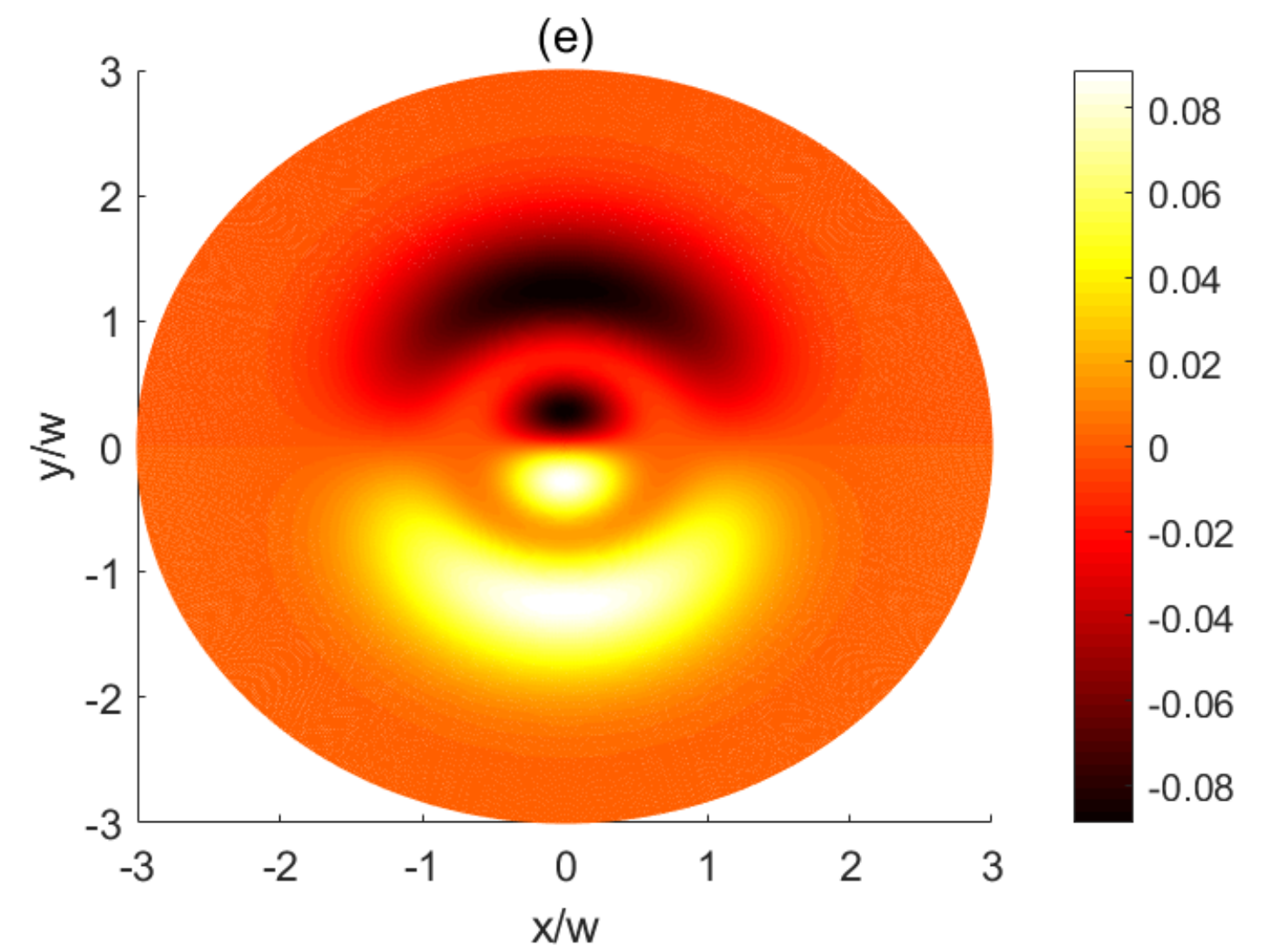} \includegraphics[width=0.2\columnwidth]{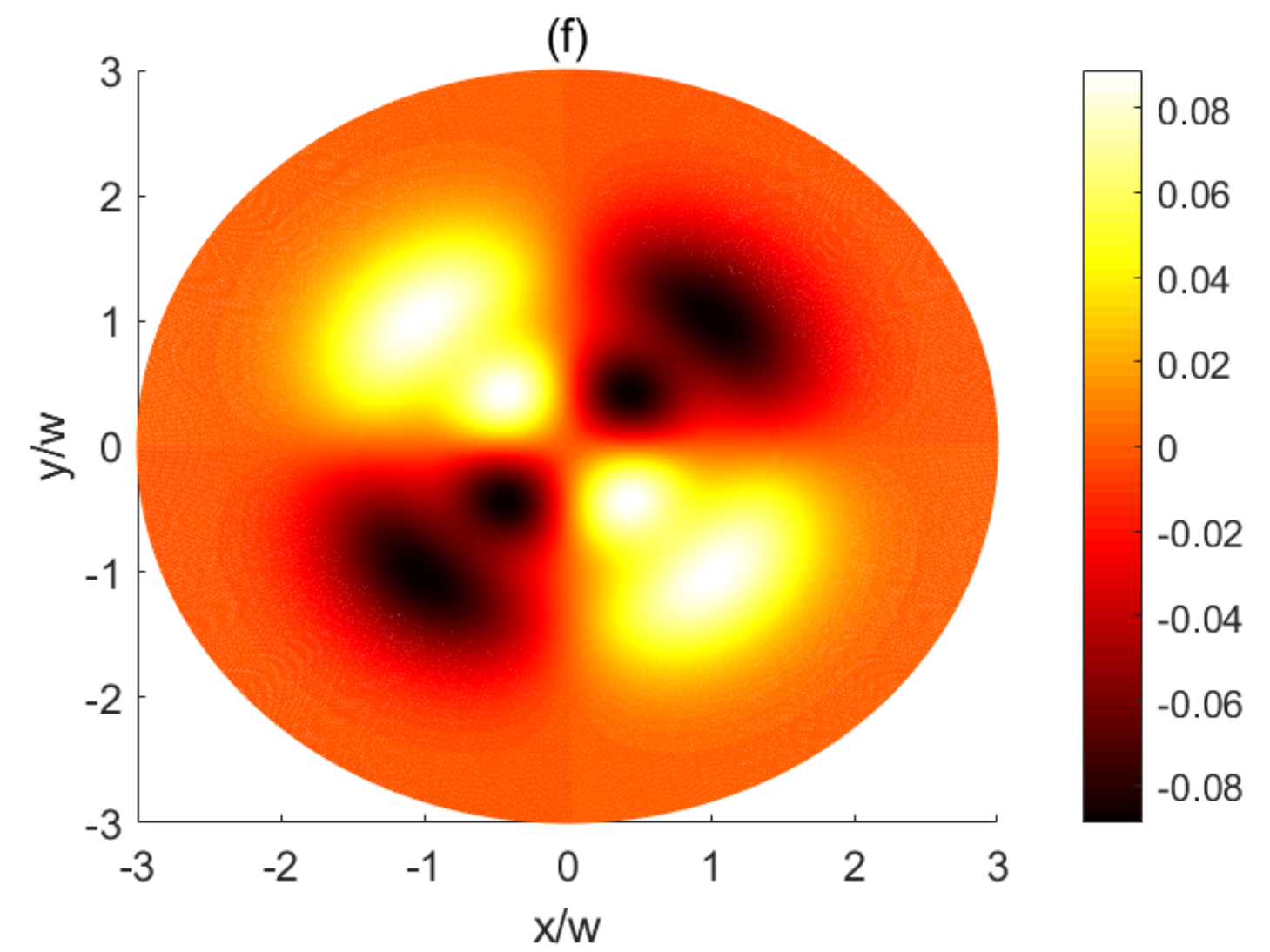}
\includegraphics[width=0.2\columnwidth]{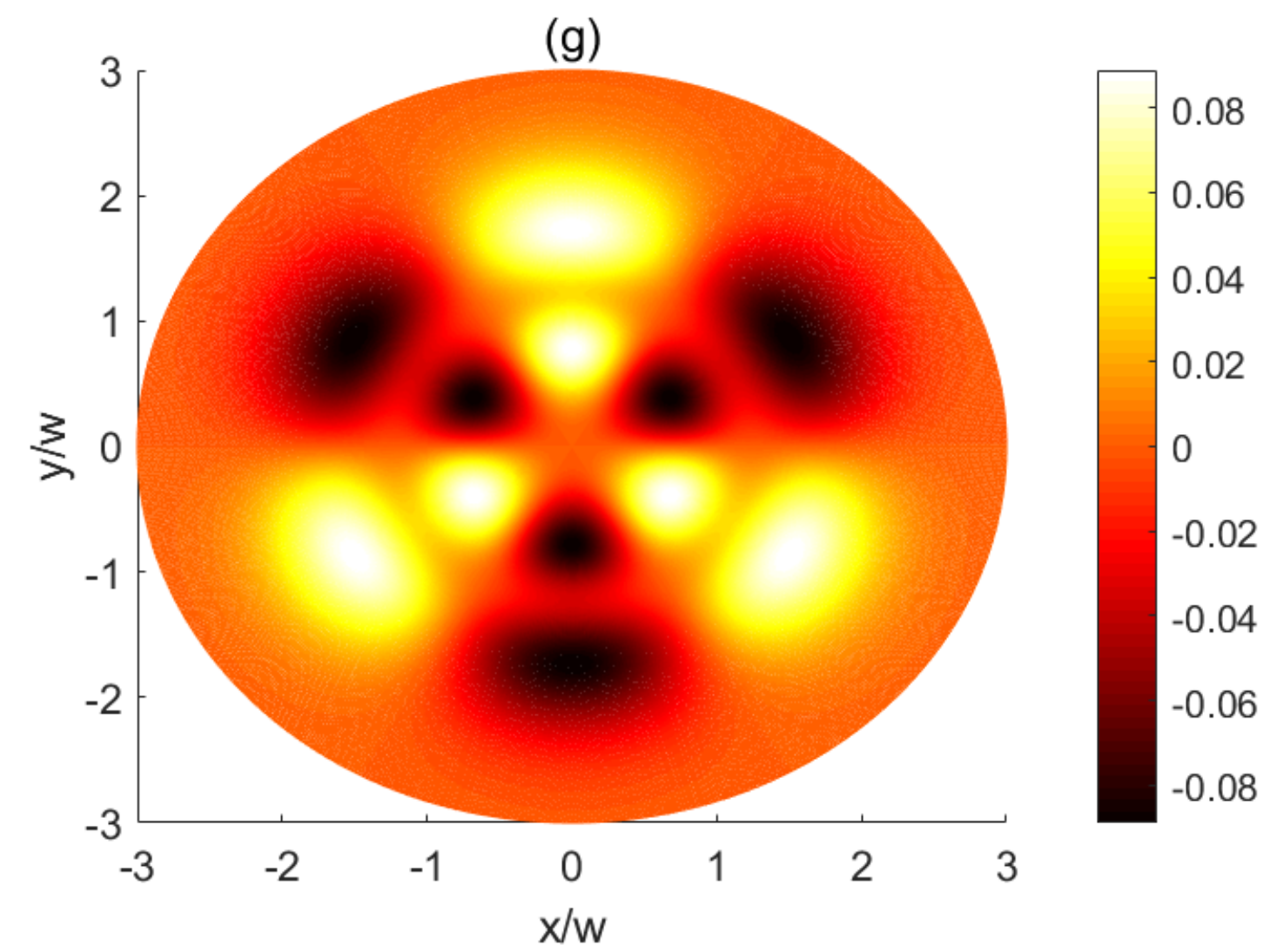} \includegraphics[width=0.2\columnwidth]{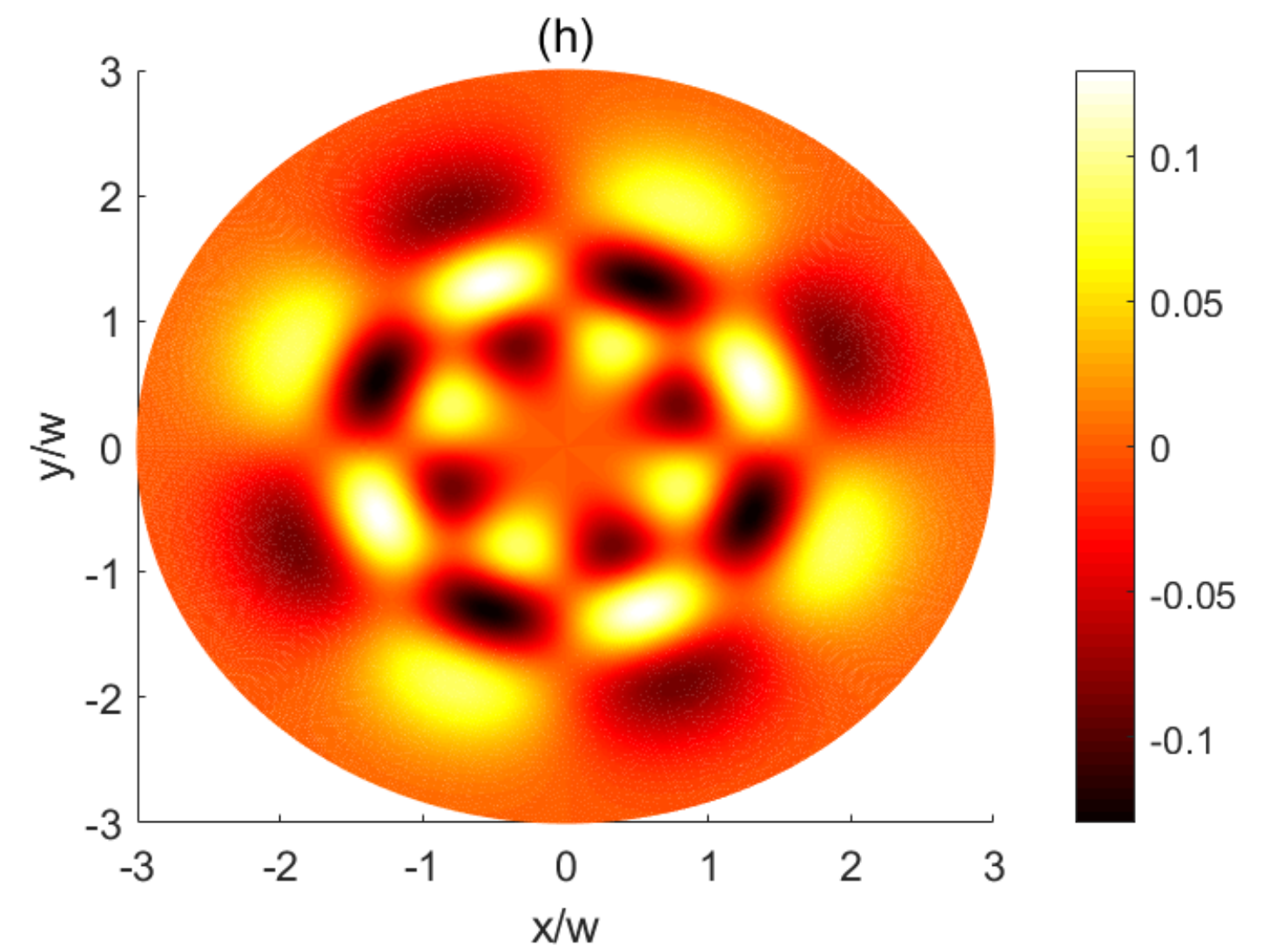}

\caption{Spatially structured Kerr nonlinearity $Re(\chi^{(3)})$ profiles
of the probe beam $\Omega_{a}$ in arbitrary units, in the presence
of the plasmonic nanostructure and for different winding $l=1$(a,e),$l=2$
(b,f), $l=3$ (c,g) and $l=4$ (d,h). Here, $d=0.4c/\omega_{p}$ (a,b,c,d),
$d=0.9c/\omega_{p}$ (e,f,g,h), and the other parameters are the same
as Fig.~\ref{fig:figs8}.}
\label{fig:figs11}
\end{figure}

In Figs.~\ref{fig:figs10} and \ref{fig:figs11}, we show the Kerr
nonlinearity of the medium as a function of the transverse
directions $x$ and $y$. As can be seen from Fig.~\ref{fig:figs10},
the spatially-dependent Kerr nonlinearity is very sensitive to the
distance from the plasmonic nanostructure. The Kerr nonlinearity is
remarkably enhanced when increasing the distance parameter $d$. In
particular, the maximal Kerr nonlinearity is enhanced by almost $10$
times when we increase $d$ from $0.1c/\omega_{p}$ [Figs.~\ref{fig:figs10}
(a,e)] to $0.3c/\omega_{p}$ [Figs.~\ref{fig:figs10} (b,f))]. Larger
values of $d$ mean higher values of the Kerr nonlinearity.
However, the maximal of Kerr nonlinearity is distributed to other regions
of the transverse plane [see (Figs.~\ref{fig:figs10} (c,d,g,f))].

Finally in Fig.~\ref{fig:figs11} we display how the winding number
affects the Kerr nonlinearity of the system. The results show a $l$-fold
symmetry of the Kerr nonlinearity. Moreover, very large position-dependent
Kerr nonlinearities can be achieved just by increasing the winding
number$l$.

Note that we have assumed the quantum system to be degenerated ($\omega_{32}=0$). We have also performed calculations with non-zero $\omega_{32}$ (not shown here).
We have observed a similar qualitative response for linear and nonlinear
susceptibilities for $\omega_{32} \neq 0$ with that presented above with $\omega_{32} = 0$. Yet, both the linear and nonlinear susceptibilities reduce in magnitude as $\omega_{32}$ increases.

\section{Concluding Remarks \label{sec:conc} }

We have studied the third-order nonlinear susceptibility behavior
of a four-level closed-loop double-V-type quantum system near a plasmonic
nanostructure. In the system under study, the lower V-type transition
interacts with the free-space vacuum, while the upper V-type transition
is affected by the interaction with localized
surface plasmons. Two orthogonal circularly polarized laser fields
with the same frequency and different phases and electric field amplitudes
act on both transitions of the lower V-type system. A 2D array of
metal-coated dielectric nanospheres is considered as a plasmonic nanostructure
for which the relevant decay rates are calculated by a rigorous electromagnetic
Green tensor technique.

We have shown that the presence of the plasmonic nanostructure
significantly modifies the nonlinear response of the system resulting
in large enhancement of the Kerr nonlinearity. In particular, the
Kerr nonlinearity can be remarkably modified by increasing the distance
of the quantum system from the plasmonic nanostructure. Phase control
of the Kerr nonlinearity has also been discussed for such a quantum
system. A wide range of tunability has been observed over the Kerr
nonlinear response through the effect of the relative phase. Such
a mechanism for phase control of the Kerr nonlinearity may be realized
by the state-of-the-art nanomethods and it may find application in
on-chip photonic nonlinear devices.

We have also analyzed the  light-matter interaction of the same system when one probe field carries an optical vortex, and
another probe field has no vortex. Because of the creation of quantum interference, the linear and nonlinear susceptibility of the nonvortex probe
beam depends on the azimuthal angle and the vorticity of the twisted
probe beam. This is different from an open double-V type quantum system interacting
with free-space vacuum, because no quantum interference occurs in
that case. Thanks to the angular dependence of the optical susceptibility
for the quantum system we can obtain regions of
high or low transmission as well as regions of large or small nonlinearity. We have then investigated the effect of different external parameters of
the system, i.e., the surface of plasmonic nanostructure and the vorticity
of twisted probe beam. The results obtained
here can be used in optoelectronics and quantum information processing
and may find potential applications in storage of high-dimensional
optical information in phase dependent quantum memories.

\section*{Acknowledgements} We acknowledge useful discussions with S.H. Asadpour.

\appendix

\section{Explicit expressions for $\rho_{ij}^{(2)}$ \label{sec:appendix-A}}

The expressions for the steady-state solutions $\rho_{ij}^{(2)}$
are:

\begin{equation}
\rho_{11}^{(2)}=-\frac{(r+s)}{2\gamma^{\prime}},\label{eq:AA1}
\end{equation}

\begin{equation}
\rho_{22}^{(2)}=\frac{2\gamma^{\prime}\kappa r+\kappa\gamma(r+s)}{4\gamma^{\prime}\kappa(\gamma+\gamma^{\prime})},\label{eq:AA2}
\end{equation}

\begin{equation}
\rho_{33}^{(2)}=\frac{2\gamma^{\prime}\kappa s+\kappa\gamma(r+s)}{4\gamma^{\prime}\kappa(\gamma+\gamma^{\prime})},\label{eq:AA3}
\end{equation}

\begin{equation}
\rho_{23}^{(2)}=\frac{\kappa(r+s)-2\gamma^{\prime}t}{2\gamma^{\prime}(i\omega_{32}-2\gamma-2\gamma^{\prime})},\label{eq:AA4}
\end{equation}

and $\rho_{00}^{(2)}=0$, where

\begin{equation}
t=\frac{i\Omega_{a}}{2}(-i\frac{\Omega_{b}}{2}e^{i\phi}S_{3}^{*}+i\kappa\frac{\Omega_{a}}{2}S_{2}^{*})-i\frac{\Omega_{b}}{2}e^{i\phi}(i\frac{\Omega_{a}}{2}S_{1}-i\kappa\frac{\Omega_{b}}{2}e^{-i\phi}S_{2}),\label{eq:AAp}
\end{equation}

\begin{equation}
r=\frac{i\Omega_{a}}{2}\left((-i\frac{\Omega_{a}}{2}S_{1}^{*}+i\kappa\frac{\Omega_{b}}{2}e^{i\phi}S_{2}^{*})-(i\frac{\Omega_{a}}{2}S_{1}-i\kappa\frac{\Omega_{b}}{2}e^{-i\phi}S_{2})\right),\label{eq:AAr}
\end{equation}
\begin{equation}
s=\frac{i\Omega_{b}}{2}\left((-i\frac{\Omega_{b}}{2}e^{i\phi}S_{3}^{*}+i\kappa\frac{\Omega_{a}}{2}S_{2}^{*})e^{-i\phi}-(i\frac{\Omega_{b}}{2}e^{-i\phi}S_{3}-i\kappa\frac{\Omega_{a}}{2}S_{2})e^{i\phi}\right).\label{eq:AAs}
\end{equation}

\section{Explicit expressions for $A$, $B$, $C$, $D$, and $f_{i}$\label{sec:appendix-B}}

The expressions for the coefficients $A$, $B$, $C$ and $D$
are:

\begin{align}
A & =e^{-i\phi}x,\label{eq:ABA}\\
B & =1,\label{eq:ABB}\\
C & =\frac{1}{8}\left[-\frac{2\gamma^{\prime}f_{1}+\gamma f_{2}}{4\gamma^{\prime}(\gamma+\gamma^{\prime})}-\frac{\kappa f_{3}+2\gamma^{\prime}f_{4}}{2\gamma^{\prime}(-i\omega_{32}-2\gamma-2\gamma^{\prime})}\right],\label{eq:ABC}\\
D & =\frac{1}{8}\left[-\frac{2\gamma^{\prime}f_{5}+\gamma f_{6}}{4\gamma^{\prime}(\gamma+\gamma^{\prime})}-\frac{\kappa f_{7}+2\gamma^{\prime}f_{8}}{2\gamma^{\prime}(i\omega_{32}-2\gamma-2\gamma^{\prime})}\right],\label{eq:ABD}
\end{align}

with

\begin{align}
f_{1} & =-x^{3}e^{-i\phi}(S_{3}+S_{3}^{*})+x^{2}\kappa e^{-2i\phi}S_{2}^{*}+\kappa x^{2}S_{2},\label{eq:ABf1}\\
f_{2} & =-xe^{-i\phi}(S_{1}+S_{1}^{*})+\kappa x^{2}(S_{2}+S_{2}^{*})-x^{3}e^{-i\phi}(S_{3}+S_{3}^{*})-\kappa x^{2}e^{-2i\phi}(S_{2}-S_{2}^{*}),\label{eq:ABf2}\\
f_{3} & =-(S_{1}+S_{1}^{*})+\kappa xe^{-i\phi}S_{2}+x\kappa e^{i\phi}S_{2}^{*}-x^{2}S_{3}+x\kappa S_{2}e^{i\phi}-x^{2}e^{i\phi}S_{3}^{*}+\kappa xS_{2}^{*}e^{-i\phi},\label{eq:ABf3}\\
f_{4} & =xe^{-i\phi}(S_{3}+S_{1}^{*})-\kappa S_{2}-x^{2}\kappa S_{2}^{*},\label{eq:ABf4}\\
f_{5} & =(S_{1}+S_{1}^{*})-x\kappa S_{2}^{*}-\kappa xS_{2}e^{-i\phi},\label{eq:ABf5}\\
f_{6} & =(S_{1}+S_{1}^{*})-\kappa x(e^{i\phi}S_{2}^{*}+S_{2}e^{-i\phi}+S_{2}^{*}e^{-i\phi}+S_{2}e^{i\phi})+x^{2}(S_{3}+S_{3}^{*}),\label{eq:ABf6}\\
f_{7} & =xe^{-i\phi}(S_{1}+S_{1}^{*})-\kappa x^{2}(S_{2}+S_{2}^{*})-\kappa x^{2}e^{-2i\phi}(S_{2}+S_{2}^{*})+x^{3}(e^{-i\phi}S_{3}+S_{3}^{*}),\label{eq:ABf7}\\
f_{8} & =-x^{2}S_{3}^{*}+x\kappa e^{-i\phi}S_{2}^{*}-x^{2}S_{1}+\kappa x^{3}S_{2}e^{-i\phi}.\label{eq:ABf8}
\end{align}

\section{Explicit expressions for the resonant coefficients $Im(\chi^{(1,3)}(\delta))$
and $Re(\chi^{(1,3)}(\delta))$\label{sec:appendix-C}}

Setting $\omega_{32}=0$ and $\delta=0$, the Eqs.~(\ref{eq:LS})
and (\ref{eq:NLS}) and their corresponding coefficients given in
Appendix~\ref{sec:appendix-B} simplify, resulting in the following analytical
expressions for the linear absorption/dispersion, and third-order (Kerr) nonlinear absorption/dispersion susceptibilities
\begin{equation}
Im(\chi^{(1)}(\delta=0))=\frac{N\mu^{\prime2}}{\varepsilon_{0}\hbar}\frac{\gamma+\gamma^{\prime}-\kappa x\cos(\phi)}{(\gamma+\gamma^{\prime})^{2}-\kappa^{2}},\label{eq:AC1}
\end{equation}

\begin{equation}
Re(\chi^{(1)}(\delta=0))=\frac{N\mu^{\prime2}}{\varepsilon_{0}\hbar}\frac{-\kappa A\sin(\phi)}{(\gamma+\gamma^{\prime})^{2}-\kappa^{2}},\label{eq:AC2}
\end{equation}

\begin{equation}
Im(\chi^{(3)}(\delta=0))=\frac{2N\mu^{\prime4}}{3\varepsilon_{0}\hbar^{3}}\frac{-m_{1}\cos(\phi)-m_{2}\cos(2\phi)-m_{3}}{32\gamma^{\prime}(\gamma+\gamma^{\prime})\left((\gamma+\gamma^{\prime})^{2}-\kappa^{2}\right)^{2}},\label{eq:AC3}
\end{equation}

\begin{equation}
Re(\chi^{(3)}(\delta=0))=\frac{2N\mu^{\prime4}}{3\varepsilon_{0}\hbar^{3}}\frac{-m_{4}\sin(\phi)-m_{2}\sin(2\phi)}{32\gamma^{\prime}(\gamma+\gamma^{\prime})\left((\gamma+\gamma^{\prime})^{2}-\kappa^{2}\right)^{2}},\label{eq:AC4}
\end{equation}

where

\begin{align}
m_{1}= & 4\gamma^{\prime}x^{3}\kappa(\gamma+\gamma^{\prime})-6x\kappa\gamma(\gamma+\gamma^{\prime})+2\gamma x^{3}\kappa(\gamma+\gamma^{\prime})-4\kappa^{3}x-\kappa^{2}x^{2}(\gamma+\gamma^{\prime})-7\gamma^{\prime}\kappa x(\gamma+\gamma^{\prime})\nonumber \\
- & 4\kappa x\gamma(\gamma+\gamma^{\prime})-2x\kappa(\gamma+\gamma^{\prime})^{2}-\kappa x^{3}(\gamma+\gamma^{\prime})^{2}-\gamma^{\prime}\kappa x^{3}(\gamma+\gamma^{\prime}),\label{eq:AC5}
\end{align}

\begin{equation}
m_{2}=2\kappa^{2}x^{2}\gamma^{\prime}+2\kappa^{2}x^{2}(\gamma+\gamma^{\prime}),\label{eq:AC6}
\end{equation}

\begin{equation}
m_{3}=2\kappa^{2}(\gamma+\gamma^{\prime})(1+2x^{2})+2\gamma^{\prime}\kappa^{2}(1-x^{2})-2\gamma^{\prime}\kappa x(\gamma+\gamma^{\prime})-\kappa x^{3}(\gamma+\gamma^{\prime})^{2}+4\gamma^{\prime}(\gamma+\gamma^{\prime})^{3}(x^{2}-1),\label{eq:AC7}
\end{equation}

\begin{equation}
m_{4}=3\kappa\gamma^{\prime}x^{3}(\gamma+\gamma^{\prime})+2\gamma x^{3}\kappa(\gamma+\gamma^{\prime})-\kappa^{2}x^{2}(\gamma+\gamma^{\prime})-5\gamma^{\prime}x\kappa(\gamma+\gamma^{\prime})-2x\kappa(\gamma+\gamma^{\prime})^{2}-\kappa x^{3}(\gamma+\gamma^{\prime}).\label{eq:AC8}
\end{equation}

\section{EM Green's tensor for a 2D periodic nanostructure\label{sec:appendix-D}}

The classical EM Green's tensor is defined through the following
equation:
\begin{equation}\label{eq00}
\nabla\times\nabla\times\mathbf{G}(\mathbf{r},\mathbf{r'};\omega)
-
k^2\mathbf{G}(\mathbf{r},\mathbf{r'};\omega)=\mathbf{1}_{3}\cdot\delta(\mathbf{r}-\mathbf{r'})\;,
\end{equation}
where $k=\sqrt{\epsilon_d}\omega / c$ is the wavevector inside the
material, $\omega$ is the angular frequency of incident light, $c$
is the speed of light in vacuum, and $\mathbf{1}_{3}$ is the
3$\times$3 unit matrix.

We deal with arrays of macroscopic spheres with 2D periodicity.
The method employed here is an EM Green's tensor formalism based
on an EM layer-multiple-scattering (LMS) method
\cite{Stefanou1998,Stefanou2000}. The LMS method is ideally suited
for the calculation of the transmission/reflection/ absorption
coefficients of an EM wave incident on slab containing a number of
planes of non-overlapping scatterers with the same 2D periodicity.
Namely, for each one plane of spheres, the method determines the
full multipole expansion of the total multiply scattered wave
field and deduces the corresponding transmission and reflection
matrices of the while slab in the plane-wave basis. Having
determined the transmission/ reflection matrices via the LMS
method one can the calculate the EM Green's tensor from
\cite{Sainidou2004,PaspalakisPRL2009}

\begin{eqnarray}
G^{EE}_{ii'}({\bf r}, {\bf r}';\omega)=&&g^{EE}_{ii'}({\bf r},
{\bf r}';\omega) - \frac{i} {8 \pi^{2}} \int \int_{SBZ} d^{2} {\bf
k}_{\parallel} \sum_{{\bf g}} \frac{1}{c^2 K^{+}_{{\bf
g}; z}} \times \nonumber \\
&&v_{{\bf g} {{\bf k}_{\parallel}}; i}({\bf r}) \exp(-{\rm i} {\bf
K}^{+}_{\bf g} \cdot {\bf r}) \hat{{\bf e}}_{i'} ({\bf
K}^{+}_{{\bf g}}) \; , \label{eq:gii}
\end{eqnarray}
with
\begin{equation}
v_{{\bf g} {{\bf k}_{\parallel}}; i}({\bf r})=\sum_{{\bf g}'}
R_{{\bf g}'; {\bf g}}(\omega, {\bf k}_{\parallel}) \exp(-{\rm i}
{\bf K}^{-}_{{\bf g}'} \cdot {\bf r}) \hat{{\bf e}}_{i} ({\bf
K}^{-}_{{\bf g}'})\; , \label{eq:refle}
\end{equation}
and
\begin{equation}
\mathbf{K}_{\mathbf{g}}^{\pm}= (
\mathbf{k}_{\parallel}+\mathbf{g},\ \pm [
q^{2}-\left(\mathbf{k}_{\parallel}+\mathbf{g}\right)^{2}]
^{1/2})\; . \label{eq:kg}
\end{equation}

The vectors ${\bf g}$ correspond to the reciprocal-lattice vectors
associated with the 2D periodic lattice of the plane of
scatterers. ${\bf k}_{\parallel}$ is the reduced wavevector which
lies within the surface Brillouin zone od the corresponding
reciprocal lattice \cite{Stefanou1998,Stefanou2000}. When
$q^{2}=\omega^2 / c^2 <(\mathbf{k}_{\parallel}+\mathbf{g})^{2}$,
$\mathbf{K}_{\mathbf{g}}^{\pm}$ defines an evanescent wave. The
term $g^{EE}_{ii'}({\bf r}, {\bf r}';\omega)$ of
Eq.~(\ref{eq:gii}) is the free-space Green's tensor and $\hat{{\bf
e}}_{i} ({\bf K}^{\pm}_{{\bf g}})$ denotes the polar unit vector
normal to ${\bf K}^{\pm}_{{\bf g}}$. $R_{{\bf g}'; {\bf
g}}(\omega, {\bf k}_{\parallel})$ is the reflection matrix which
provides the sum (over ${\bf g}$'s) of reflected beams generated
by the incidence of plane wave from the left of the plane of
scatterers and is calculated via the LMS method
\cite{Stefanou1998,Stefanou2000}. We note that the above
expression [Eqs.~(\ref{eq:gii})] is derived from the transverse
part of the general classical-wave Green's tensor
\cite{Sainidou2004}. Also, in Eq.~(\ref{eq:gii}), the terms
corresponding to $s$-polarized waves (those containing components
with the azimuthal unit vector $\hat{{\bf e}}_{i} ({\bf
K}^{\pm}_{{\bf g}})$ normal to ${\bf K}^{\pm}_{{\bf g}}$) have
very marginal contribution to the decay rates and have been,
justifiably, neglected.

\end{document}